\documentclass[useAMS,usenatbib]{mn2e}
\usepackage{times}
\usepackage{graphicx}
\usepackage{amsmath}
\usepackage{fmtcount}
\usepackage{amssymb}
\usepackage{natbib}
\usepackage{color}
\bibpunct{(}{)}{;}{a}{}{,}
\usepackage{psfig}
\title[Globular Clusters in Early-type Galaxies]{Gemini/GMOS Imaging of 
Globular Cluster Systems in Five Early-type Galaxies\thanks{
Based on observations obtained at the Gemini Observatory, which is operated by the 
Association of Universities for Research in Astronomy, Inc., under a cooperative agreement 
with the NSF on behalf of the Gemini partnership: the National Science Foundation (United 
States), the Science and Technology Facilities Council (United Kingdom), the 
National Research Council (Canada), CONICYT (Chile), the Australian Research Council (Australia), 
Minist\'{e}rio da Ci\^{e}ncia e Tecnologia (Brazil) 
and Ministerio de Ciencia, Tecnolog\'{i}a e Innovaci\'{o}n Productiva (Argentina).}
}
\author[Faifer et al.]{Favio R. Faifer$^{1,2}$\thanks{E-mail:favio@fcaglp.unlp.edu.ar}, Juan C. Forte$^{1}$, Mark A. Norris$^{3}$, Terry Bridges$^{4}$, Duncan A. Forbes$^{5}$,\newauthor Stephen E. Zepf$^{6}$, Mike Beasley$^{7}$, Karl Gebhardt$^{8}$, David A. Hanes$^{9}$, Ray M. Sharples$^{10}$\\
$^{1}$Facultad de Cs. Astron\'omicas y Geof\'isicas, UNLP, Paseo del Bosque 1900, La Plata, and CONICET, Argentina\\
$^{2}$ Instituto de Astrof\'isica de La Plata (CCT La Plata - CONICET - UNLP)\\
$^{3}$Dept. of Physics and Astronomy, University of North Carolina, Chapel Hill\\
$^{4}$Department of Physics, Engineering Physics, and Astronomy, Queen's University, Kingston, ON K7L 3N6, Canada\\
$^{5}$Centre for Astrophysics \& Supercomputing, Swinburne University, Hawthorn, VIC 3122, Australia\\
$^{6}$Department of Physics and Astronomy, Michigan State University, East Lansing, MI 48824, USA\\
$^{7}$Instituto de Astrof\'isica de Canarias, La Laguna 38200, Tenerife, Spain\\
$^{8}$Astronomy Department, University of Texas, Austin, TX 78712, USA\\
$^{9}$Department of Physics, Queen's University, Kingston, ON K7L 3N6, Canada\\
$^{10}$Department of Physics, University of Durham, South Road, Durham DH1 3LE}
\begin{document}

\date{Accepted 2011 May 5.  Received 2011 April 29; in original form 2010 December
17 }
\pagerange{\pageref{firstpage}--\pageref{lastpage}} \pubyear{}
\maketitle

\label{firstpage}

\begin{abstract}
This paper presents deep high quality photometry of globular cluster 
(GC) systems belonging to five early-type galaxies covering a range of mass
and environment. 
Photometric data were obtained with the Gemini North and Gemini 
South telescopes in the filter passbands {\it g$'$}, {\it r$'$}, 
and {\it i$'$}. The combination of these filters with good 
seeing conditions allows an excellent separation between GC
candidates and unresolved field objects. 
In fact, our previously published spectroscopic data indicates a 
contamination level of only $\sim$10 percent in our sample of GC candidates.
Bimodal GC colour distributions are found in all five galaxies. Most 
of the GC systems appear bimodal even in the {\it (g$'$-r$'$)} vs {\it (r$'$-i$'$)} 
plane. A population of resolved/marginally resolved GC and 
Ultra Compact Dwarf candidate{\bf s} was found in all the galaxies. 
A search for the so-called ``blue tilt'' in the colour-magnitude 
diagrams reveals that NGC 4649 clearly shows that phenomenon 
although no conclusive evidence was found for the other 
galaxies in the sample. This ``blue tilt'' translates into a 
mass-metallicity relation given by {\it Z $\propto$ $M^{0.28\pm0.03}$}.
This dependence was found using a new empirical {\it (g$'$-i$'$)} vs [Z/H] 
relation which relies on an homogeneous sample of GC colours and metallicities.
This paper also explores the radial trends in both colour and 
surface density for the blue (metal-poor) and red (metal-rich) GC 
subpopulations. As usual, the red 
GCs show a steeper radial distribution than the blue ones. 
Evidence of galactocentric colour gradients is found in 
some of the GC systems, being more significant for the two S0 
galaxies in the sample. Red GC subpopulations show similar
colours and gradients to the galaxy halo stars in their inner region.
A GC mean colour-galaxy luminosity relation, consistent with 
{\it [Z/H] $\propto L{_B}^{0.26\pm0.08}$}, is present for the red GCs.
An estimate of the total GC populations and specific frequency S$_N$ 
values is presented for NGC 3115, NGC 3923 and NGC 4649.

\end{abstract}

\section{Introduction}
\label{intro}

The idea that globular clusters (GCs) are good tracers of very early events in
the lives of galaxies was put forward many years ago 
(e.g. \citealt*{ELBS,SZ78}). A thorough description of all the
important aspects is given, for example, in \citet{BS06}. However, some
key issues, such as the relation between GCs and galaxy field stars, still 
remain as open questions (see \citealt{K2009}). If such a 
relation does exist,
some galaxy properties could be described in terms of the
characteristics of their globular cluster systems (GCS). A tentative approach
in this direction (\citealt*{FFG07,FVF09})
assumes that GCs formation efficiency depends on chemical abundance, and
leads to a quantitative connection between field stars and GCs. In
turn, this type of connection allows further analysis of the different 
GCs-galaxy formation scenarios found in the 
literature  (e.g. \citealt{AZ92,FBG97,BBFSF02,MG10}). 
   
In this work we present new photometry for the GCS of five 
early-type galaxies, using the imaging capabilities of the GMOS instruments 
on the Gemini telescopes. This is part of a long term study to obtain 
both imaging and spectra for these GCS. Our first spectroscopic 
results were published for NGC 3379  \citep{PBFBGFFZSHP2006}, 
NGC 3923 \citep{NSBGFPFFBZH2008} and NGC 4649 (\citealt{PBFPBGFFZSH2006,BGSFFBZFHP2006}). 
In particular, we aim  at a detailed characterization 
of the most relevant photometric features of each GCS, 
such as colour bimodality, galactocentric colour gradients, spatial 
distributions, and the presence (or absence) of the so called ``blue tilt'' in the 
colour-magnitude diagrams of the GCS. All of these results will be later 
analyzed in combination with the final spectroscopic sample. Additionally, 
a search for Ultra Compact Dwarf (UCD) candidates was performed in each galaxy.

The paper is organized as follows. The galaxy sample is described in section 2,
photometric observations and data handling in section 3, results in terms of
colour-magnitude and colour-colour diagrams, colour histograms, spatial 
distributions, galactocentric colour gradients and GC integrated 
luminosity functions are presented in section 4 and, finally, a 
summary is given in section 5.

\section{The galaxy sample}
\label{Gsample}

Our sample consists of five nearby early-type galaxies that cover a
range of mass. These galaxies have been
imaged in several filters over multiple pointings with the 
GMOS instrument on the Gemini telescopes. 
Basic data for the sample studied in this paper, ordered by decreasing 
luminosity (and presumably decreasing mass) are listed in Table
\ref{Tsample}. The sample comprises two S0 and three 
ellipticals (E), with {\it B} 
band luminosities in the range {\it M$_B$} 
= --19.94 to --21.43. They are located in field (NGC 3115), group 
(NGC 3923, NGC 524 and NGC 3379) and cluster (NGC 4649) 
environments. In this work we have adopted the distance modulus 
from the surface brightness fluctuation method presented by \citet{TDB2001}. 
This provides an homogeneous source which can easily
be converted to other distance scales, such as that of \citet{JTBTLRAB03}. 
In the following we give a short description of each 
galaxy with references to earlier work pertaining to their GCS.

\begin{table*}
\centering
\caption{Galaxy sample. Coordinates and Hubble types are taken from the
NASA Extragalactic Database, {\it B} magnitudes from the RC3 
catalogue \citep{dV91}, distance modulus from \citet{TDB2001}, extinction 
from \citet{SFD98} and X-ray luminosity is from \citet{OFP2001}.}
\label{Tsample}
\scriptsize
\begin{tabular}{lcccclcccccc}
\hline
\hline
\multicolumn{12}{c}{} \\
\multicolumn{1}{c}{\textbf{Galaxy}} &
\multicolumn{1}{c}{$\mathbf{\alpha}_{\mathbf{J2000}}$} &
\multicolumn{1}{c}{$\mathbf{\delta}_{\mathbf{J2000}}$} &
\multicolumn{1}{c}{\textbf{l}} &
\multicolumn{1}{c}{\textbf{b}} &
\multicolumn{1}{c}{\textbf{Type}} &
\multicolumn{1}{c}{\textbf{B${}_{\mathbf T}^{\mathbf 0}$}} &
\multicolumn{1}{c}{\textbf{M$_{\mathbf{B}}$} } &
\multicolumn{1}{c}{\textbf{A$_B$}} &
\multicolumn{1}{c}{\textbf{(m--M)$_0$} } &
\multicolumn{1}{c}{\textbf{Log L$_X$} } &
\multicolumn{1}{c}{\textbf{Environ} } \\
\multicolumn{1}{c}{} & 
\multicolumn{1}{c} {\it {(h:m:s)}} &
\multicolumn{1}{c}{($^\circ:':''$)} &
\multicolumn{1}{c} {\it {($^\circ:':''$)}} &
\multicolumn{1}{c}{($^\circ:':''$)} &
\multicolumn{1}{c}{} &
\multicolumn{1}{c}{({\it mag})} &
\multicolumn{1}{c}{({\it mag})} & 
\multicolumn{1}{c}{({\it mag})} &
\multicolumn{1}{c}{({\it mag})} & 
\multicolumn{1}{c}{({\it erg/s})} & 
\multicolumn{1}{c}{} \\
\hline \multicolumn{12}{c}{}\\
 NGC 4649 & 12:43:39.7 & +11:33:09.4      &  295:52:10.2 & $+$74:19:3.36  & E2 & ~9.70 & $-$21.43 & 0.116 & 31.13$\pm$0.15 & 41.28 & Cluster\\
 NGC 3923 & 11:51:01.8 & $-$28:48:22.0    &  287:16:34.0 & $+$32:13:21.4  & E4-5 & 10.62 & $-$21.18 & 0.362 & 31.80$\pm$0.28 & 40.66 & Group\\ 
 NGC 524 & 01:24:47.7 & +09:32:20.0       &  136:30:20.2 & $-$52:27:05.4  & S0 & 11.17 & $-$20.73 & 0.362 &31.90$\pm$0.20 & 40.53 & Group\\ 
 NGC 3115 & 10:05:14.1 & $-$07:43:07.0    &  247:46:58.8 & $+$36:46:53.0  & S0 & ~9.74 & $-$20.19 & 0.207 & 29.93$\pm$0.09 & 39.72 & Field\\
 NGC 3379 & 10:47:49.6 & +12:34:54.0      &  233:29:24.7 & $+$57:37:58.4  & E1 & 10.18 & $-$19.94 & 0.107 & 30.12$\pm$0.11 & $<$ 39.52 & Group\\ 
 Comparison & 22:37:24.2 & $+$22:48:20.5  & 86:44:10.8 & $-$30:29:07.5    &    &       &          &       &                &  \\
 WHDF       & 00:22:32.8 & $+$00:21:07.5  & 107:34:49.5 & $-$61:39:20.7   &    &       &          &       &                &  \\
\multicolumn{10}{l}{}\\
\hline
\multicolumn{10}{l}{}\\ 
\end{tabular}
\end{table*}

\vspace{0.5cm}
\noindent {\bf NGC 4649}
Also known as M60, this Virgo Cluster member is a giant elliptical (E2) 
within the Virgo
cluster and located in a sub-group to the East of the main cluster
concentration. NGC 4647, a spiral  some 2.5 arcmin to the
NW of NGC 4649, appears to be a background object since there is no
evidence of reddening produced by this spiral on NGC 4649
(\citealt{WKC2000}) and no evidence of interaction between them.
The GMOS fields presented here, and in our earlier work 
(\citealt{FFFBBGHSZG04}), were chosen to minimize the effect of 
possible contamination of the NGC 4649 GC system by NGC 4647 GCs.

With a luminosity of {\it M$_B$} = --21.43, NGC 4649 is the most massive galaxy 
in our sample. Based on spectra of the central region, \citet{TF2002}
used Lick indices to measure an old
age of 11 Gyr and a metallicity of [Fe/H] = 0.3. NGC 4649 shows
strong X-ray emission arising from a hot gas halo (\citealt{OFP2001}). 
Chandra observations reveal the presence
of numerous discrete sources that are mostly identified as low
mass binaries (LXMBs) within NGC 4649's globular clusters
(\citealt{SKISBR2003,RSI2004}). Using
the XMM satellite, \citet{RSI2006} present a hot gas
density profile of slope --1.5 $\pm$ 0.1, determined between radii 
of 10 and 300 arcsec. \citet{FBPOK2006} included this galaxy 
in their study of 53 ellipticals 
 using archival Chandra X-ray data. Fitting a $\beta$ 
model to the X-ray surface brightness profile they obtained a 
value of $\beta$=0.512, corresponding to a power-law slope of --2.07.

The NGC 4649 GCS was studied by \citet{CHA91} in the  
{\it B} and {\it V} bands. Within the small CCD (2.1 x 3.4 arcmin)
area, no bimodality was detected and the authors derived a mean
colour of {\it (B-V)} = 0.75 based on 82 globular cluster candidates. 

Bimodality in the GCs colour distribution was detected by 
\citet{LBHFG2001} on the basis of HST/WFPC2
observations that found colour peaks at {\it (V-I)$_0$} = 0.95 
and 1.26. This bimodality was confirmed by \citet{FFFBBGHSZG04} 
in the {\it g$'$} and {\it i$'$} bands using Gemini/GMOS imaging. These 
authors estimated a total GC population of 3700 $\pm$ 900, 
giving a specific frequency  S$_N$ = 4.1 $\pm$ 1. These values are 
in good agreement with those presented by \citet{LPKHKG2008} 
in their Washington CT$_1$ photometric study.
\citet{PBFPBGFFZSH2006} found that most of their 38 spectroscopically-confirmed GCs are old ($>$ 10 Gyr), but 
they found some young and super-solar metallicity GCs. This sample was used by 
\citet{BGSFFBZFHP2006} to analyze the GC kinematics and dark matter content
of NGC 4649 within 3.5 effective radii (20 kpc). No rotation of the GCS
 was detected, and both isotropic and axisymmetric orbit-based models 
gave support for the presence of a dark matter halo in NGC~4649.
 
\vspace{0.5cm}
\noindent {\bf NGC 3923}
This is an E4-5 galaxy within a small group and exhibits a well known
shell structure (\citealt{MC1980}). These kind of structures are frequent in 
ellipticals and S0 galaxies not belonging to galaxy clusters
(\citealt{SchwS92}), and they are usually identified as the
result of an interaction or merger (e.g. \citealt{TW1990,Q1984}). 
These shell structures may be long-lived and therefore an interaction/merger 
event may have occurred long time ago.

From spectroscopic data, \citet{TMBO05} and \citet{DTT2005} derive
a central galaxy age of 3.4 and 2.6 Gyrs respectively. Thus about 3 Gyrs ago,
the galaxy appears to have experienced a central starburst (perhaps
induced by the same event that created the shells). The galaxy has a
halo of hot gas (e.g. \citet{OFP2001}). Using the ROSAT satellite,
\citet{BC1999} fit a $\beta$ model to the X-ray surface density 
profile out to 500 arcsec and derived a power-law 
slope of --1.76 $\pm$ 0.02. \citet{FBPOK2006} fitted a double $\beta$ model 
to the X-ray surface brightness and found $\beta$ values of 0.598/0.314 for 
the less/more extended component. These values correspond to power-law slopes of --2.54/--0.88.  
  
Two previous photometric studies (\citealt{ZGA94,ZAG95})
using ground-based Washington {\it C} and {\it T$_1$} photometry, noted the presence of 
colour peaks at
{\it (C-T1)} = 1.47 and 1.87, somewhat redder than typical values. This led
these authors to suggest a higher than average metal content
for the NGC 3923 GCs. The total globular cluster population was estimated
to range from 2000 to 7000 and the S$_N$, from 4.4 to 7.3.
More recently \citet{SPCVB2006} using HST/ACS imaging measured similar blue and
red GC projected density slopes of --0.87. However, their fits were 
obtained within the central
100 arcsec and could be affected by the flattening of the inner GC profiles
as seen in many other galaxies. They obtained a S$_N$ value of 5.6, and 
concluded that NGC 3923 has the highest S$_N$  of any isolated elliptical.

\citet{NSBGFPFFBZH2008} found that the GCs spectroscopically examinated 
are old ($>$ 10 Gys) with $[Z/H]=$ -1.8 to 0.35. Additionally, they 
show that the diffuse light of the galaxy at $3R_e$  
is found to have ages, metallicities and [$\alpha$/Fe] abundance ratios 
indistinguishable from those of the red GCs. 

There are two spectroscopically confirmed UCDs in this galaxy, 
found by \cite{NK2011}.

\vspace{0.5cm}
\noindent {\bf NGC 524}
This galaxy is part of a group that includes at least eight smaller
galaxies known as CfA 13 (\citealt{GH83}). This group was 
detected in the X-ray study by \citet{MDMB2003}. The $\beta$ parameter 
fitted to this galaxy is 0.41$\pm$0.02, which translates into a power-law slope of --1.46$\pm$0.12.
The galaxy nucleus appears to be chemically decoupled from its bulge, 
being about 0.5--0.6 dex more metal-rich according to \citealt{SAV92} 
and \citet{S2000}, who also find evidence for a dusty disk and ionized 
gas within 3 kpc of the galaxy centre.

\citet{HH85} found a rich and extended GCS with
some 2830 $\pm$ 880 clusters. \citet{LBHFG2001} used a KMM
analysis on HST/WFPC2 data to find colour peaks at {\it (V-I)$_0$} = 0.98
and 1.19 mag. More recently, \citet{BFBK2004} found a broad {\it (V-I)} colour
distribution and the possible existence of peaks at {\it (V-I)$_0$} = 0.9 and
1.1. However, none of these results are conclusive.
Beasley et al.'s Keck/LRIS spectra of a small sample of GCs points to a 
metallicity range from [Fe/H] = --2.0 to 0.0 and a decreasing
trend of the [$\alpha$/Fe] ratio with metallicity. The GCs are
generally consistent with being old. 
Regarding UCDs, there is one confirmed object by 
Norris \& Kannappan (private communication).

\vspace{0.5cm}
\noindent {\bf NGC 3115}
This is a relatively isolated S0 galaxy with a dominant bulge, located
to the South of the Leo group.  The X-ray emission from NGC 3115 is low, 
contained within 10 kpc, and probably dominated by discrete 
sources (\citealt{FBPOK2006}).
Its GC system is bimodal with colour peaks at {\it (V-I)$_0$} = 0.96 and 1.17, 
and has a low local S$_N$ value of $\sim$ 1.3 within the inner region of the
galaxy (\citealt{KW1998}). These authors also find that while the 
red GCs subpopulation
has a spatial distribution comparable to that of the thick disk, the
blue GCs exhibit a distribution more similar to the bulge/halo.

Selecting from the \citet{KW1998} GCs sample, \citet{KZSWF2002} 
derived spectroscopic ages and metallicities for 17 GCs, finding both 
subpopulations to be $\sim$12 Gyr old and coeval within 2 to 3 Gyr. This 
result is consistent with \citet{PZKHMG2002} who also find coeval 
ages within 3 Gyr. 

From GMOS spectroscopy of the galaxy halo, \citet{NSK2006} calculated an
age of 10 to 12 Gyr, similar to that of the GCs. They also obtain 
 an [$\alpha$/Fe] ratio of $\sim$ 0.2--0.3, consistent with the typical GC
values. 

\vspace{0.5cm}
\noindent {\bf NGC 3379}
This galaxy is the dominant object of the nearby Leo group.
\citet{TF2002} derive an old central age of $\sim$9.3 Gyr, consistent with ages 
larger than 8 Gyr determined through HST-NICMOS observations of resolved 
stars by \citet{GFMTC2004}. The galaxy contains a very modest hot gas 
halo, and \citet{RCSFJ2008} suggest that a significant contribution to 
the observed X-ray flux of the galaxy comes from unresolved point sources.

From a CCD study of the NGC\,3379 GC system, \citet{AT94} detected $\sim$  60 
GCs but no evidence of bimodality. Using HST/WFPC2, \citet{LBHFG2001}
confirmed that NGC 3379 contains a poor GCS.
Both \citet{WFB2003} and \citet{RZ2004}
found evidence of bimodality using broad band colours.
These last authors estimated a total GC population of 270 clusters giving 
a low S$_N$ value of 1.2 $\pm$ 0.3.
 
In their spectroscopic study, \citet{PBFBGFFZSHP2006} found the GCs to be 
consistent with old ages, i.e. 10 Gyr, with a wide range of metallicities.

\vspace{0.5cm} 
\begin{figure}
\caption{R-band images of the sample galaxies from the DSS2. Each image 
has a field-of-view of  $15 \times 15$ arcmin. From top to bottom we show 
NGC 4649, NGC 3923, NGC 524, NGC 3115 and NGC 3379. For 
each galaxy, we display the red GC candidates in the right panels, and the
blue GC candidates in the left panels. The GMOS fields are also depicted.  
North is up and East, to the left.}
\resizebox{0.5\hsize}{!}{\includegraphics{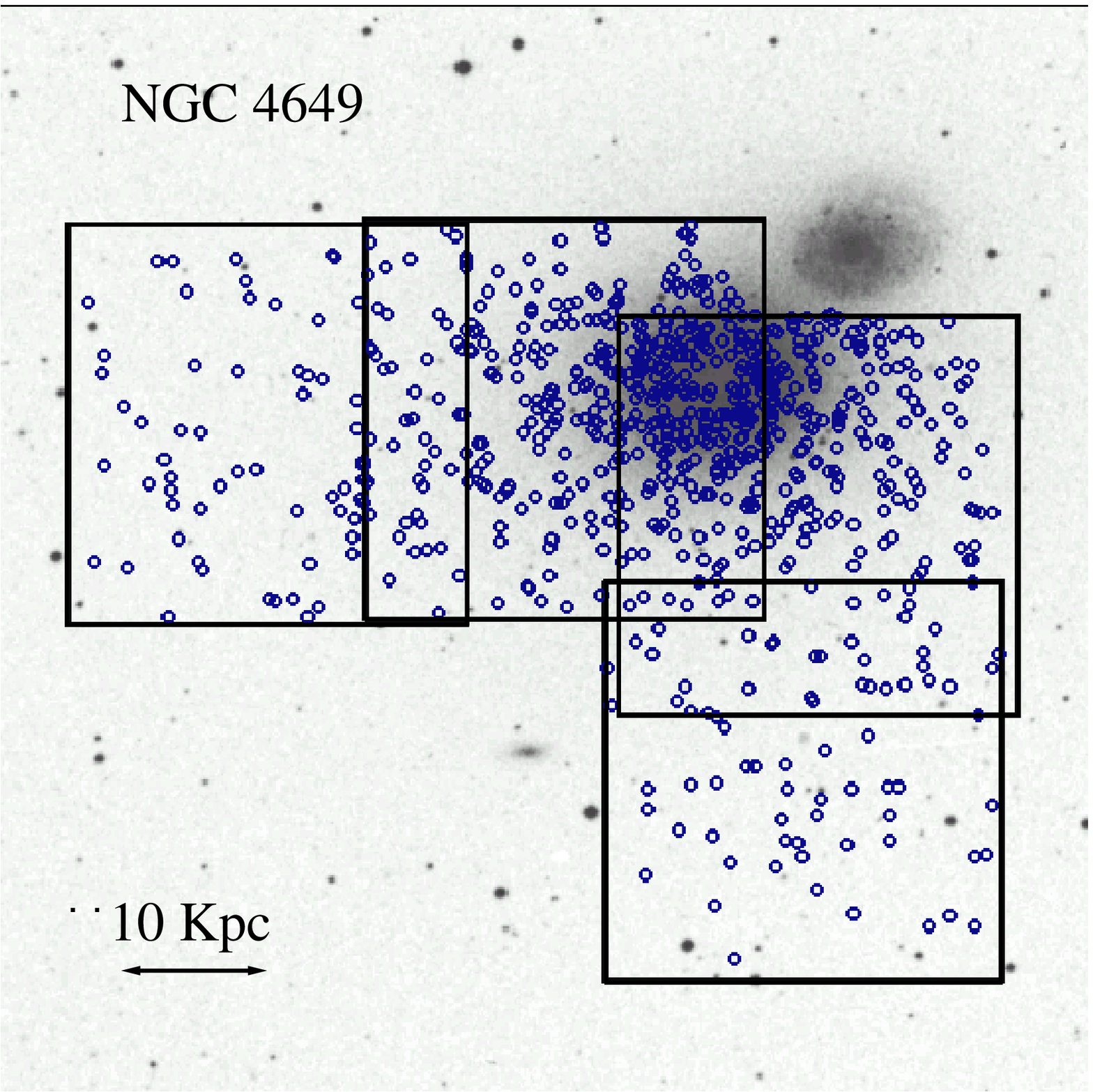}}
\resizebox{0.5\hsize}{!}{\includegraphics{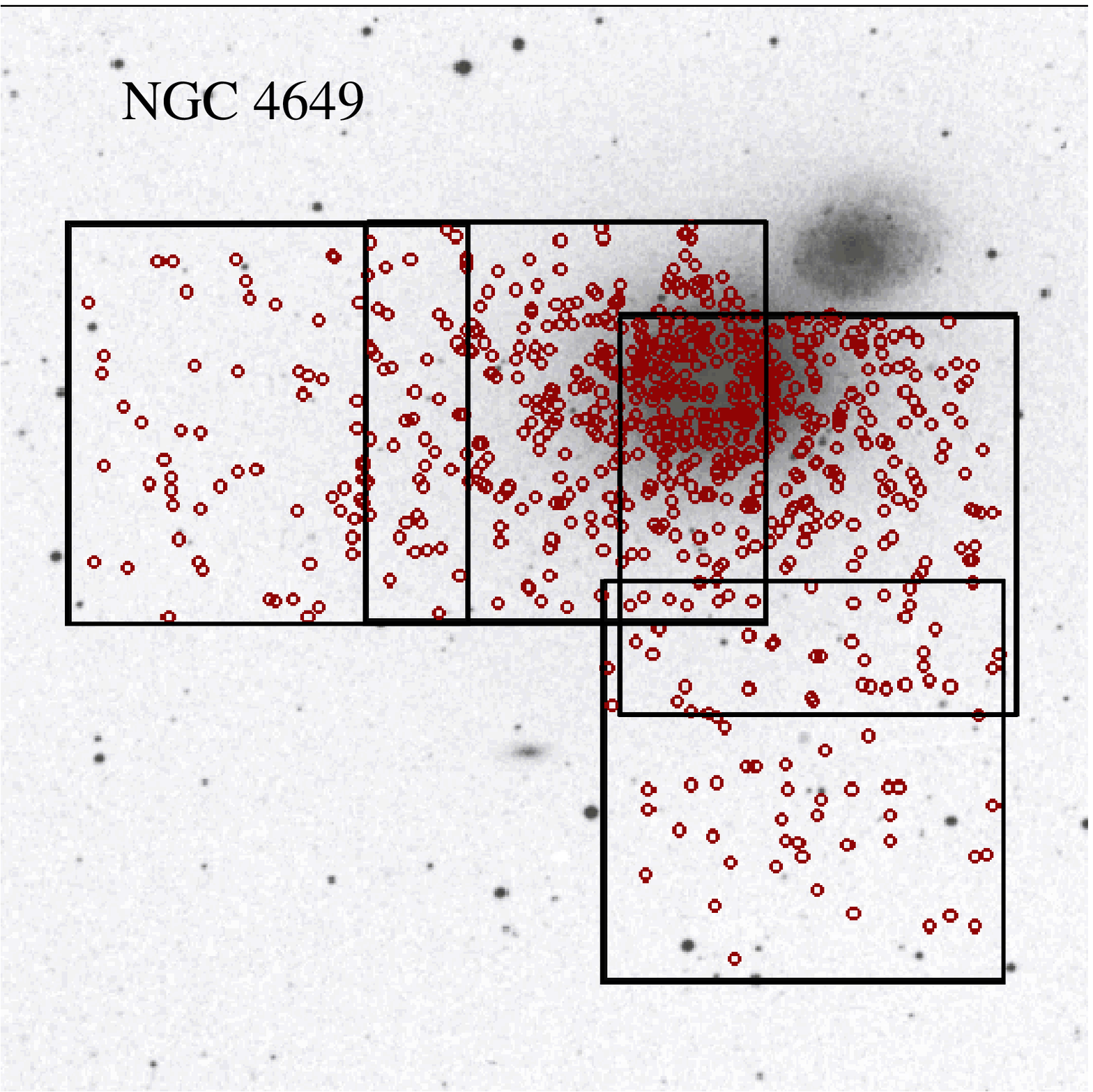}}\\
\resizebox{0.5\hsize}{!}{\includegraphics{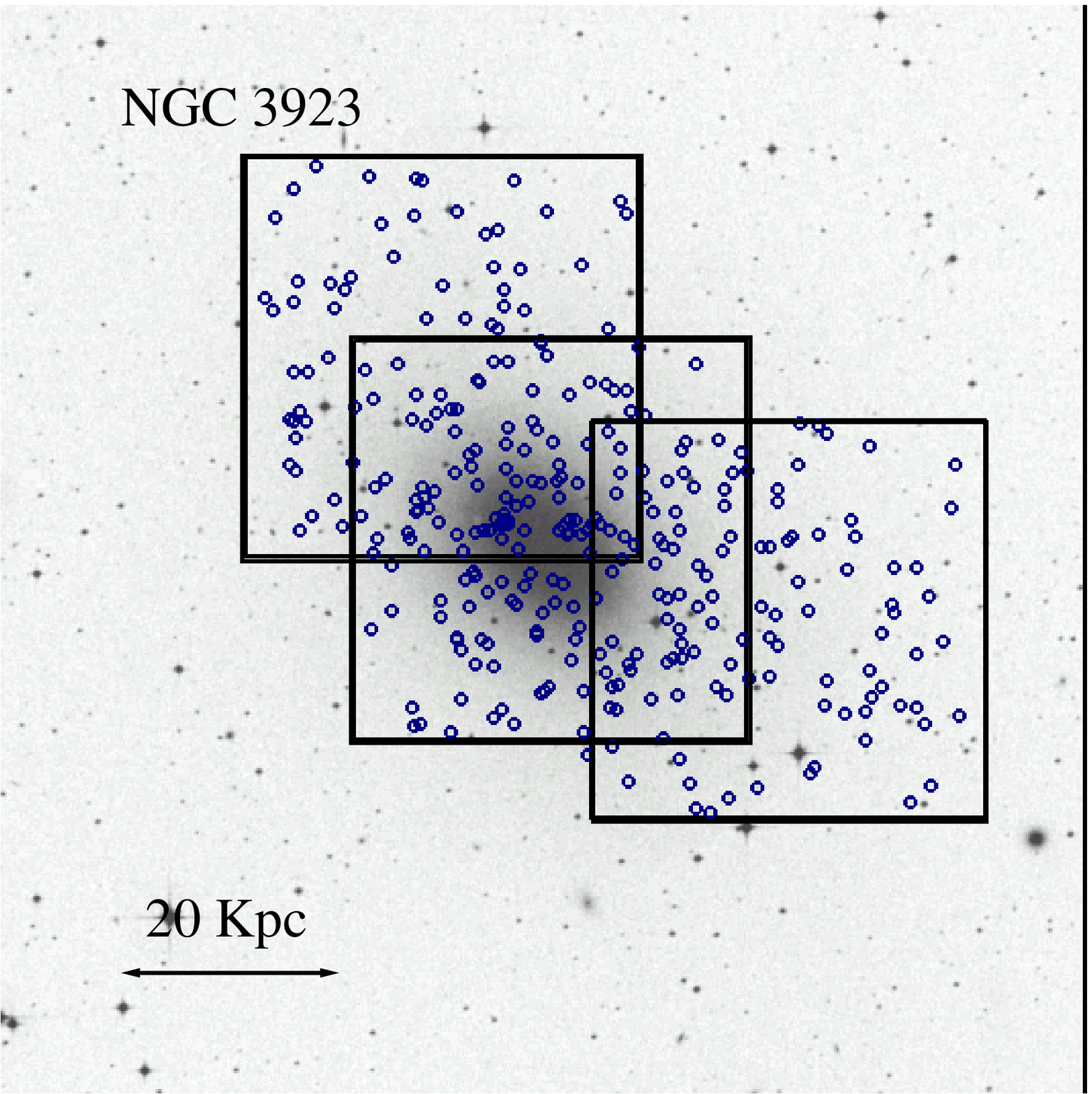}}
\resizebox{0.5\hsize}{!}{\includegraphics{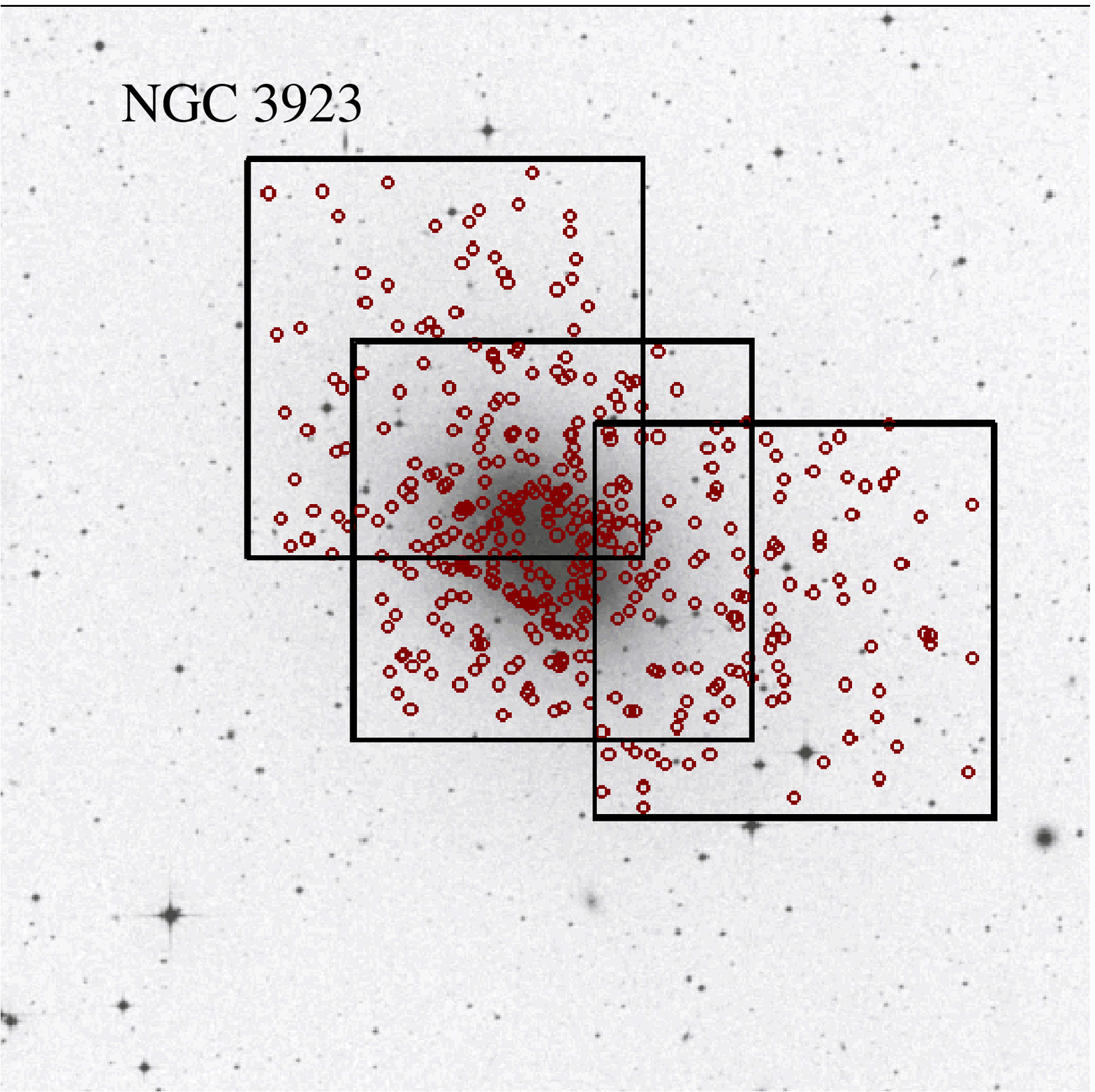}}\\
\resizebox{0.5\hsize}{!}{\includegraphics{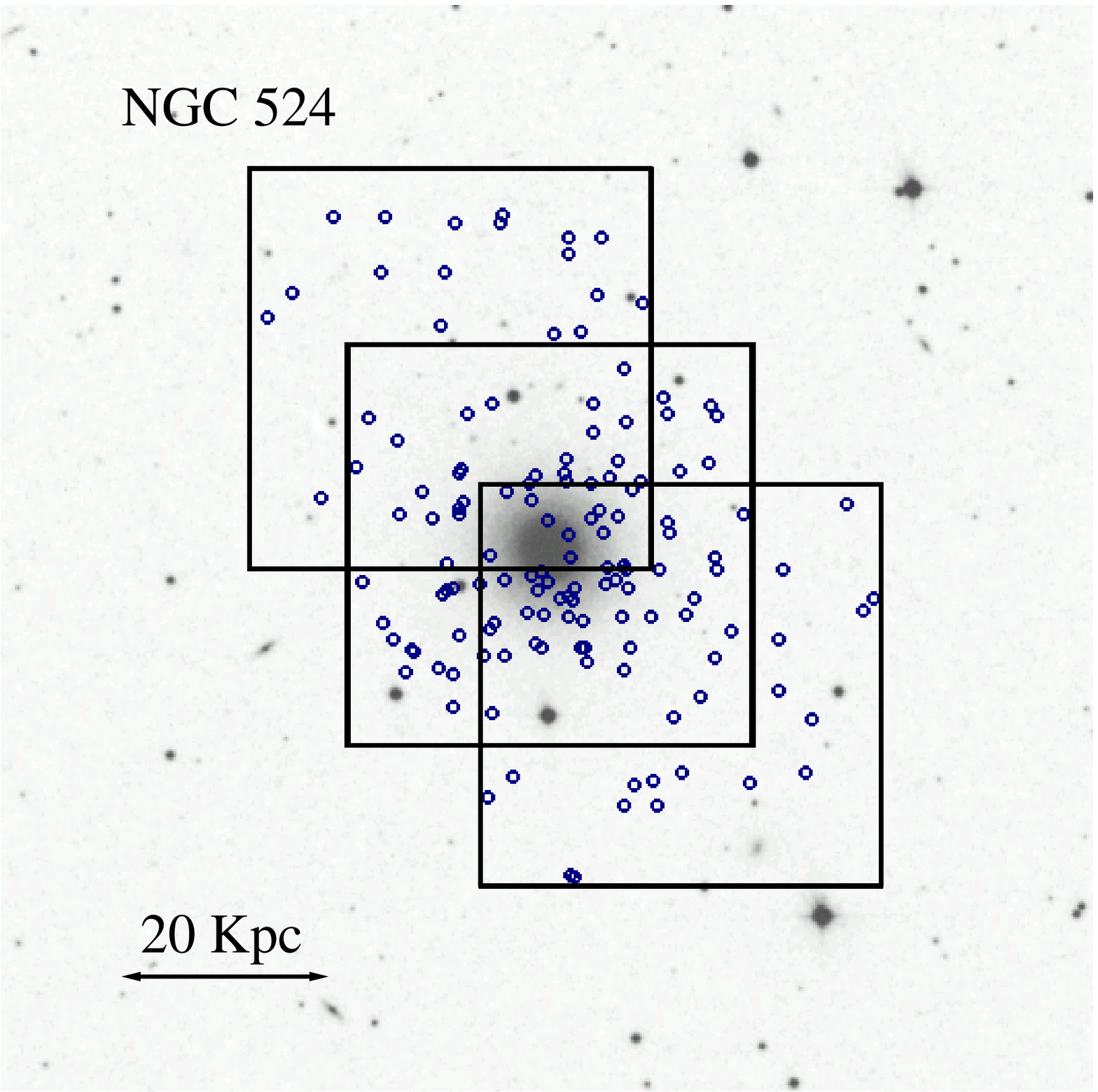}}
\resizebox{0.5\hsize}{!}{\includegraphics{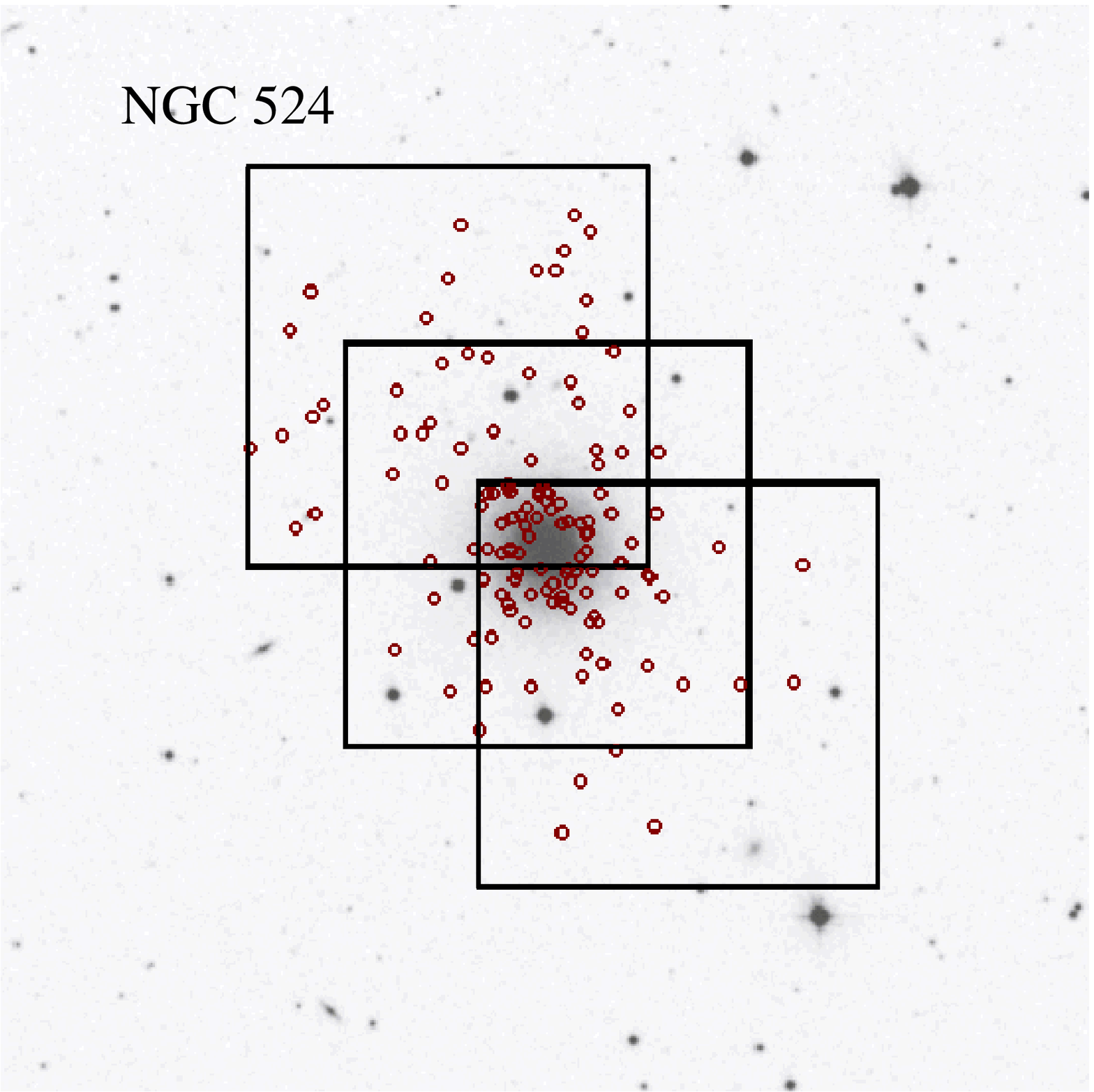}}\\
\resizebox{0.5\hsize}{!}{\includegraphics{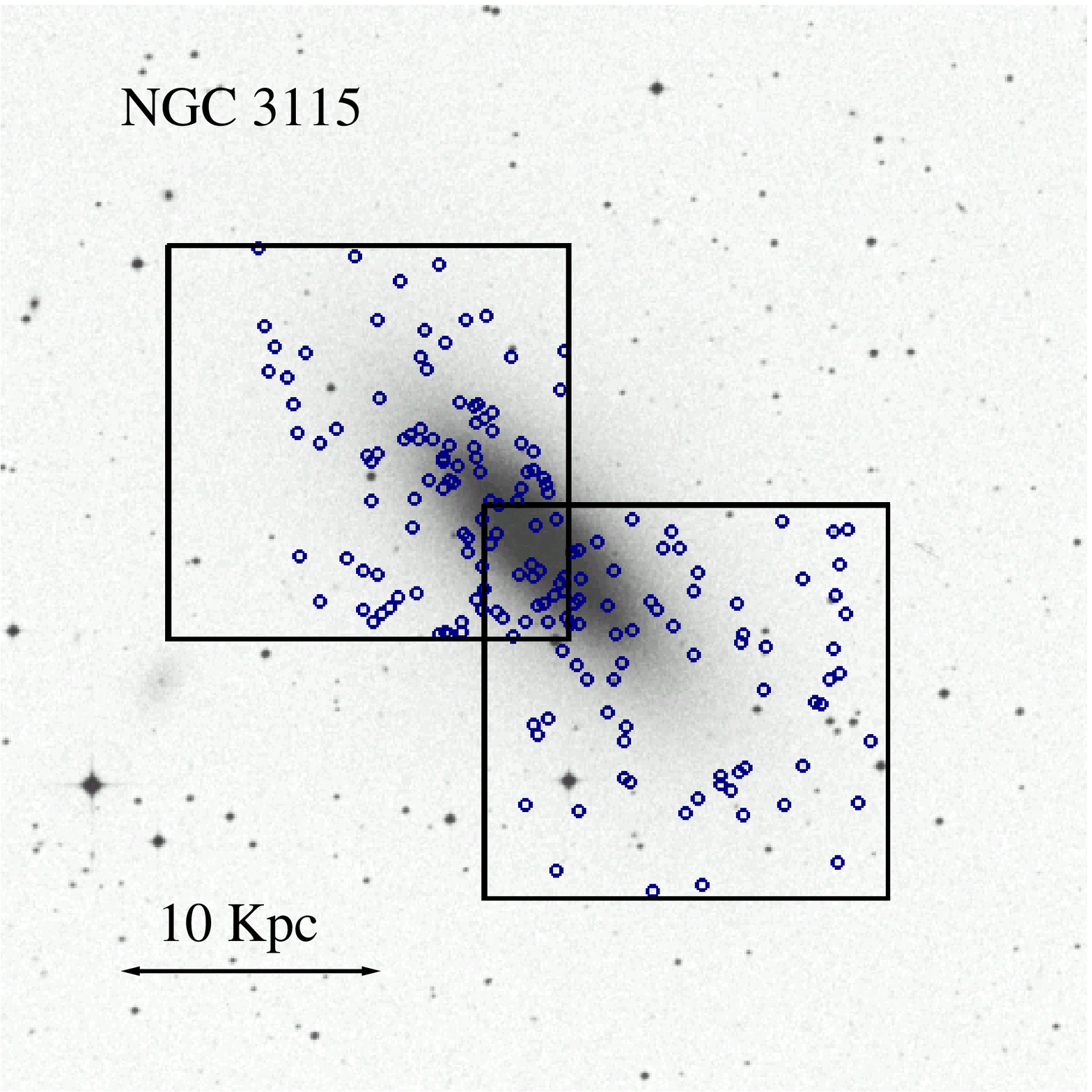}}
\resizebox{0.5\hsize}{!}{\includegraphics{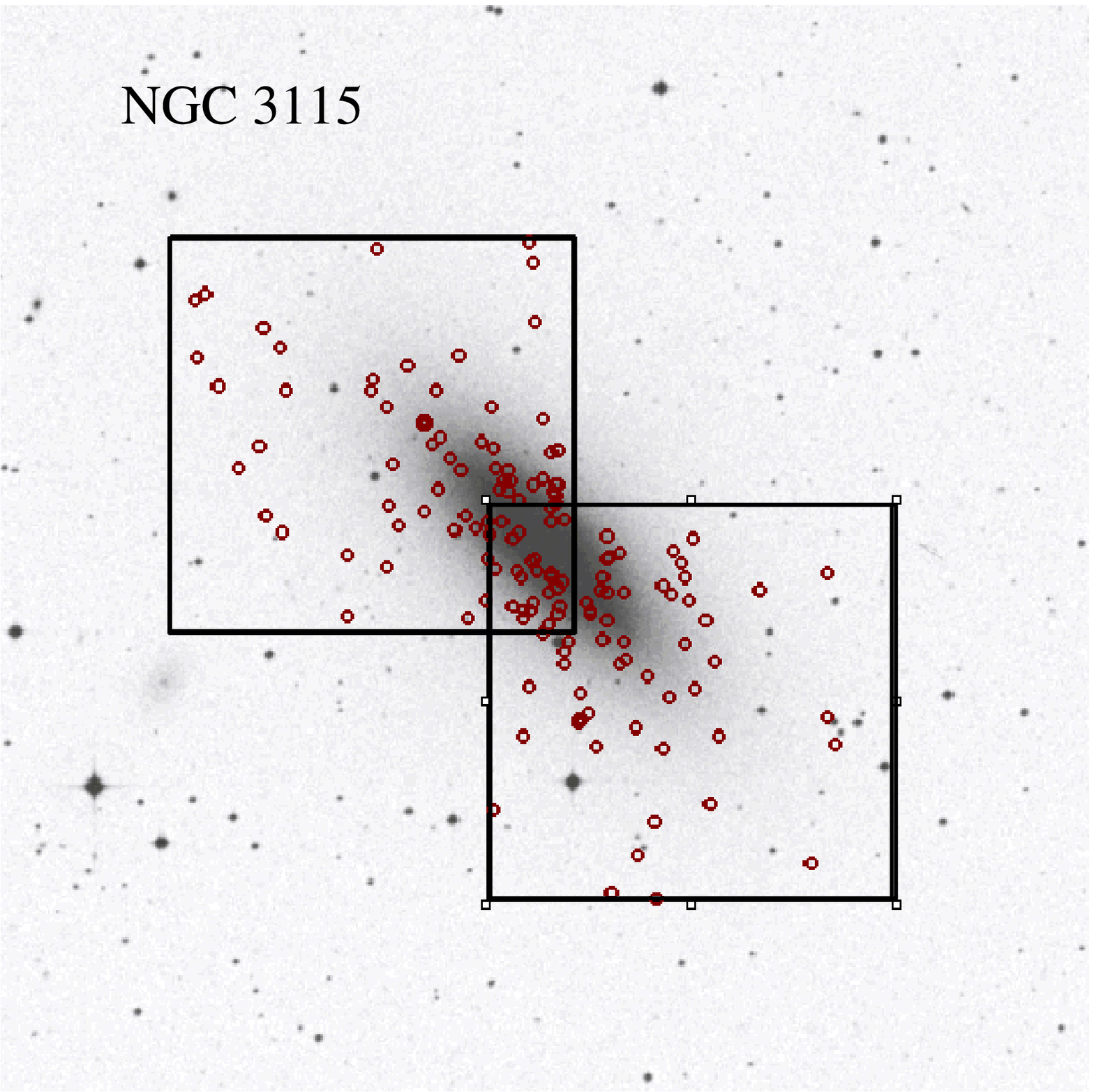}}\\
\resizebox{0.5\hsize}{!}{\includegraphics{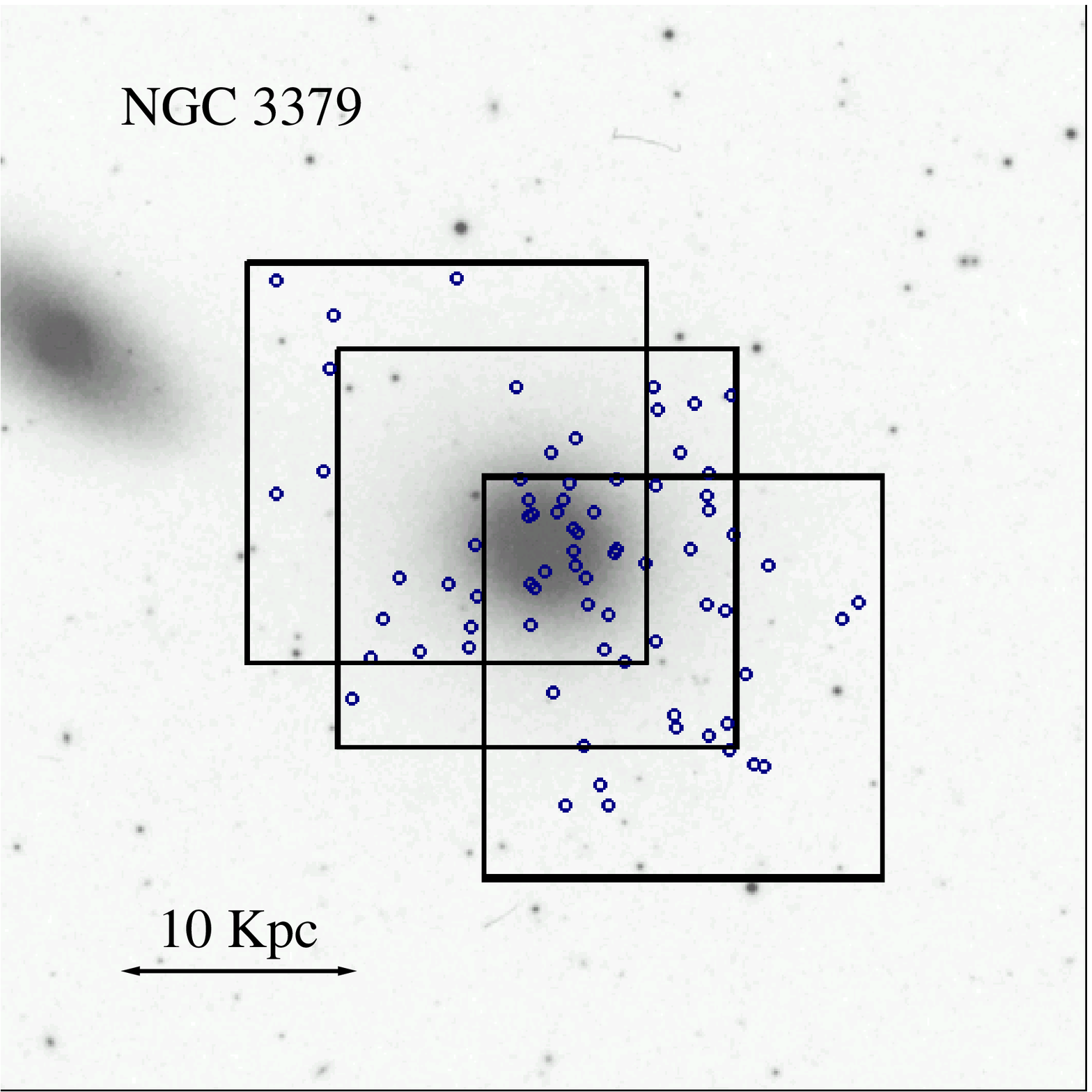}}
\resizebox{0.5\hsize}{!}{\includegraphics{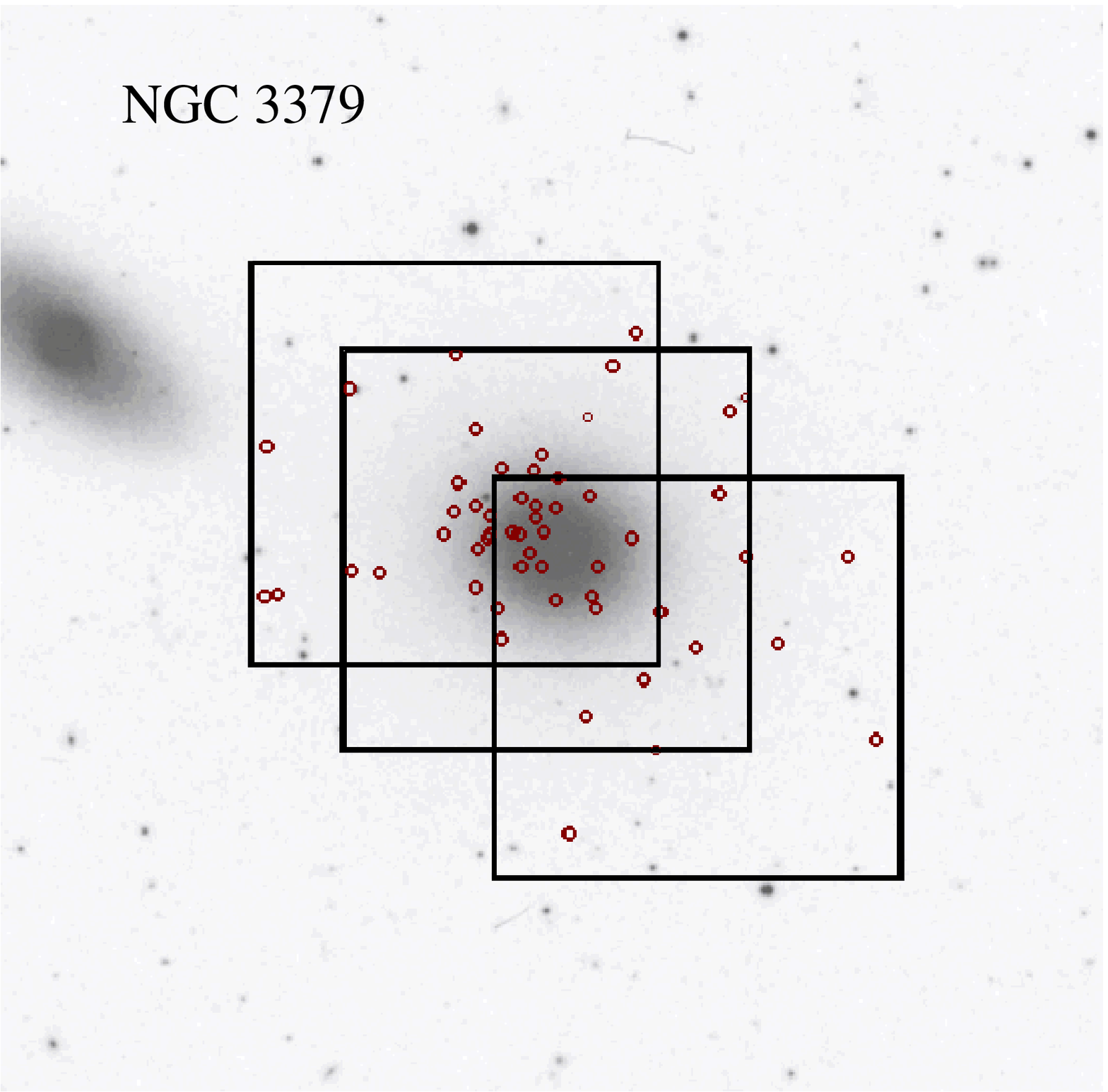}}
\label{DSS2}
\end{figure}

\section{Data}
\label{data}
\subsection{Observations and Data Reduction}
\label{obs}

The images presented here were taken using GMOS \citep{HJADMMC04}\footnote{Further information on GMOS can be found on the instrument
homepage (http://www.gemini.edu/sciops/instruments/gmos/)} in imaging mode 
on both the Gemini South and Gemini North telescopes. The instruments 
consist of 
three $2048 \times 4608$ pixel CCDs, with a scale of 0.0727 arcsec
pixel${}^{-1}$. We used $2\times 2$ binning, yielding a 0.146 arcsec pixel${}^{-1}$
scale. The GMOS camera has a field-of-view of $5.5'\times 5.5'$. Except for
NGC 524 and our Comparison Field (where we had to reject 
and repeat some bad images), four images per field were taken for each of the 
three SDSS filters {\it g$'$, r$'$} and {\it i$'$} \citep{FIGSS96}. The telescope was 
dithered between exposures to facilitate cosmic-ray removal and to fill the 
gaps between the CCD chips. Details of the observational parameters are given in
Table \ref{Tobs}. In addition, we make use of the William Herschel Deep Field
to help estimate contamination (see Sec \ref{CompF}). Fig. \ref{DSS2} 
shows the positions of the GMOS 
fields around each galaxy superimposed on Digital Sky Survey images. 

With the aim of obtaining a homogeneous photometric sample, we 
re-reduced the images provided by Gemini. The raw images were 
processed using the Gemini GMOS package within IRAF 
\footnote{IRAF is distributed by the National Optical
Astronomical Observatories,which are operated by the Association of
Universities for Research in Astronomy, Inc., under cooperative
agreement with the National Science Foundation} (GPREPARE, GBIAS,
GIFLAT, GIREDUCE, GMOSAIC). In all cases we chose suitable sets of raw
bias and flat-field frames from the available data in the Gemini 
Science Archive (GSA).
The images taken with the {\it i$'$} band filter on Gemini South show
considerable fringing. We therefore built
suitable fringe images from seven exposures of blank sky regions taken 
as part of the baseline calibration the night after the 
observations of NGC 3923 and NGC3115. These frames
were used to correct {\it i$'$} images (listed in table \ref{Tobs}) by 
means of the GIFRINGE and GIRMFRINGE tasks. The resulting images 
for each filter were
then co-added and cosmic-ray cleaned using IMCOADD. Because the images
were taken in queue mode, some of them were obtained under
slightly different sky conditions. As a consequence, we set
``fl\_scale = yes''  in order to combine images with different FWHM and signal. 
In this step, we tuned some IMCOADD parameters such as ``aperture'', 
``statsec'', and set ``scalenoise $\sim 3$'' in order to avoid losing 
good pixels and spurious zero points offsets during 
the adding process. These final co-added images were then used for all the 
subsequent data analysis.

\subsection{Photometry}
\label{phot}

Galaxy light subtraction was done using an iterative combination of the 
SExtractor background modelling approach (\citealt{BA96}) and median 
filtering, in a similar way to that presented in \citet{PKTMSBRGH2004}. This 
procedure was implemented in a script which yields a catalogue of all the 
objects detected by SExtractor and a galaxy light subtracted image. This 
software gives better results than IRAF tasks such as DAOFIND. As 
in \citet{FFFBBGHSZG04}, the {\it i$'$} band objects list was adopted as 
input for the subsequent photometry. At the distance of the targets 
listed in Table \ref{Tsample} most of the GCs would be expected to be 
unresolved (starlike) sources.
 The Daophot package \citep{S87} within IRAF was used to obtain psf magnitudes for all
objects detected by SExtractor, and they were then separated into resolved and unresolved 
 objects  following the outline presented in Section \ref{clas}. After that, aperture
photometry was obtained for all the resolved objects. For this purpose 
we have used the PHOT task with a fixed 2 arcsecond aperture. This means that 
we have good photometry for unresolved and marginally resolved objects.
A ``master'' catalogue of resolved and unresolved objects was built 
by combining photometry of objects successfully measured in all three filters.

We searched in the GSA for standard star fields observed during 
the same nights 
as our targets and the Comparison Field. They were re-reduced using the 
same biases and flats applied to our science frames. The synthetic 
transformation from \citet{FIGSS96} was adopted to obtain the {\it g$'$, r$'$, i$'$}  
standard magnitudes. Because of the small number of the available 
standards, usually 3-4 stars, and their very limited range 
in airmass, this calibration 
was suitable only for establishing a mean zero-point value 
for our frames, and not for a more complete evaluation of the 
coefficients in airmass or colour (see Table \ref{Tphot}). Thus, we 
measured these zero points, and we used the atmospheric extinction 
coefficients given by the Gemini web page. 

NGC 4649 is an exception because standards were not observed at the same 
night as our program fields. So, in this case, we adopted the mean zero 
points obtained from the three sets of standards (about five stars 
per field) observed during the Gemini North GMOS run of March 2007. 

For each galaxy, the field with the best overall seeing in the
three filters and having a corresponding standard stars field, was adopted
as the reference for the photometry. A list of common objects in each field was 
used to obtain any small zero point differences among the pointings.
Finally, we applied the galactic extinction coefficients given 
by \citet{SFD98}, and used their Table 6 to transform 
them into {\it A$_{g'}$, A$_{r'}$} and {\it A$_{i'}$}. Thus we 
quote extinction corrected magnitudes and colours in this paper.

\begin{table}
\caption{Summary of observations. This table lists the program ID, numbers 
of the fields, exposures and final seeing of the co-added images.}
\label{Tobs}
\scriptsize
\begin{tabular}{lllll}
\multicolumn{1}{c}{ } \\
\hline
\hline
\multicolumn{1}{c}{ } \\
\multicolumn{1}{c}{\textbf{Galaxy}} &
\multicolumn{1}{c}{\textbf{Gemini ID}} &
\multicolumn{1}{c}{\textbf{Fields}} &
\multicolumn{1}{c}{\textbf{T$_{exp.}$(s)}} &
\multicolumn{1}{c}{\textbf{FWHM (arcsec)}} \\
\multicolumn{3}{c}{} &
\multicolumn{1}{c}{{\it g$'$\hspace{5mm}r$'$\hspace{5mm}i$'$}} &
\multicolumn{1}{c}{{\it g$'$\hspace{3mm}r$'$\hspace{3mm}i$'$}} \\
\hline 
NGC 4649    &  GN-2007A-Q-37  & 1 &  4x120~~4x120~~4x120  & 0.48~~0.49~~0.48\\
            &                 & 2 &  4x120~~4x120~~4x120  & 0.48~~0.45~~0.45\\
            &  GS-2007A-Q-49  & 3 &  4x120~~4x120~~4x120  & 1.11~~1.05~~1.12\\
            &  GN-2007A-Q-37  & 4 &  4x120~~4x120~~4x120  & 0.51~~0.51~~0.47\\
NGC 524     &  GN-2002B-Q-25  & 1 &  4x150~~4x100~~4x100  & 0.67~~0.60~~0.65\\
 	    &                 & 2 &  6x150~~4x100~~5x100  & 0.85~~0.76~~0.72\\
 	    &                 & 3 &  4x100~~4x100~~5x100  & 0.79~~0.81~~0.84\\
NGC 3379    &  GN-2003A-Q-22  & 1 &  4x200~~4x100~~4x100  & 1.03~~0.99~~0.84\\
	    &                 & 2 &  4x200~~4x100~~4x100  & 0.54~~0.54~~0.52\\
	    &                 & 3 &  4x200~~4x100~~4x100  & 0.93~~0.98.~~0.90\\
NGC 3115    &  GS-2004A-Q-9   & 1 &  4x200~~4x100~~4x100  & 0.77~~0.65~~0.53\\
	    &                 & 2 &  4x200~~4x100~~4x100  & 0.80~~0.84~~0.70\\
NGC 3923    &  GS-2004A-Q-9   & 1 &  4x200~~4x100~~4x100  & 0.79~~0.68~~0.67\\
            &                 & 2 &  4x200~~4x100~~4x100  & 0.72~~0.62~~0.55\\
	    &                 & 3 &  4x200~~4x100~~4x100  & 0.72~~0.63~~0.59\\
Comp. Field &  GN-2003A-Q-22  & 1 &  3x200~~3x100~~3x100  & 0.64~~0.55~~0.50\\
WHDF        &  GN-2001B-SV-104& 1 &  6x300~~6x300~~6x300  & 0.86~~0.67~~0.67\\
Blank-sky & GS-CAL20040120 & 1 & {\hspace{12.5mm}}7x300  &{\hspace{9.5mm}}0.47\\
\hline
\end{tabular}
\end{table}

\subsection{Object classification and completeness}
\label{clas}

\begin{figure*}
\resizebox{0.4\hsize}{!}{\includegraphics{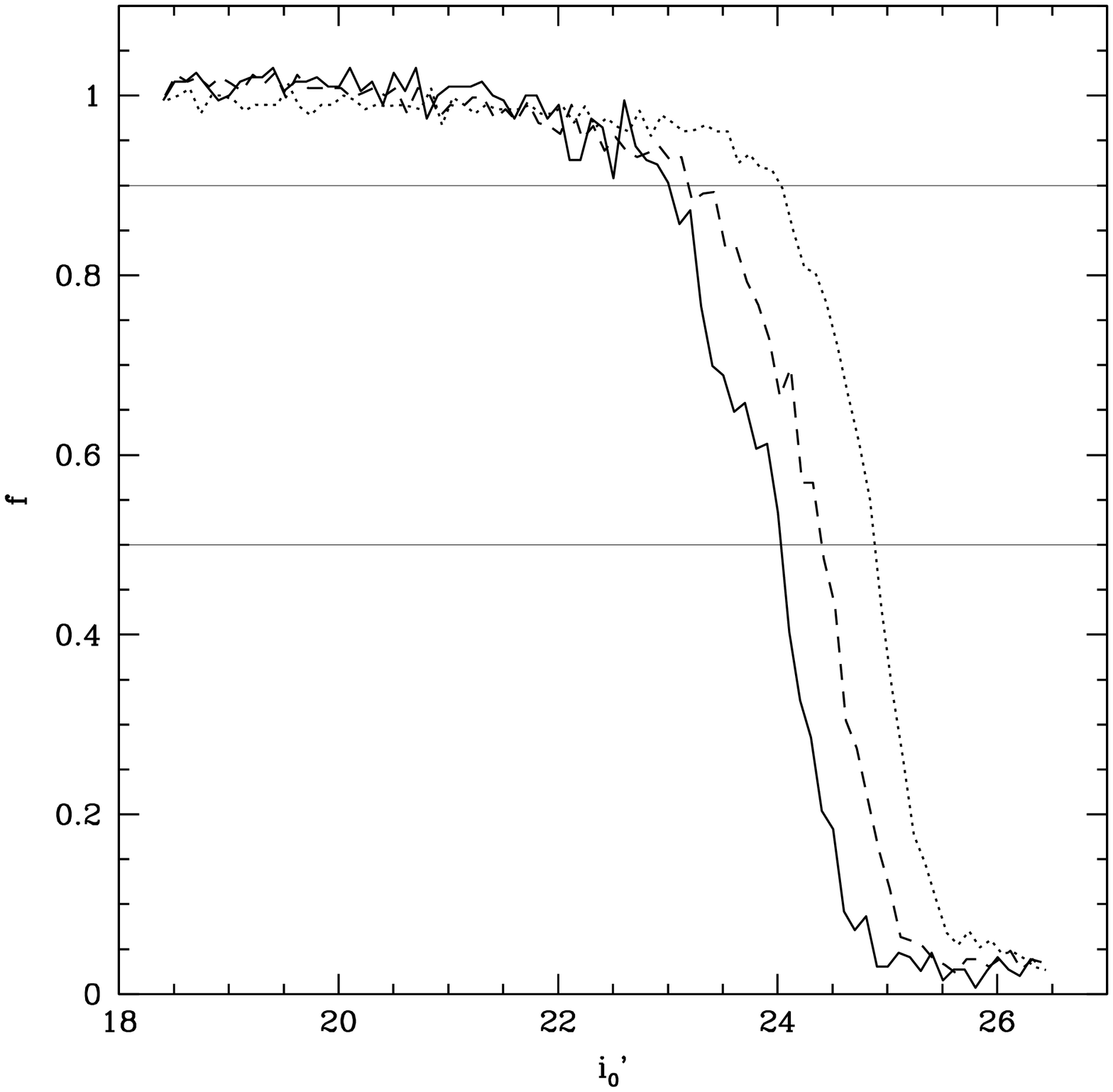}}
\resizebox{0.4\hsize}{!}{\includegraphics{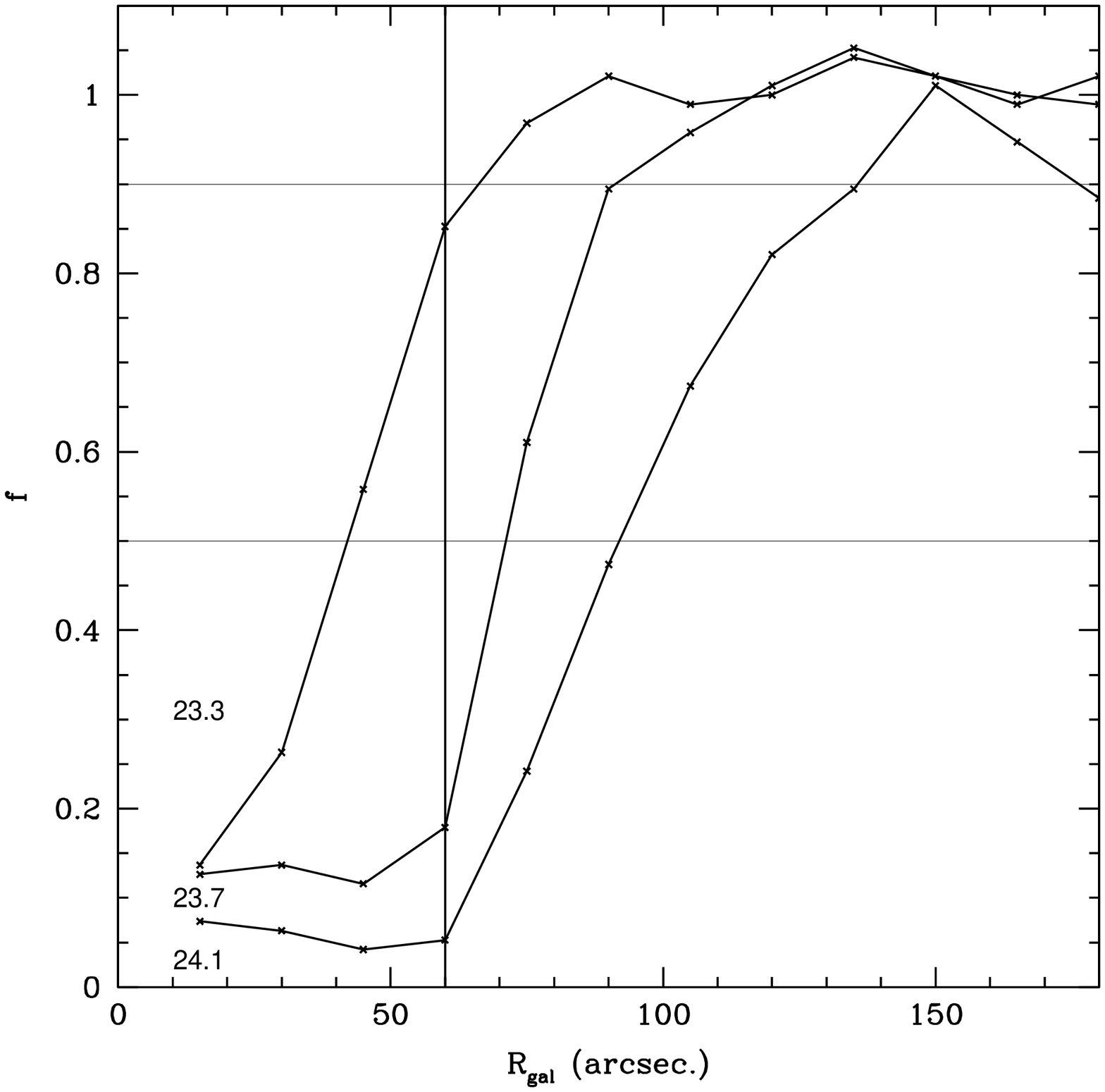}}
\caption{Completeness factors for the i images of NGC 3923. The solid, dotted
and dashed lines correspond to fields 1, 2 and 3, respectively.
Left panel:  Fraction of recovered objects versus the input
magnitudes.  Field 1 (centered in the galaxy) shows a
smaller overall completeness than fields 2 and 3. The solid lines are the 90\%
and 50\% levels. Right panel: 
Completeness for field 1 versus galactocentric radius for three different 
magnitude bins. The fraction of lost objects is 
strongly dependent on galactocentric radius for R$<$100 arcsec.}
\label{Comple}
\end{figure*}
 
The classification between resolved and unresolved sources was carried
out using the same procedure as in \citet{FFFBBGHSZG04}. Briefly, we used
a combination of SExtractor Stellarity Index (0 for resolved objects 
and 1 for unresolved ones), and the positions of each object in the aperture 
minus PSF magnitude vs. PSF magnitude diagram (\citealt{FGOPG01}, and see 
Fig. 1 and 2 from \citealt{FFFBBGHSZG04}). This approach has been tested 
using our follow-up spectroscopic studies  (\citealt{PBFBGFFZSHP2006}; 
\citealt{BGSFFBZFHP2006}; \citealt{NSBGFPFFBZH2008}),
and combined with three-colour selection,
 has been very successful in  giving a low contamination rate by 
background galaxies.

In order to quantify the detection limits of our photometry a series of 
completeness tests were carried out on each field using the ADDSTARS task. 
The {\it i$'$} band magnitude range from 18.5 to 26.5 mag was divided into 
intervals of 0.1 mag. Then, trying to avoid crowding effects, 200 point 
sources were added to the original {\it i$'$} 
images in each magnitude bin and the same procedure of galaxy light subtraction,
detection and classification was followed to recover
and classify the added sources. The positions of the artificial objects 
were created 
using a stochastic-generating Fortran program that follows the slope of 
the GC system spatial distribution. A total of 32000 artificial objects
were added in two independent experiments per field.

As an example, Fig. \ref{Comple} shows a typical result of this
procedure. In this case, the  results for each of the {\it i$'$} fields of NGC 3923 
are plotted. The left panel shows the fraction of artificial 
objects recovered versus the input
magnitude for each field. The right panel shows the 
completeness as a function of R$_{gal}$ for three different magnitude 
bins in field 1. If a global completeness level is selected from the 
left plot, the right one shows that most of the objects are lost 
at small R$_{gal}$. That is, the completeness has a strong spatial 
dependence in the sense that the loss of objects is always larger 
at small radii. Taking this into account, and in order 
to ensure completeness levels greater than 50\% and 90\% for R$_{gal} > 45$ 
arcsecs, we selected the {\it i$'$} band limits listed in Table \ref{Tphot}. The 50\% 
level was used to define a sample of GC candidates, but we used the 90\% level 
in the subsequent analysis.

\subsection{Comparison Fields}
\label{CompF}

\begin{table*}
\centering
\caption{Ranges adopted for GC candidates selections, zero 
points, and completeness levels in the $i'_0$ band.}

\label{Tphot}
\scriptsize
\begin{tabular}{lcccclccccc}
\hline
\hline
\multicolumn{10}{c}{} \\
\multicolumn{1}{c}{\textbf{Galaxy}} &
\multicolumn{1}{c}{\textbf{$i'_0$}} &
\multicolumn{1}{c}{\textbf{$(g'-i')_0$}} &
\multicolumn{1}{c}{\textbf{$(g'-r')_0$}} &
\multicolumn{1}{c}{\textbf{$(r'-i')_0$}} &
\multicolumn{1}{c}{\textbf{$Z_p$ $g'$}} &
\multicolumn{1}{c}{\textbf{$Z_p$ $r'$}} &
\multicolumn{1}{c}{\textbf{$Z_p$ $i'$}} &
\multicolumn{2}{c}{\textbf{Completeness}} & \\
\multicolumn{1}{c}{} & 
\multicolumn{1}{c}{} &
\multicolumn{1}{c}{} &
\multicolumn{1}{c} {} &
\multicolumn{1}{c}{} &
\multicolumn{1}{c}{} &
\multicolumn{1}{c}{} &
\multicolumn{1}{c}{} & 
\multicolumn{1}{c}{90\%} &
\multicolumn{1}{c}{50\%} &
\multicolumn{1}{c}{} \\
\hline \multicolumn{11}{c}{}\\
 NGC 4649    & 19.5-24.3 & 0.40-1.45 &   0.35-0.95    &  0.0-0.60  & 27.885$\pm$0.005 & 28.268$\pm$0.004 & 28.190$\pm$0.006 & 	23.90  &  24.30  \\
 NGC 3923    & 20.0-23.7 & 0.40-1.40 &   0.30-0.90    &  0.0-0.60  & 28.239$\pm$0.006 & 28.128$\pm$0.008 & 27.711$\pm$0.009 & 	23.30  &  23.70  \\ 
 NGC 524     & 20.4-23.5 & 0.40-1.40 &   0.30-0.90    &  0.0-0.60  & 27.630$\pm$0.018 & 27.992$\pm$0.028 & 27.777$\pm$0.030 & 	23.00  &  23.50  \\ 
 NGC 3115    & 18.4-23.5 & 0.40-1.40 &   0.30-0.90    &  0.0-0.60  & 28.239$\pm$0.006 & 28.128$\pm$0.008 & 27.711$\pm$0.009 & 	23.00  &  23.50  \\
 NGC 3379    & 18.6-23.2 & 0.40-1.40 &   0.30-0.90    &  0.0-0.60  & 27.985$\pm$0.006 & 28.215$\pm$0.005 & 27.961$\pm$0.004 &   22.90  &  23.20   \\ 
 Comp. Field &           &           &                &            & 27.905$\pm$0.017 & 28.189$\pm$0.028 & 27.977$\pm$0.017 &   24.00  &  24.33   \\ 
 WHDF        &           &           &                &            & 27.934$\pm$0.020 & 28.184$\pm$0.029 & 27.942$\pm$0.030 &   24.75  &  25.05   \\
\multicolumn{10}{l}{}\\
\hline                                                                             
\multicolumn{10}{l}{}\\ 
\end{tabular}
\end{table*}

In addition to the galaxy sample listed in Table \ref{Tsample},
we include two comparison fields. One of them is the same 
as in \citet{FFFBBGHSZG04}. However, here we present re-reduced and
re-calibrated data for this field. 

\begin{figure}
\resizebox{1.0\hsize}{!}{\includegraphics{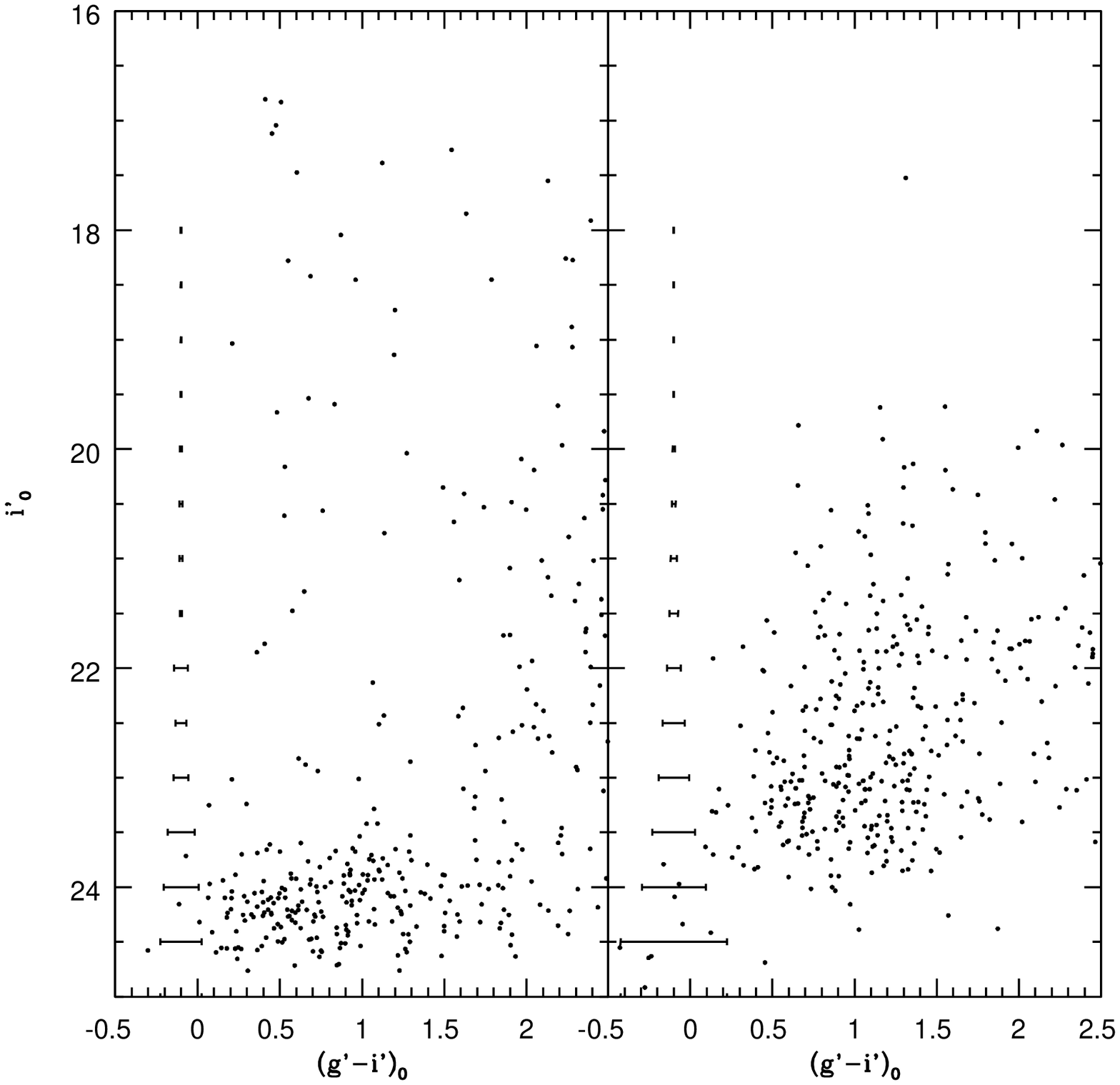}}\\
\resizebox{1.0\hsize}{!}{\includegraphics{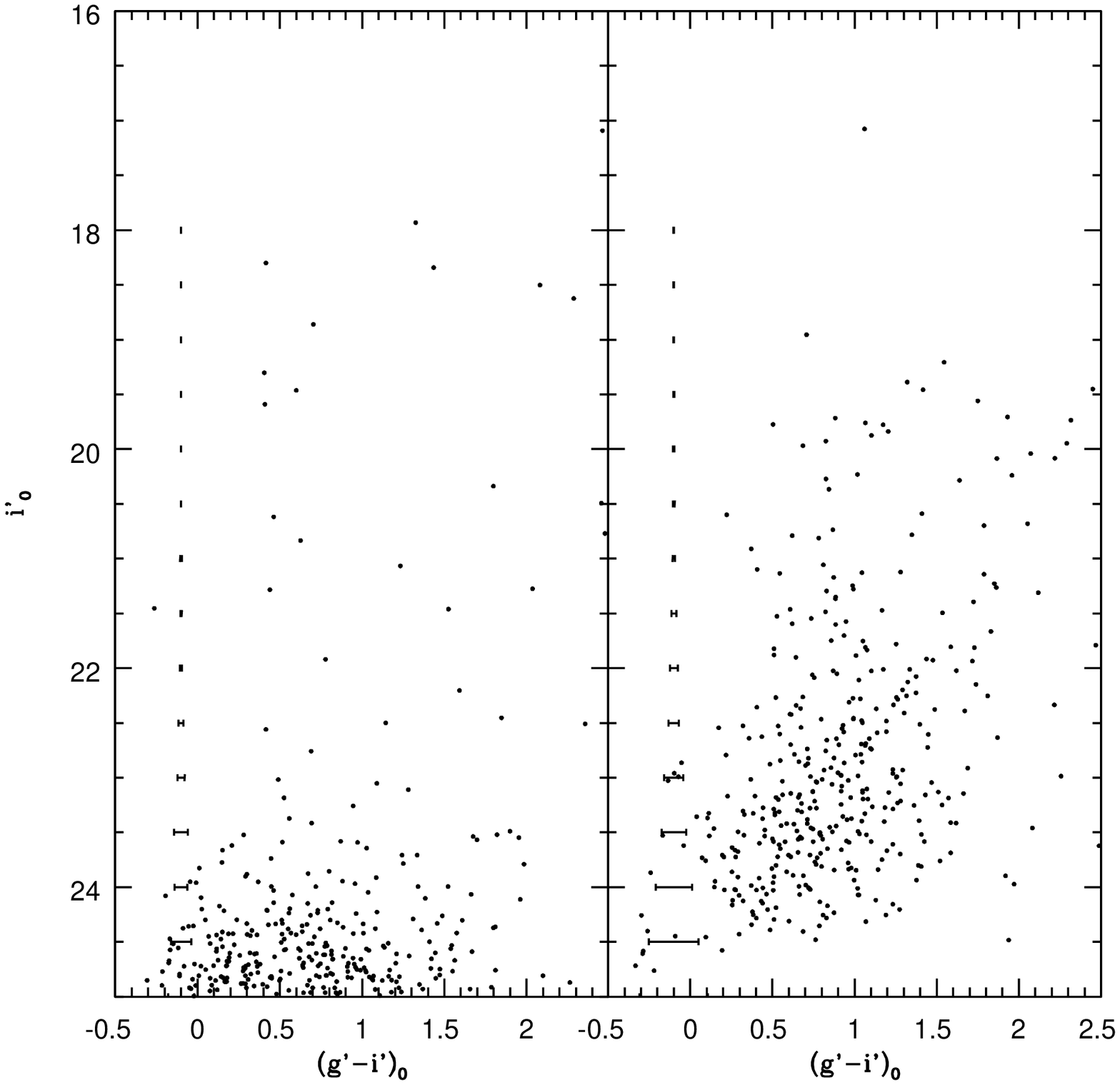}}
\caption{Photometric properties of the sources 
detected in our comparison field (top panel) and in the WHDF (lower panel). 
In each panel, the left/right  plot is showing the  colour-magnitude 
diagram for unresolved/resolved sources. Mean 
colour errors are shown by the horizontal bars in each panel. It is  
possible to efficiently separate background galaxies from Milky Way (MW) stars only
for objects brighter than {\it i$'_0$} $\sim 23.5$.}
\label{Comparation_f} 
\end{figure}

We have applied the same object detection and classification procedures as 
for our program images. Fig. \ref{Comparation_f} shows, in the upper panel, 
the colour-magnitude diagrams for the unresolved and resolved objects 
in this field. For unresolved objects, we used psf magnitudes, while for 
resolved objects we used aperture photometry with 2 arcsec apertures. Most 
of the objects with magnitudes {\it i$'_0$} $<23.5$ and 
colours within the ranges adopted for GC candidates are resolved and 
were correctly classified (see Sec \ref{CMD}). However, for {\it i$'_0$} $\sim 23.5$ the 
situation is very different and there are  many point 
sources displaying {\it(g$'$--i$'$)} colours similar to those of GCs. Most of them 
are probably distant background galaxies.

In order to test the results obtained with our Comparison Field (CF),
we applied our classification recipe on the William Herschel Deep Field
(WHDF, \citealt{MSCMcF2001}). This field was observed with the same instrument 
and filters as our CF. The seeing in both fields is similar, so the 
only difference are the total exposure times (see Table \ref{Tobs}).
The 90\% and 50\% completeness levels obtained for our comparison
field and the WHDF are listed in Table \ref{Tphot}.

Fig. \ref{Comparation_f} shows, in the bottom panel, the colour-magnitude
diagrams for unresolved and resolved
objects detected in the WHDF. Clearly, there are two main differences 
between the two CMDs in Fig \ref{Comparation_f}. First, the numbers of 
star-like objects brighter than {\it i$'_0$}$ < 23.5$ present in these fields
are different. There 
are more stars in our CF than in the WHDF. This is easily  seen in the red
side of the CMD of the unresolved objects ({\it (g$'$--i$'$)} $>1.5$).

The other difference is the appearance of the CMD at the low 
brightness extreme.  The number of unresolved 
objects in the CF drops steeply at {\it i$'_0$} $\sim 23.5$, whereas in 
the WHDF it does not 
happen until {\it i$'_0$} $\sim 24.2$. The reason for this behaviour 
is that the 
process of classification improves with better S/N (i.e. with the deeper 
WHDF exposures we can make a better classification), and that
the stellar content is not the same in both fields, as is to be
expected given their different galactic latitudes. However, we note that 
the number of stars with colours adopted here for GC 
candidates is only 30\% larger in the CF than in the WHDF.
Comparing the star counts in this field with the 
number from the Besancon Galaxy models \citep{RRD2003} we found a very good 
agreement for {\it i$'_0$} $< 23.5$. For fainter magnitudes the 
Besancon star counts 
are always smaller than the observed counts, possibly indicating a very strong 
contamination by unresolved background galaxies.  

The analysis given in \citet{FSI95} suggests that low redshift S, S0 
and E galaxies and medium redshift Irr galaxies can show colours in the 
ranges adopted for the GC candidates. However, under good imaging 
conditions (i.e., sub-arcsecond FWHM) and making a careful image 
analysis, most of them are correctly classified as extended sources.  
The medium and high redshift galaxies are difficult to resolve from 
ground-based observation. However, most of these objects can be rejected by 
using {\it(g$'$--i$'$)}, {\it(g$'$--r$'$)} and {\it(r$'$--i$'$)} colours. 
 
This means that, for {\it i$'_0$}$<23.5$,  the nature of the 
contamination is mainly stellar. At fainter magnitudes, the contamination 
arising from faint background galaxies increases steeply.

In the subsequent analysis, we adopt an average of the counts in
both fields to obtain the contamination levels of our photometry in 
NGC 3923, NGC 524, NGC 3115 and NGC 3379. In the case of NGC 4649, we chose 
to use only the WHDF as a comparison field for two reasons. The first is 
that the photometry in the CF is not deep enough. The second reason 
is that, as we will see in Section \ref{CMD_CCD_GC}, the colour-magnitude 
diagrams of the objects detected in our four NGC 4649 fields show a 
very small number of objects outside the regions 
occupied by GC candidates. The main difference between the WHDF and 
our CF is the red side of the colour distribution.

\section{Results}

\subsection{Color-magnitude and colour-colour diagrams}
\label{CMD}

The colour-magnitude (CMDs) and colour-colour diagrams are presented in
 Fig. \ref{CMD_CCD_GC} for all point sources detected in our 
GMOS fields (small dots), ordered by decreasing {\it B$_T$} brightness of 
the host galaxy. Magnitudes and colours were corrected for galactic 
extinction as indicated in Section \ref{phot}. The mean {\it (g$'$--i$'$)} 
photometric errors are shown as 
small bars at {\it (g$'$--i$'$)$_0$}$=0$. The dotted and long-dashed lines indicate 
the 50\% and 90\% completeness limits from our completeness tests, respectively. 

In all diagrams the GCs can be easily seen as a group 
of objects clustered around {\it (g$'$--i$'$)$_0$} $\sim 0.8$,
{\it (g$'$--r$'$)$_0$} $\sim 0.6$ and {\it (r$'$--i$'$)$_0$} $\sim 0.3$. In order to 
obtain a clean sample of GC candidates (rejecting MW stars and unresolved
background galaxies), but allowing a reasonable range of GCs colours, 
we adopted the limits listed in Table \ref{Tphot}. Regarding the magnitude 
limits, cuts were applied only on the {\it i$'$} band magnitudes. The low 
brightness ends were 
defined by the 50\% completeness level in the case of our photometric 
sample definition, and by the 90\% completeness level for the analysis 
sample (that was used in the analysis presented in the following Sections).

 For the high brightness end, the cuts were chosen in order to include 
all GC candidates with absolute magnitudes fainter 
than {\it M$_I$} $\approx -12$. The most massive galactic 
GC-like object, $\omega$ Cen, has an integrated magnitude of {\it M$_I$} $\sim -11$. However,  
M31's most massive GC, G1, is considerably more luminous. 
Also, this upper limit is in agreement with the {\it M$_V$} $=-11$ value 
suggested by \citet{MHIJ2006} to separate bright GCs from UCDs, and is 
similar to that found by \citet{WH2007} for the Hydra UCD 
candidates. Adopting this upper limit, and utilizing common 
objects between our GMOS photometry and that of \citet{LBHFG2001} from HST 
data, we obtain the {\it (g$'$,i$'$)}--{\it (V,I)} transformations:

\begin{equation}
I_0=i'_0-0.491(\pm0.003)
\label{eq1}
\end{equation}

\begin{equation}
V_0=g'_0-0.34(\pm0.02) (g'-i')_0-0.03(\pm0.02)
\end{equation}

\begin{equation}
(V-I)_0=0.80(\pm0.02) (g'-i')_0+0.32(\pm0.02)
\label{VIgi}
\end{equation}

\noindent Taking into account the distance modulus from 
Table \ref{Tsample} we obtain the upper cut values 
listed in Table \ref{Tphot}, which roughly correspond 
to {\it M$_I$} $= -12$ in each galaxy. Fig. \ref{CMD_CCD_GC} 
shows as filled circles all the point sources falling in the 
colour and magnitude ranges listed in Table \ref{Tphot}
which we take as our GC candidates.

Most of the unresolved objects in our sample define clear sequences
consistent with the locus of MW stars. The sparser appearance of 
the colour diagrams corresponding to NGC 524 is 
probably due to the slightly lower quality of the photometry. In the case 
of NGC 4649, we have higher quality and deeper photometry. The 50\% 
completeness level is fainter than that in any other galaxy 
and therefore we expect a
slightly higher level of contamination by faint background 
objects (as shown in Section \ref{CompF}, the number of 
background objects grows steeply for {\it i$'_0$}$> 23.5$). 

In both colour-colour diagrams the GC candidates form a short sequence 
which merges with MW stars in the blue extreme. Thus, a small fraction
 of the  candidates with extreme colours {\it (g$'$-i$'$)$_0$}$\sim 0.5$, 
could in fact be Galactic stars. 

As mentioned before, at the distances of the targets listed in 
Table \ref{Tsample} we can expect 
that the normal GCs will look like unresolved objects. However, in the case of 
NGC 4649 and NGC 3923, \citet{BGSFFBZFHP2006} and \citet{N2011} have 
found marginally resolved objects with colours similar to blue GCs which were 
confirmed as members of the NGC 4649 and NGC 3923 GC systems. Considering that 
most of the images in Table \ref{Tphot}
were taken with sub-arcsecond image quality, and because two of 
our targets are relatively nearby systems, we have checked the possibility 
that other marginally resolved GC-like objects exist in the lists 
of extended ones. Therefore, we inspected by eye all objects 
classified as resolved, but having 
SExtractor shape parameters indicating roundness similar to that of 
point sources in each image. As noted in 
Section \ref{clas}, good S/N is necessary to robustly determine 
these parameters, therefore we only analyzed objects 
brighter than {\it i$'_0$}$=22-23$ mag. Any object showing a complex 
structure was rejected and aperture photometry was performed on 
the remaining ones. They are 
indicated as blue open circles in the CMDs, and with the same symbols 
as GC candidates in the colour-colour diagrams. This sample of marginally 
resolved objects is expected to have a higher level of contamination 
by background galaxies than our final sample of GC candidates.

Since the half-light radii of UCD candidates (5-100 pc, according 
to \citealt{MHJI2008}) are larger than those typical for GCs, we 
have extended the inspection described
in the last paragraph to higher luminosities, including both unresolved
sources and marginally resolved ones. Plotting the point source UCD 
candidates in the colour-colour 
diagrams (see {\it{(g$'$--r$'$) vs. (r$'$--i$'$)}} panels for NGC 4649 and NGC 3923 in 
Fig. \ref{CMD_CCD_GC}), they seem to define two distinct groups. One of 
them includes objects falling exactly on the GCs sequence. The other one 
falls on the MW stellar sequence, which merges with that of the GCs at the
blue extreme. Considering this, we have classified our unresolved 
UCD candidates as follows: ``type I'' if they belong to the first 
group (the most secure, shown as red filled squares in Fig. \ref{CMD_CCD_GC}), ``type III'' if 
they belong to the second group (less secure, and probably MW 
stars, shown as green filled pentagons in Fig. \ref{CMD_CCD_GC}). 
Additionally we have defined ``type II'' as those objects that
seem to overlap both groups (red circles in Fig. \ref{CMD_CCD_GC}).
Some marginally resolved UCD candidates, shown as blue filled 
triangles in Fig. \ref{CMD_CCD_GC}, were also found. The full
 photometric tables are presented in the Appendix.
In what follows we give a summary of each GCs system:\\

\noindent {\bf NGC 4649} This galaxy shows a well-defined GC 
sequence in the colour-colour diagram. The bimodality 
in  {\it (g$'$--i)} is clear and detectable even in the
{\it{(g$'$--r$'$) vs. (r$'$--i$'$)}} plane, where we can see two slightly 
superimposed clumps of GCs. There are bright blue 
and red GC candidates ({\it M$_I$} $< -11$), so the brightest objects 
suggest a non  unimodal colour distribution. However, the 
most striking characteristic in NGC 4649 is that the bluest GCs seems to 
become redder with increasing luminosity, i.e., they show 
a ``blue tilt'' as will be discussed later.

We have found 63 marginally resolved GC candidates (shown as blue circles), 
some of which were confirmed as belonging to the NGC 4649 system 
by \citet{BGSFFBZFHP2006} (see Table \ref{GCphoT} in the Appendix). These 
objects have a broad 
color distribution but they seem to follow that of the GCs. 

Regarding UCD candidates, we have found 15 unresolved objects and 
one marginally resolved bright object. All of them have absolute magnitudes 
$-14.5<$ {\it M$_I$} $< -12$, well in the UCD luminosity range.  We have 6 
type I UCD candidates (including the one marginally 
resolved), 3 type II and 6 type III.\\

\noindent {\bf NGC 3923} The GC bimodality is easily detectable, but only 
for objects with {\it M$_I$} $> -11$. In contrast with NGC 4649, the brightest 
objects show a broad and nearly unimodal distribution. This 
phenomenon has previously been observed in other giant galaxies, such as 
NGC 1399 \citep{OFG98,DRGFBG03}.
 
In this galaxy we find 12 marginally resolved objects, all of them with blue
colours: {\it (g$'$--i$'$)$_0 < 1.0$}. A few of them have been confirmed by \citet{N2011} as members of the NGC 3923 system. In particular, object ID\#285 is a clearly 
extended object. We have also found 32 unresolved
objects with  $ -15 < $ {\it M$_I$} $< -12$, (3 type I and 10 type II). The remaining 19
type III UCD candidates are very probably MW stars. One of our type I
candidates, ID\#336, was spectroscopically confirmed by \citet{NK2011}
 as a real UCD (UCD1 in their Table 3). Regarding marginally resolved
UCD candidates, we have found 7, a relatively large number. Two of them,
ID\#760 and ID\#1030, were confirmed as members of the NGC 3923 system by
\citet{N2011} and ID\#243 by \citet{NK2011} (UCD2 in their
Table 3).\\

\noindent {\bf NGC 524} This galaxy shows a broad GC colour distribution 
without clear evidence of bimodality. This is similar to 
that found in some other GCSs like NGC 1427 \citep{FGOPG01}. We will 
return to this point in 
Section \ref{Histo}. The GC candidates show a large spread in the 
{\it{(g$'$--r$'$) vs. (r$'$--i$'$)}} plane. We have found 8 marginally 
resolved objects with colours consistent with GCs. Curiously all of 
these objects are very red  and only found in Field 1.

Regarding UCD candidates, there are 8 candidates with 
luminosities {\it M$_I$} $ > -15$. Two are type I and 6 type II. The 
classification, however, 
is tentative due to the large spread of these objects in the 
colour-colour diagrams. We have also found one marginally 
resolved UCD candidate. Norris \& Kannappan (2011, in preparation) have
confirmed one UCD object which was not included in our sample
because it is redder than our colour cuts.\\

\noindent {\bf NGC 3115} This galaxy shows the most obvious bimodality 
in our sample except for the brightest GCs  ({\it M$_I$}$<-10.5$). 
Curiously, the colour magnitude diagram shows that the red GCs
do not reach the same low luminosities as the blue ones do. 
In turn, the colour-colour diagram 
defines a very thin sequence with two GC clumps (as in the case of NGC 4649).
 
We also find a noticeable number of resolved GC candidates. A total 
of 38 objects fall in this category, and some of them 
clump at {\it M$_I$}$<-11$. It is interesting to note that most of 
these objects are confirmed GCs by \citet{KZSWF2002} and 
\citet{PKTMSBRGH2004}. There is one resolved UCD 
candidate (presented as a GC in \citealt{PKTMSBRGH2004}) 
and 7 unresolved ones. One of them is type I, and the remaining are type III.\\

\noindent {\bf NGC 3379} This is the poorest GCS and the least 
massive galaxy in our sample. However, bimodality is 
detectable even in the {\it{(g$'$--r$'$) vs. (r$'$--i$'$)}} plane.

There are three marginally resolved GC candidates and 
three possible UCDs but all of them were classified as 
type III objects. One object, with {\it M$_I$}$\sim -12$ and and 
{\it (g$'$-i$'$)$_0$}$\sim 0.8$ (which was spectroscopically confirmed 
by \citet{PBFBGFFZSHP2006} as a GC), 
lies in the transition region between bright GCs and UCDs. We do not 
find any resolved UCDs. Some of the other bright GC candidates 
are indeed confirmed as such by \citet{BZRSR2006} 
and \citet{PKTMSBRGH2004}. Therefore, this 
low-mass system also shows a detectable population of very massive GCs.

\begin{figure*}
\resizebox{0.33\hsize}{!}{\includegraphics{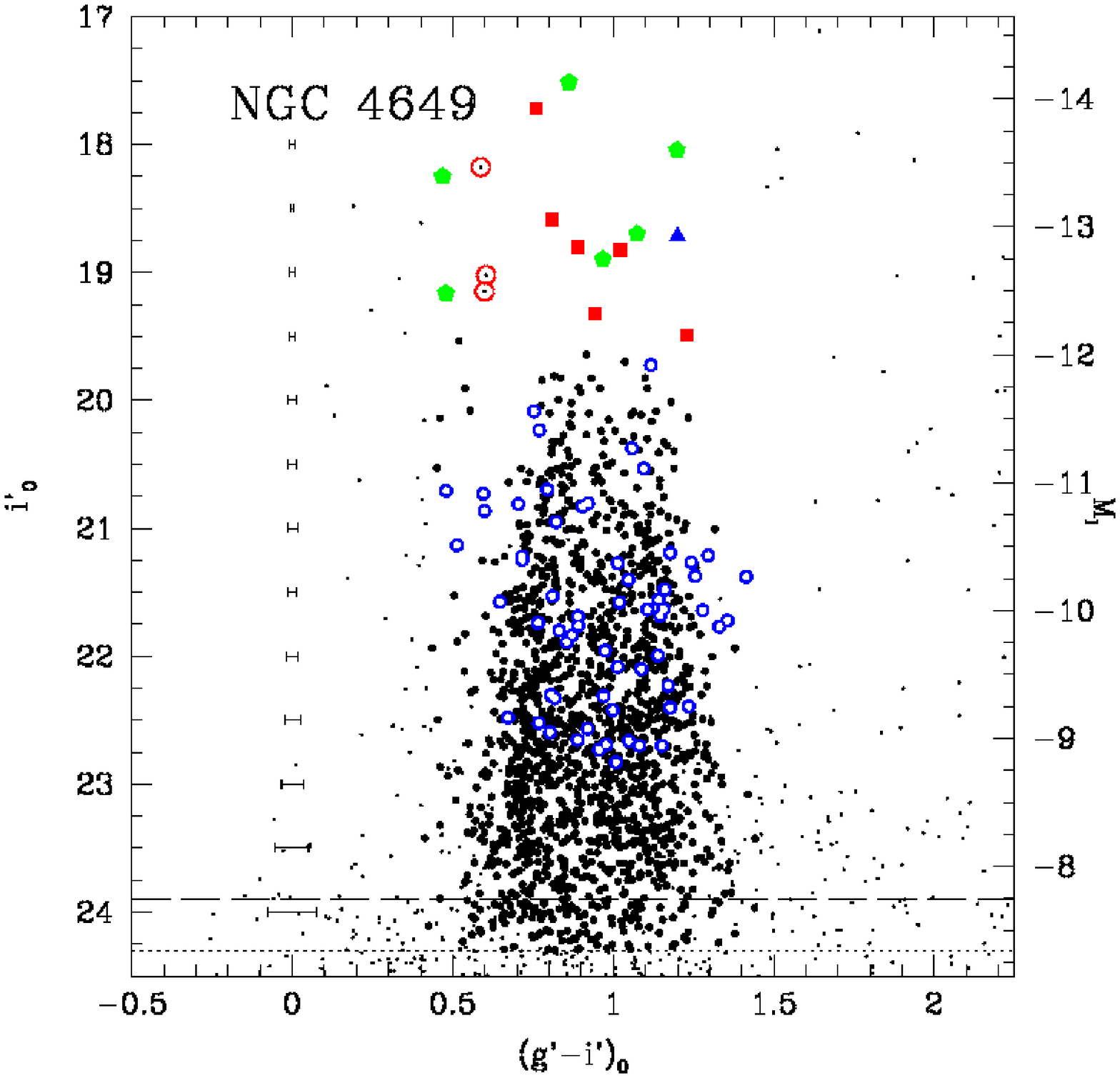}}
\resizebox{0.33\hsize}{!}{\includegraphics{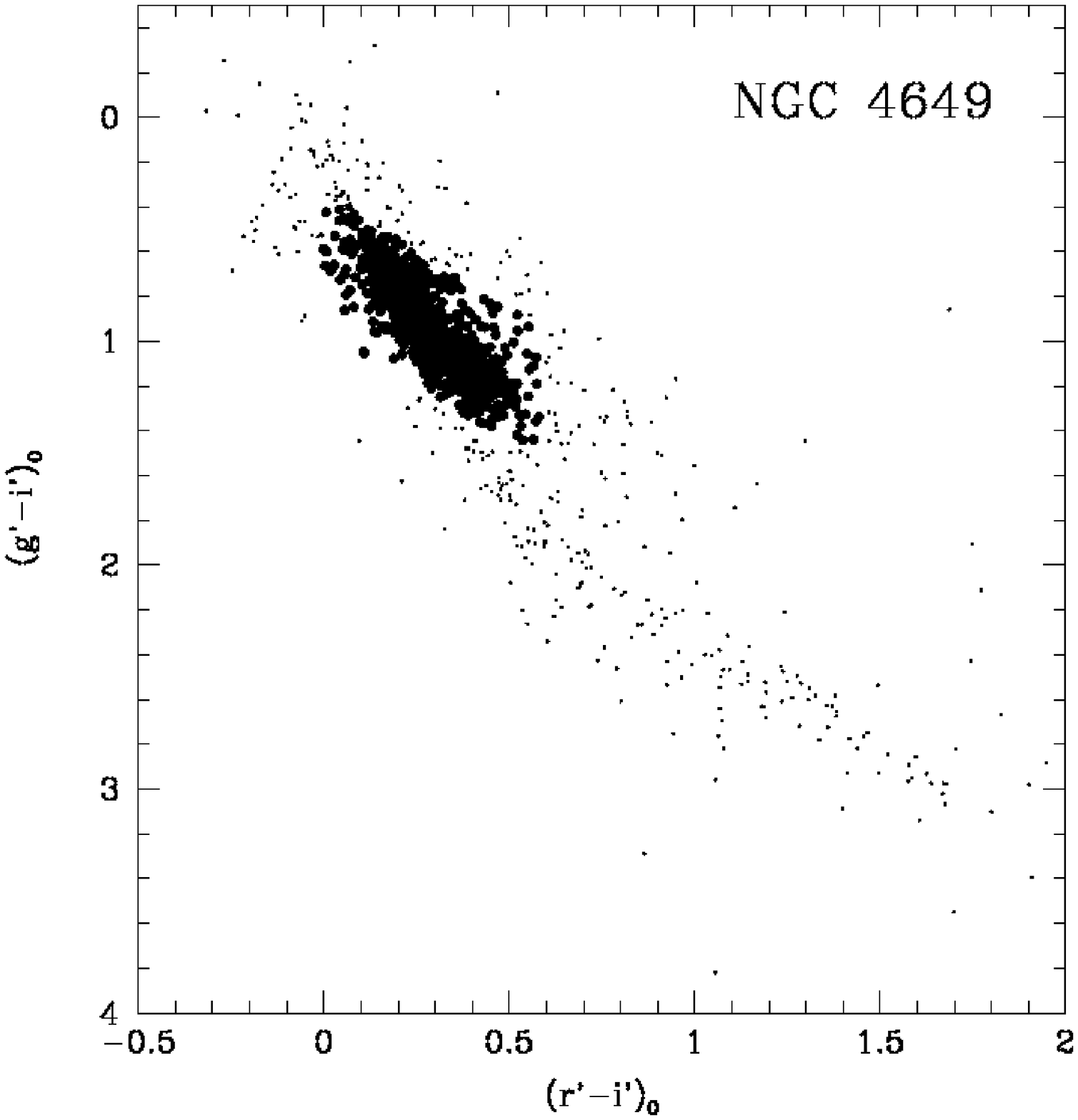}}
\resizebox{0.33\hsize}{!}{\includegraphics{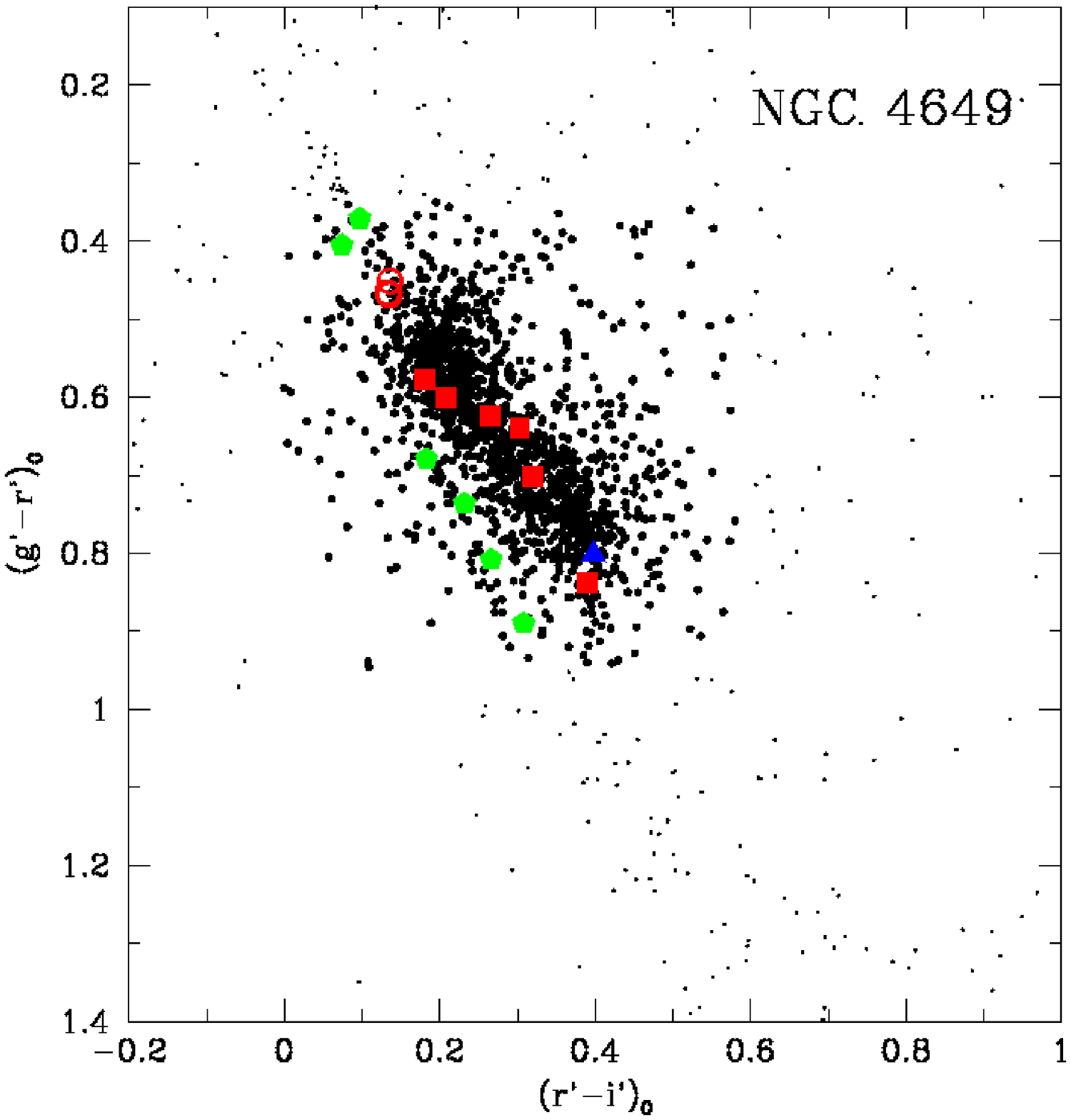}}

\resizebox{0.33\hsize}{!}{\includegraphics{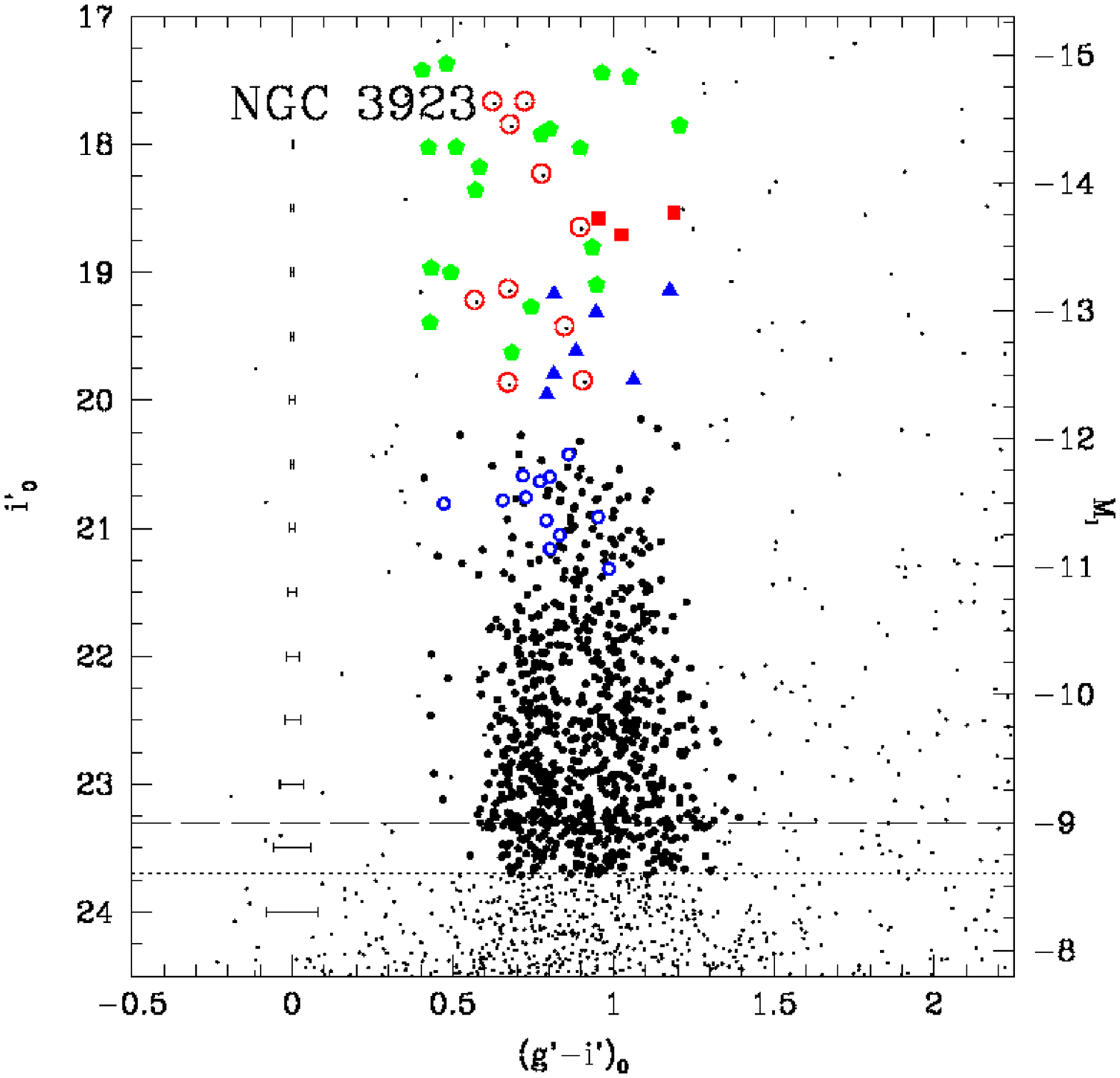}}
\resizebox{0.33\hsize}{!}{\includegraphics{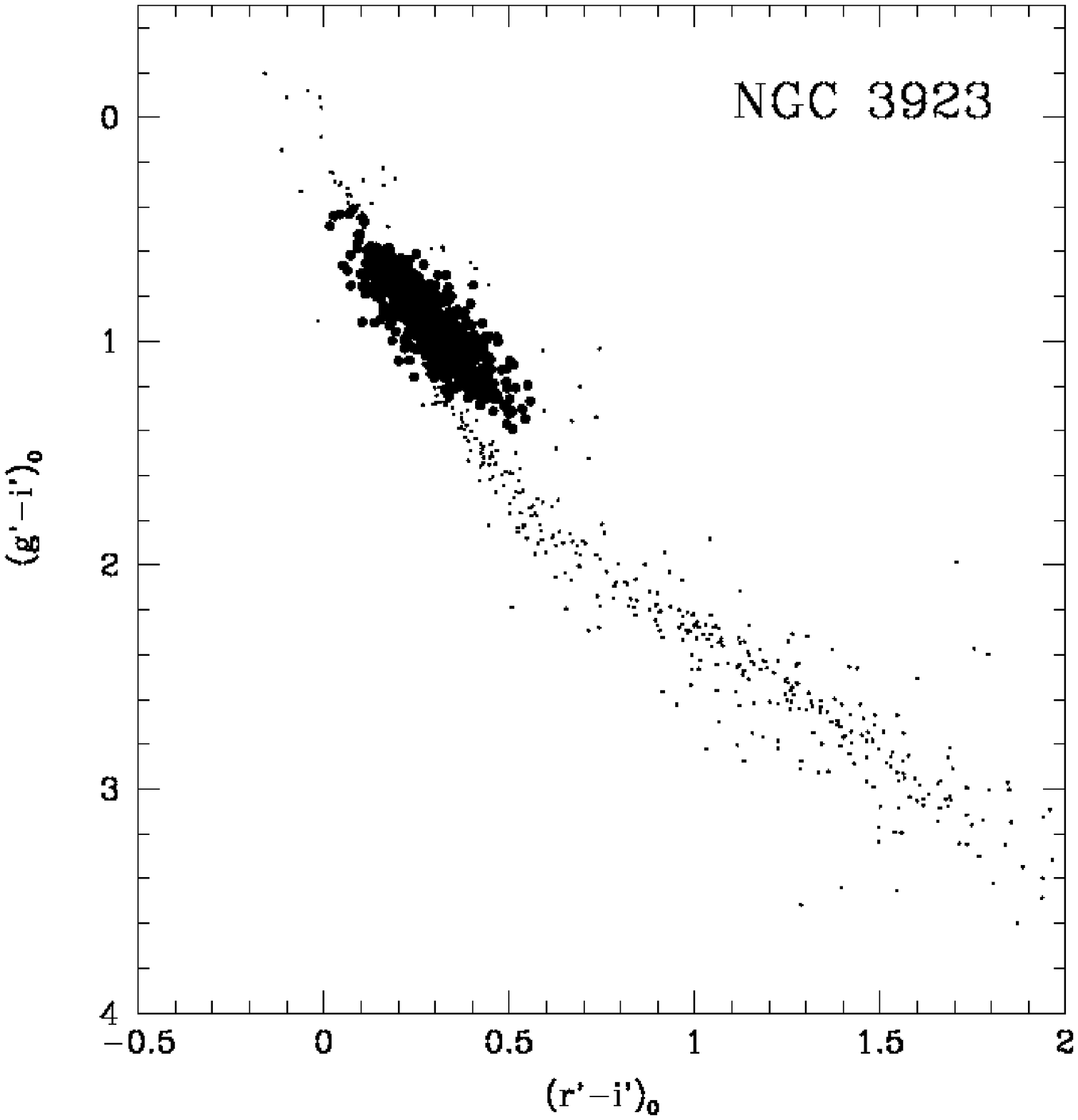}}
\resizebox{0.33\hsize}{!}{\includegraphics{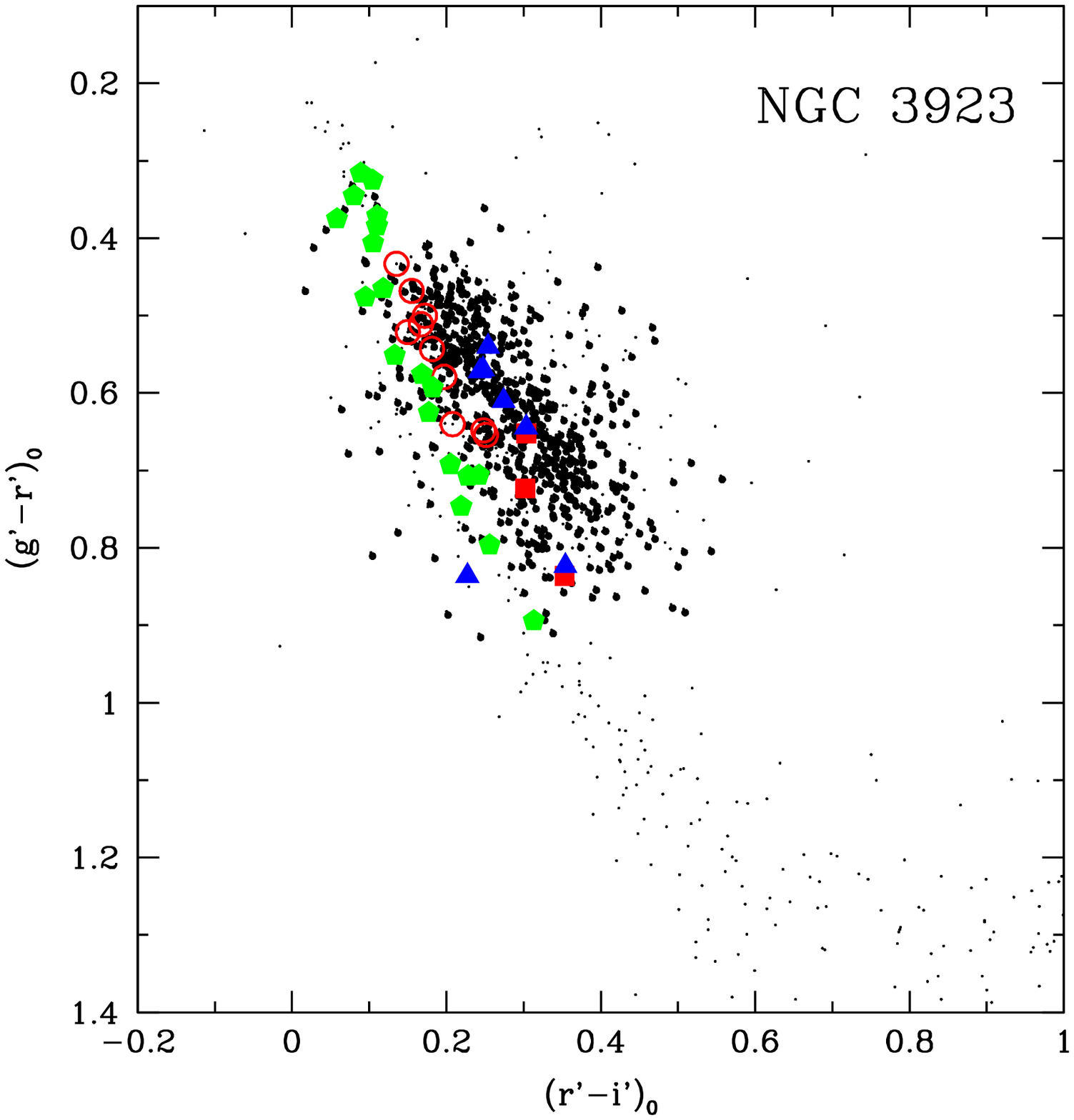}}

\resizebox{0.33\hsize}{!}{\includegraphics{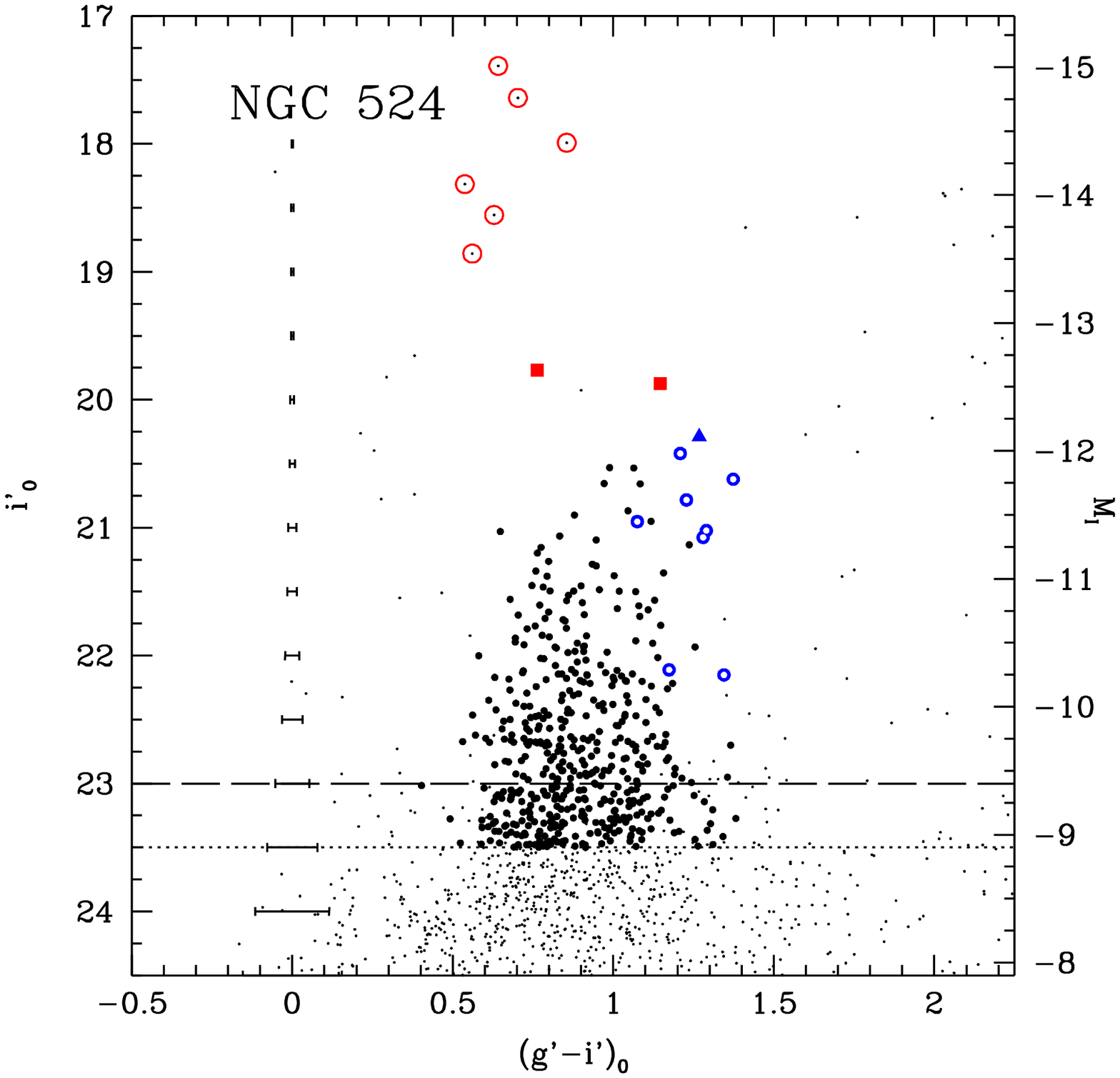}}
\resizebox{0.33\hsize}{!}{\includegraphics{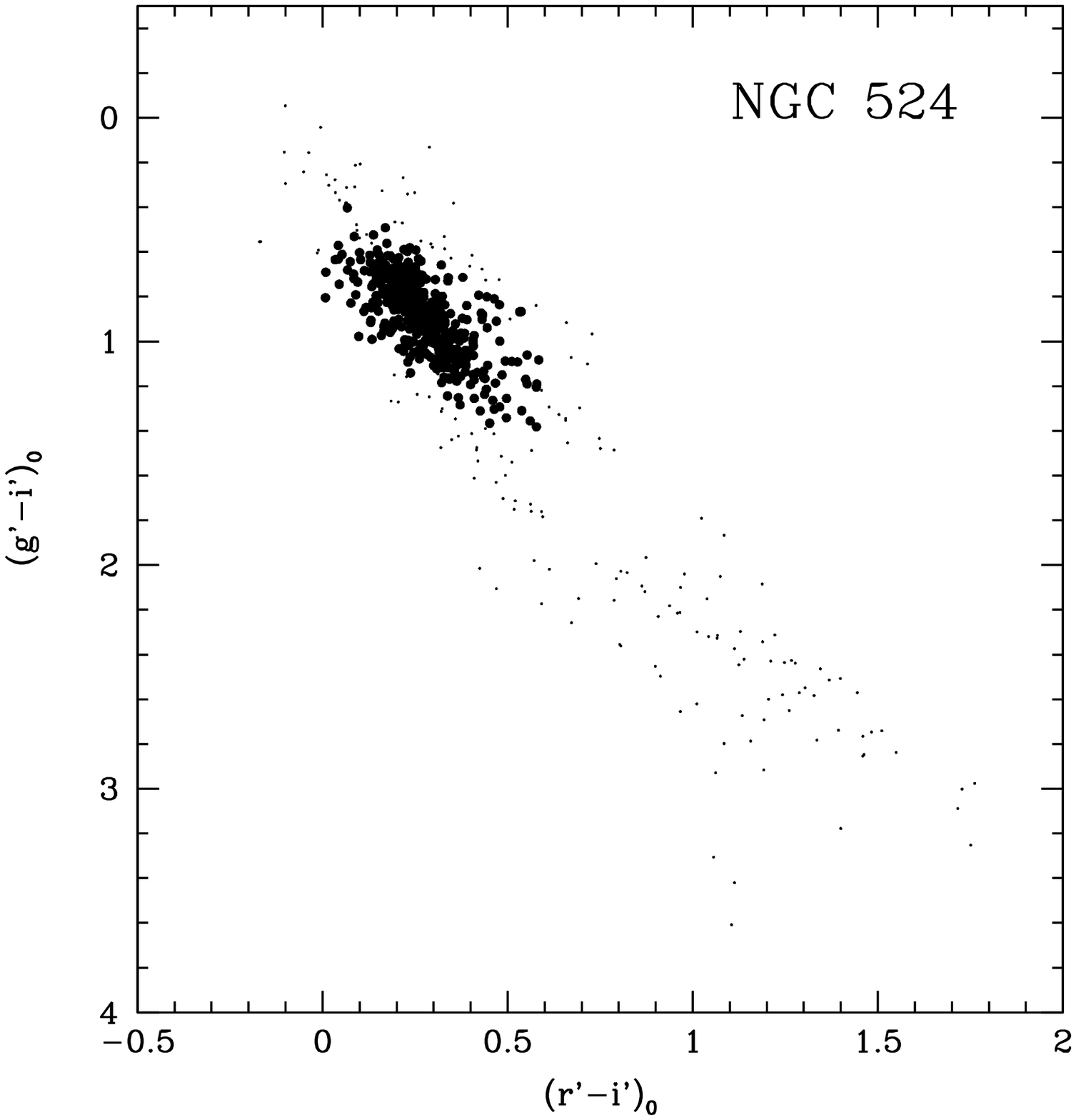}}
\resizebox{0.33\hsize}{!}{\includegraphics{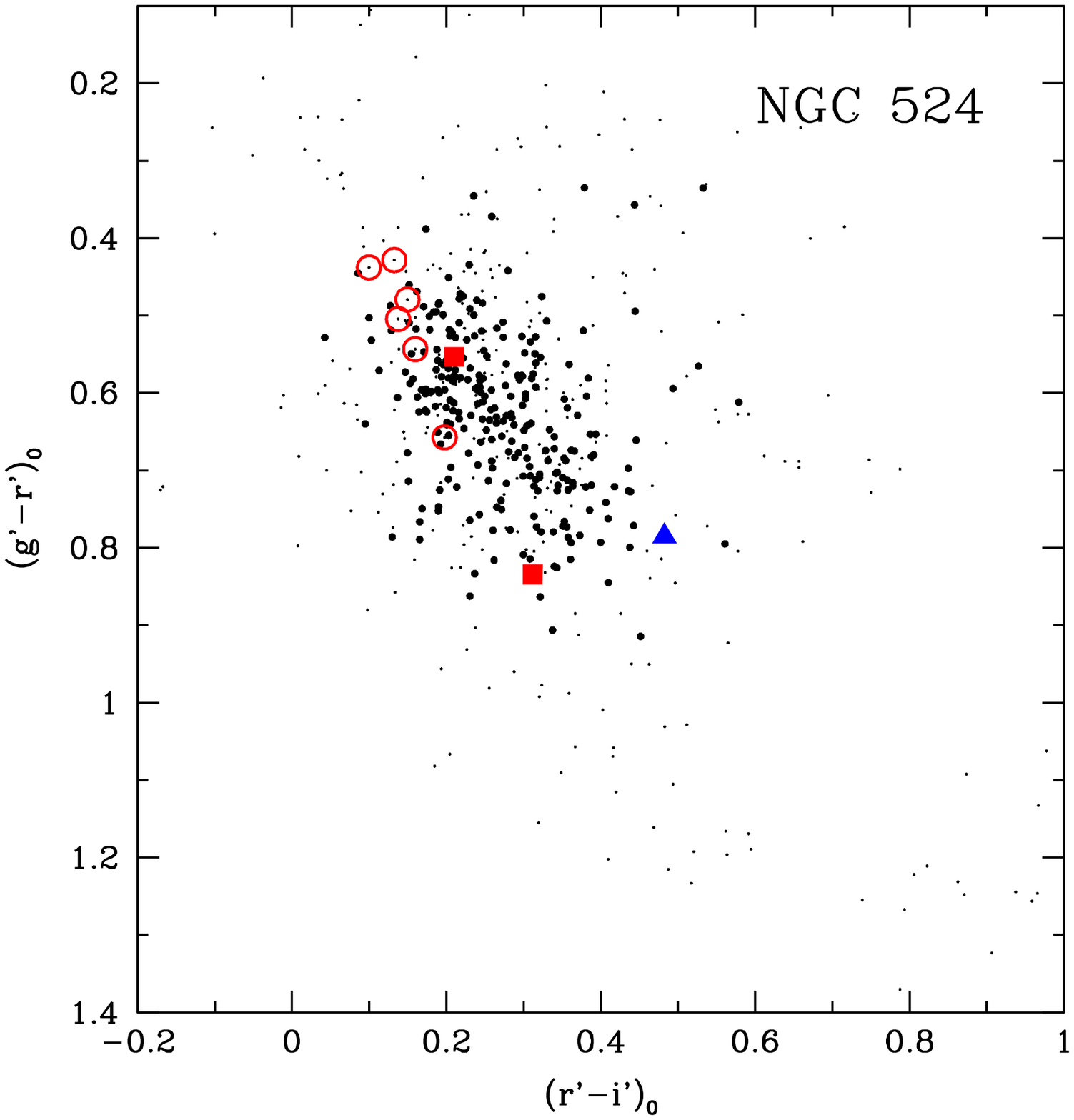}}

\caption{Colour-magnitude and colour-colour diagrams for all the point 
sources detected in our GMOS fields (small dots). The doted and 
long-dashed lines indicate the 50\% and 90\% completeness levels, 
respectively. The mean errors in {\it (g$'$--i$'$)} are plotted as 
small bars in the CMDs. Big dots are our GC candidates 
selected as indicated in the text. In the colour-magnitude and in 
the {\it{(g$'$--r$'$) vs. (r$'$--i$'$)}}  diagrams we show the UCD candidates: type I 
(red filled squares), type II (red circles) and type III (green pentagons). The marginally 
resolved GCs are plotted as big blue circles and the resolved UCDs ones as blue filled triangles.}

\end{figure*}

\setcounter{figure}{3}
\begin{figure*}
\resizebox{0.33\hsize}{!}{\includegraphics{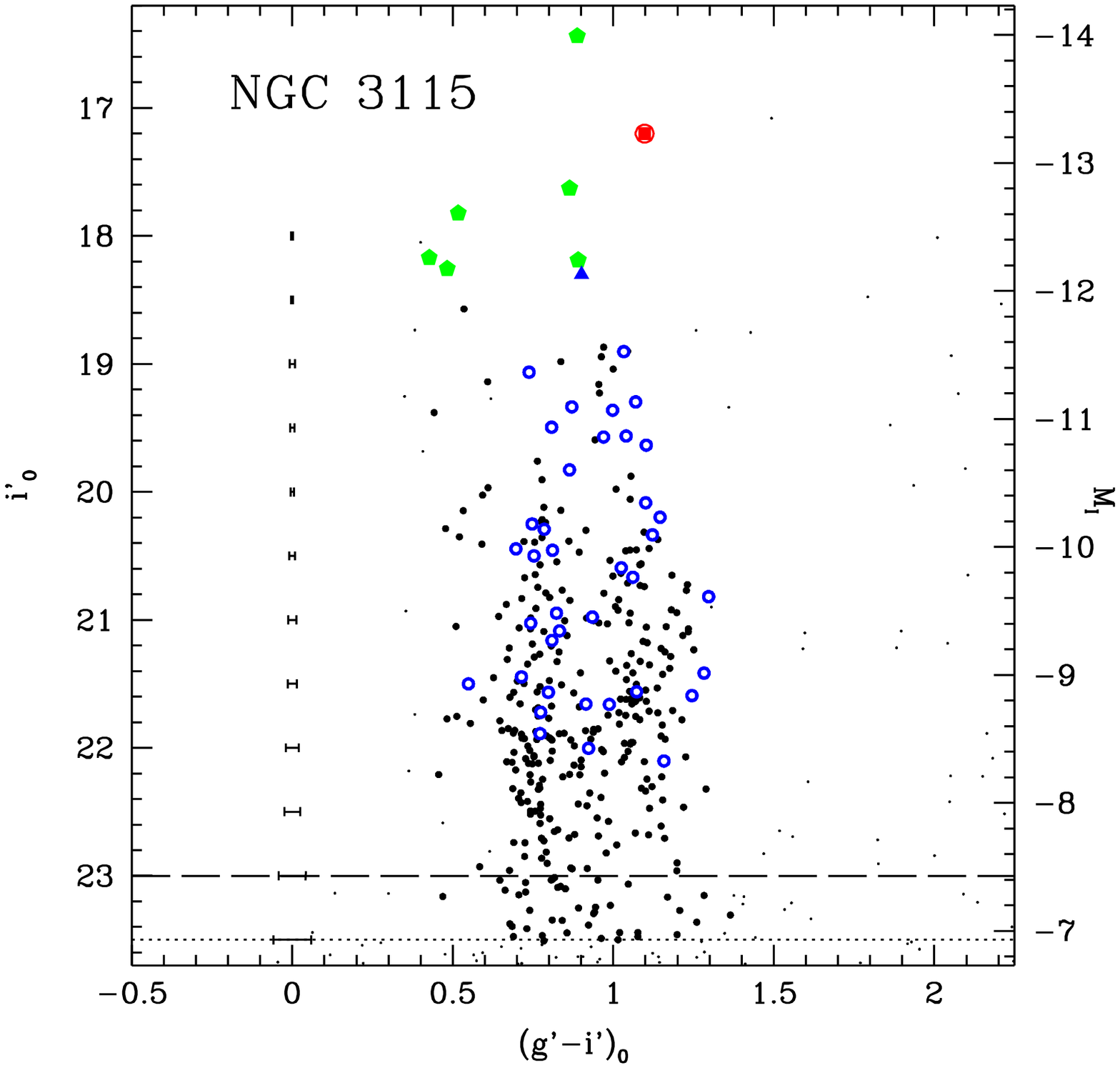}}
\resizebox{0.33\hsize}{!}{\includegraphics{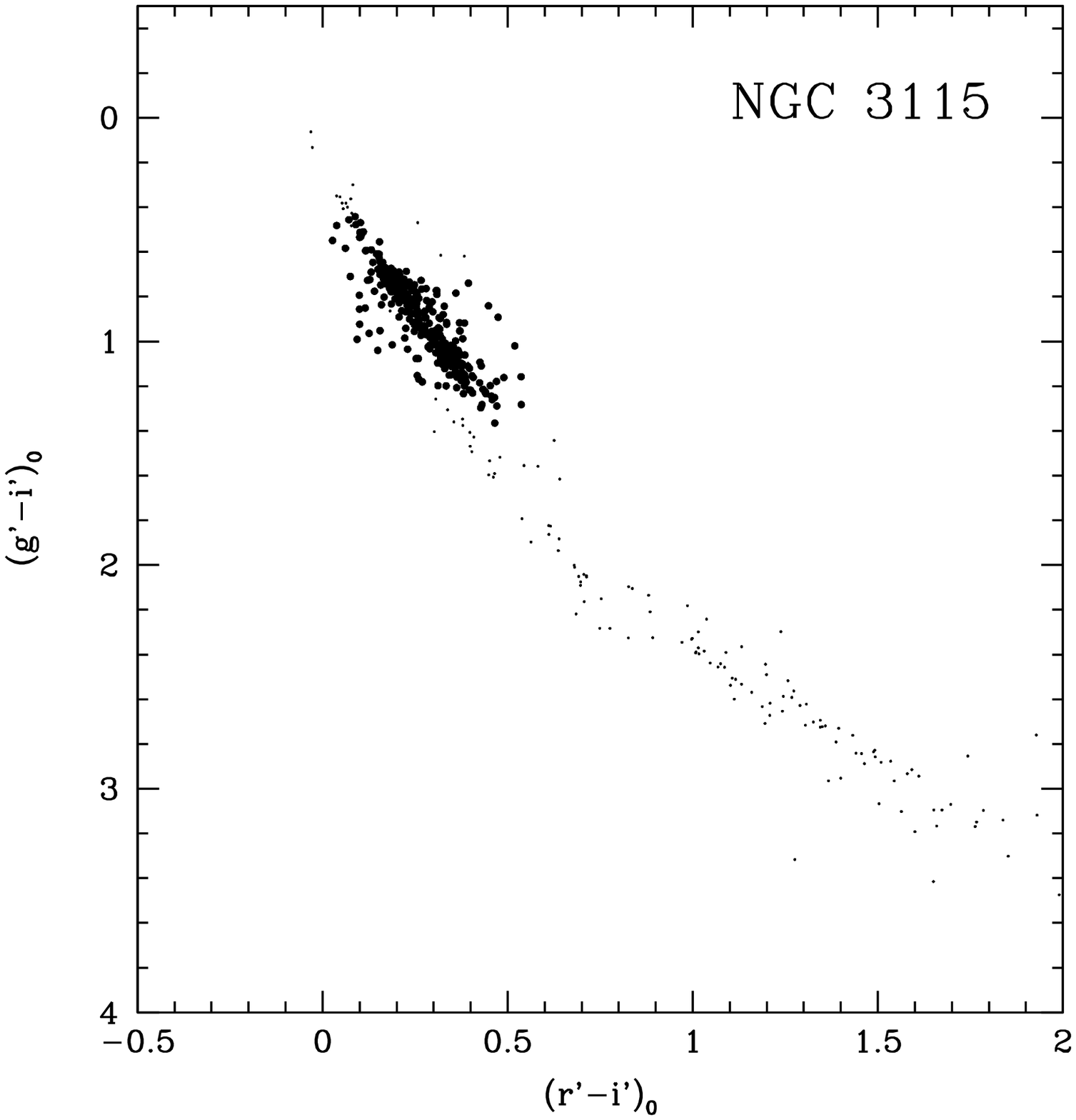}}
\resizebox{0.33\hsize}{!}{\includegraphics{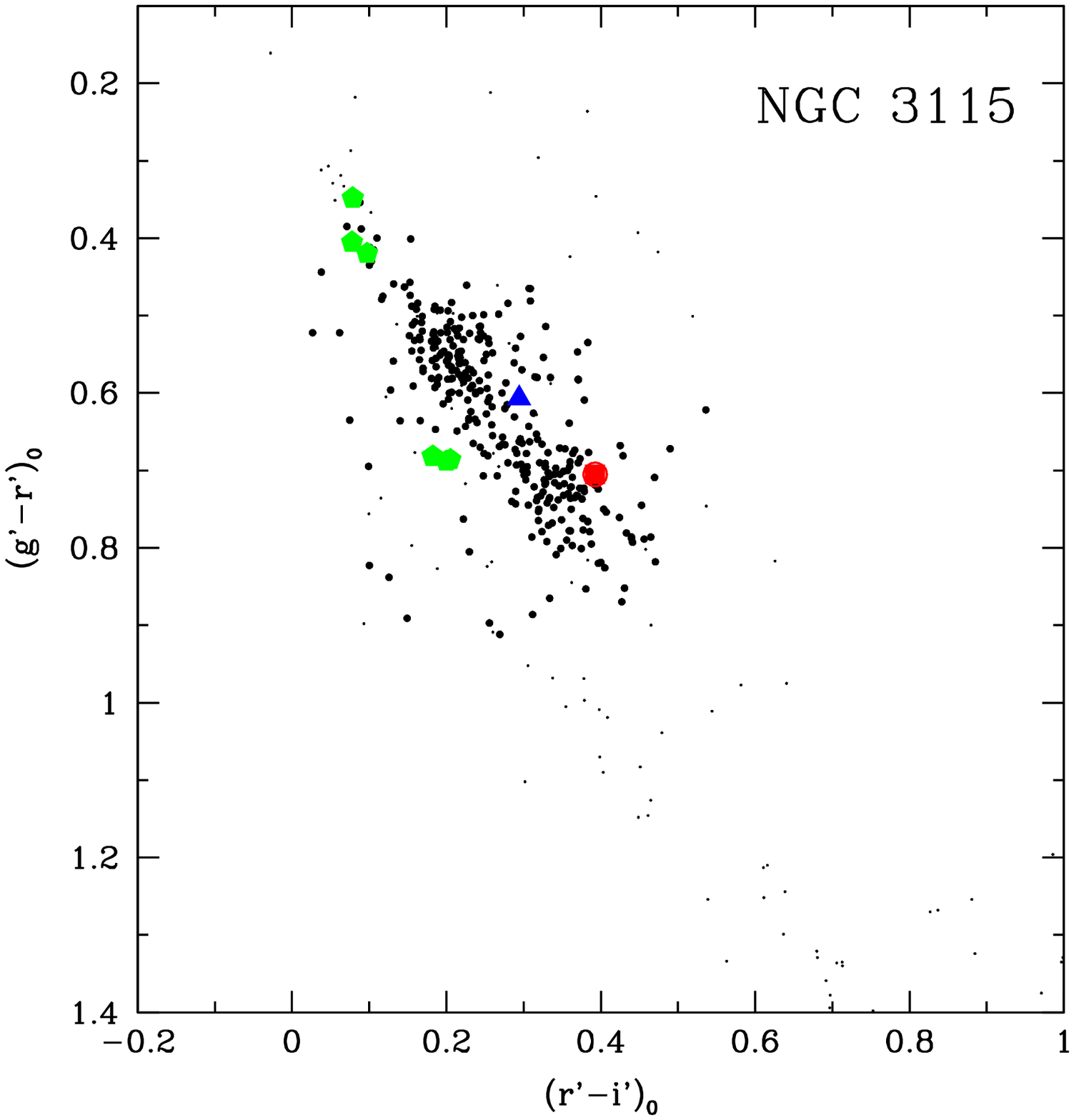}}

\resizebox{0.33\hsize}{!}{\includegraphics{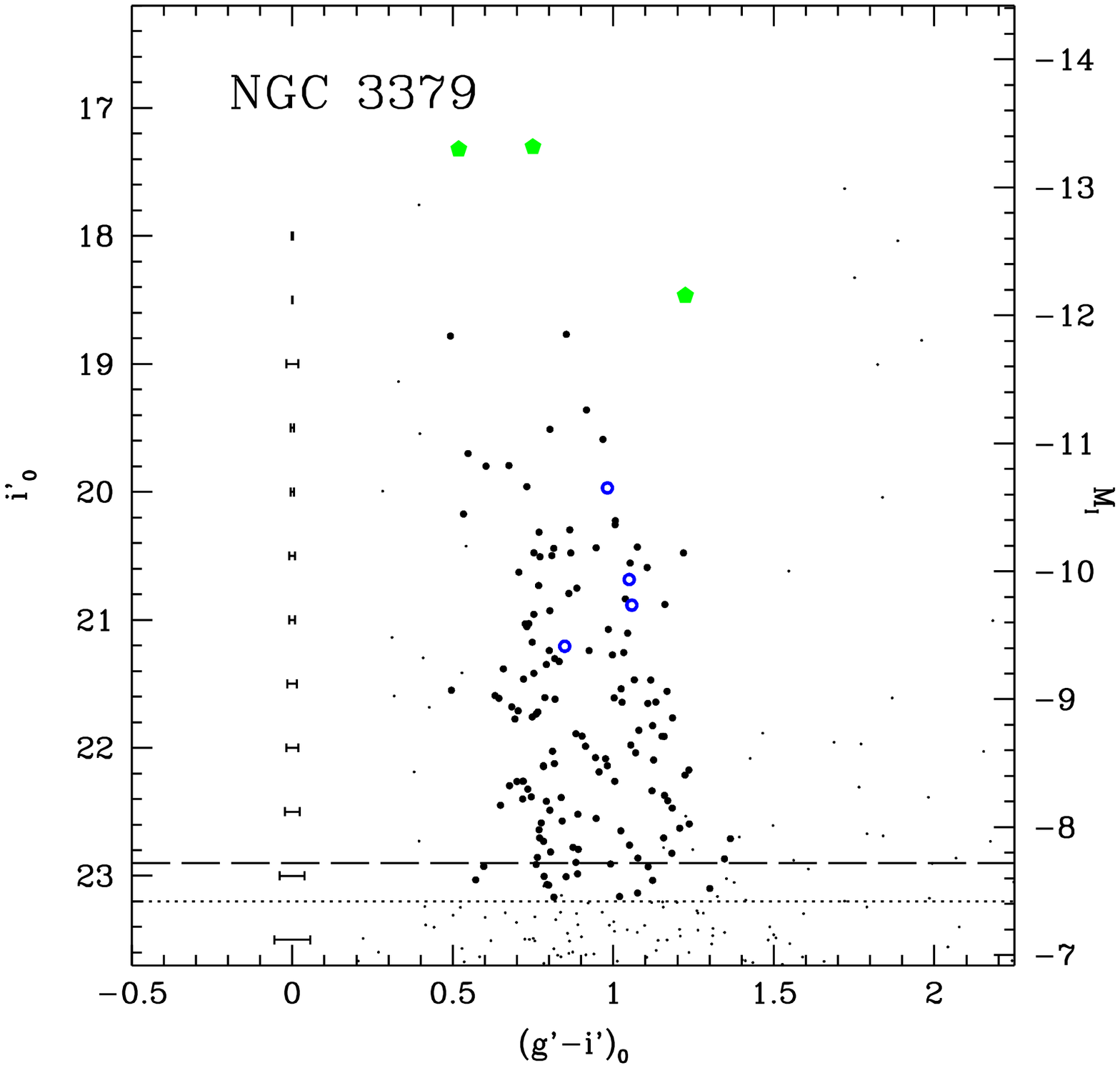}}
\resizebox{0.33\hsize}{!}{\includegraphics{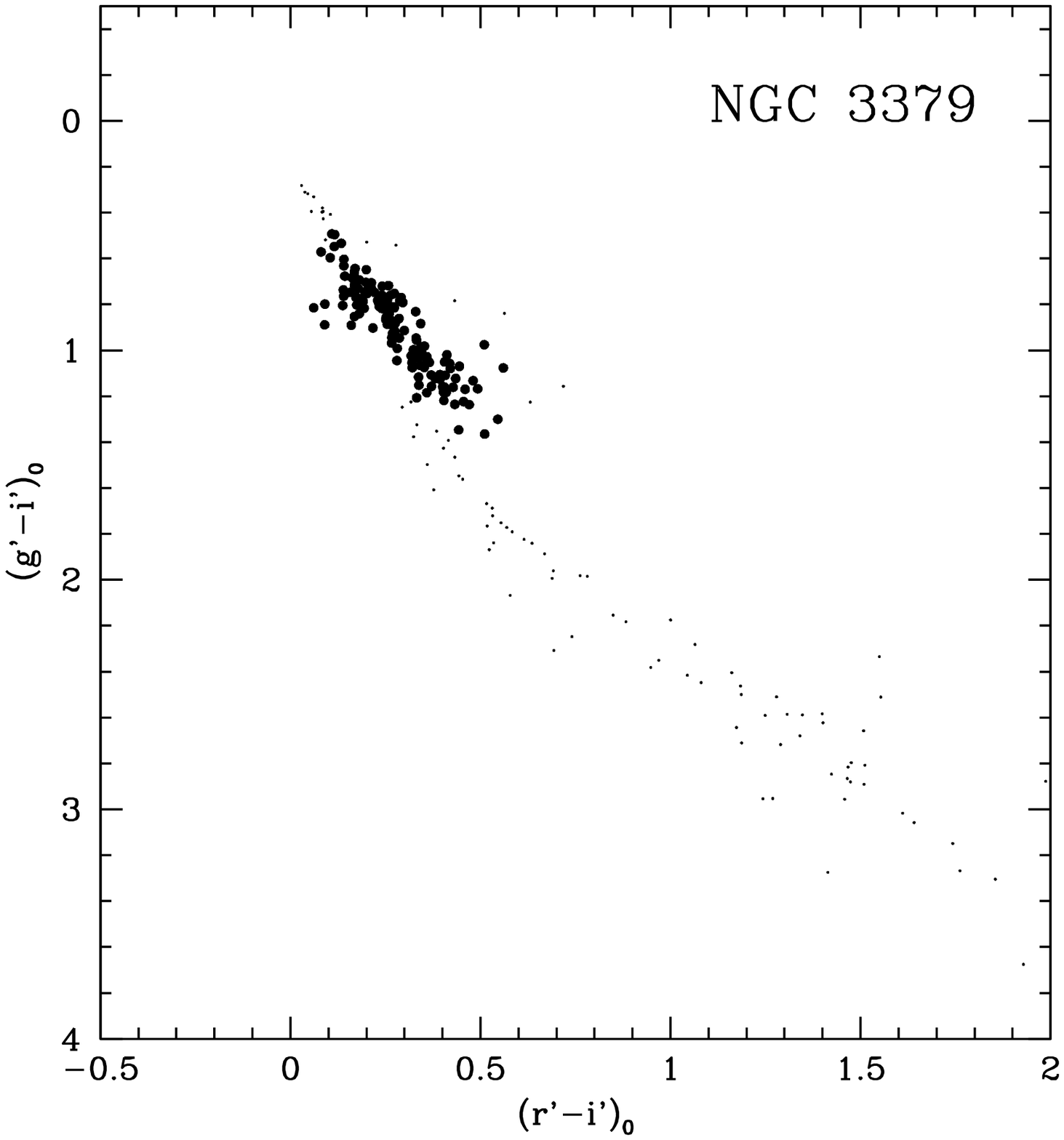}}
\resizebox{0.33\hsize}{!}{\includegraphics{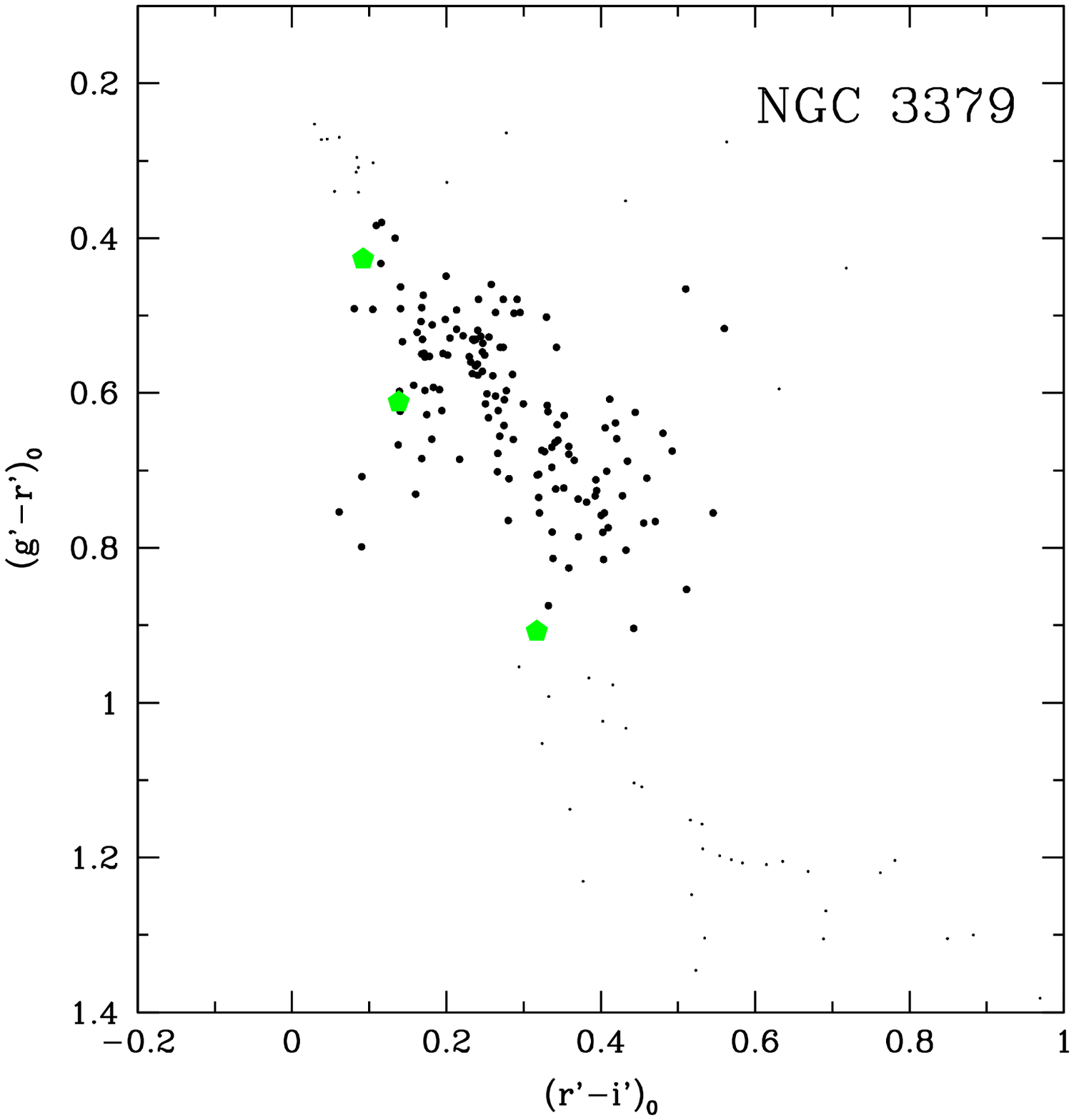}}

\caption{Continued.}

\label{CMD_CCD_GC}
\end{figure*}

\subsection{Colour Histograms}
\label{Histo}

\begin{figure}
\resizebox{0.5\hsize}{!}{\includegraphics{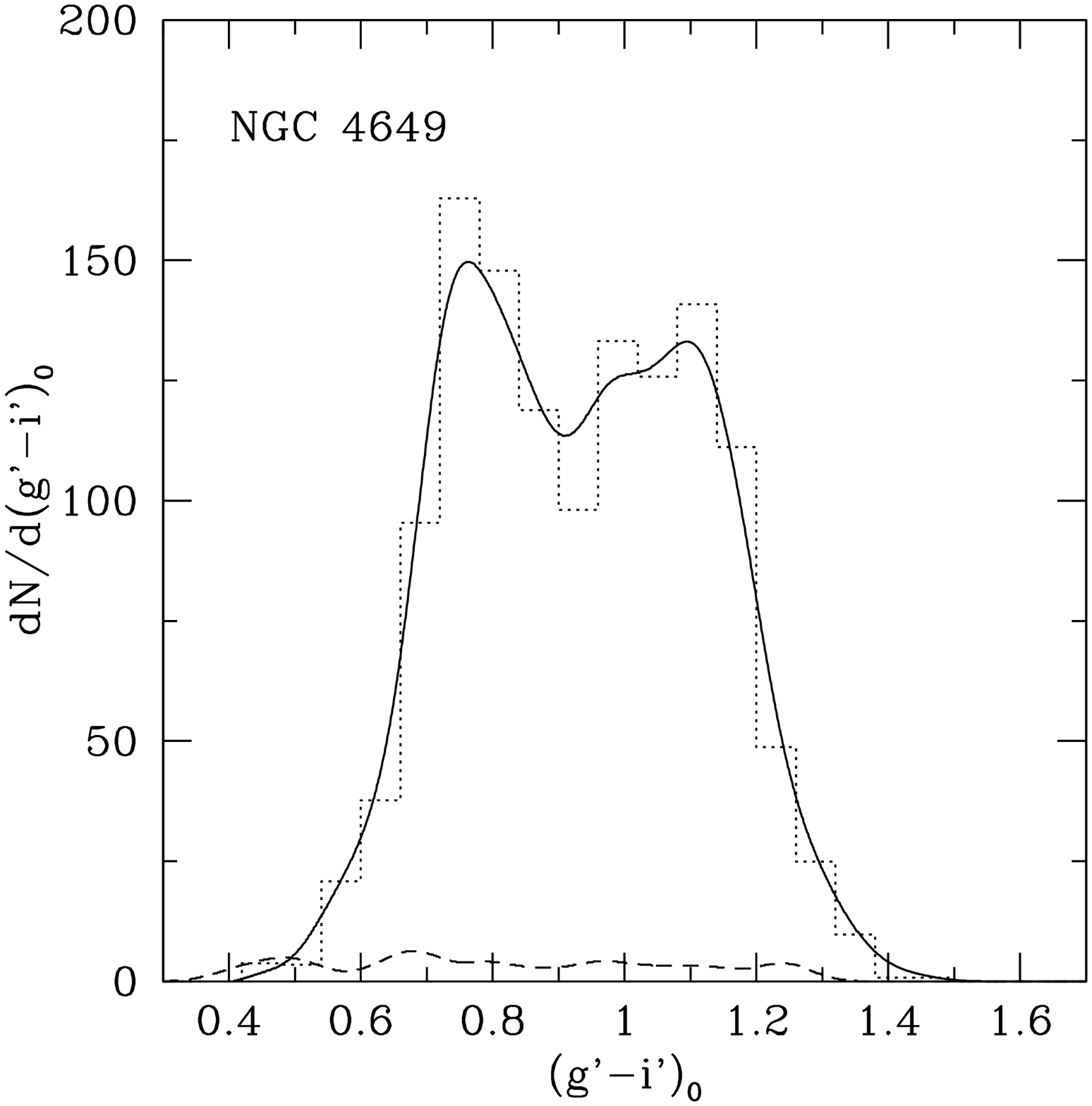}}
\resizebox{0.5\hsize}{!}{\includegraphics{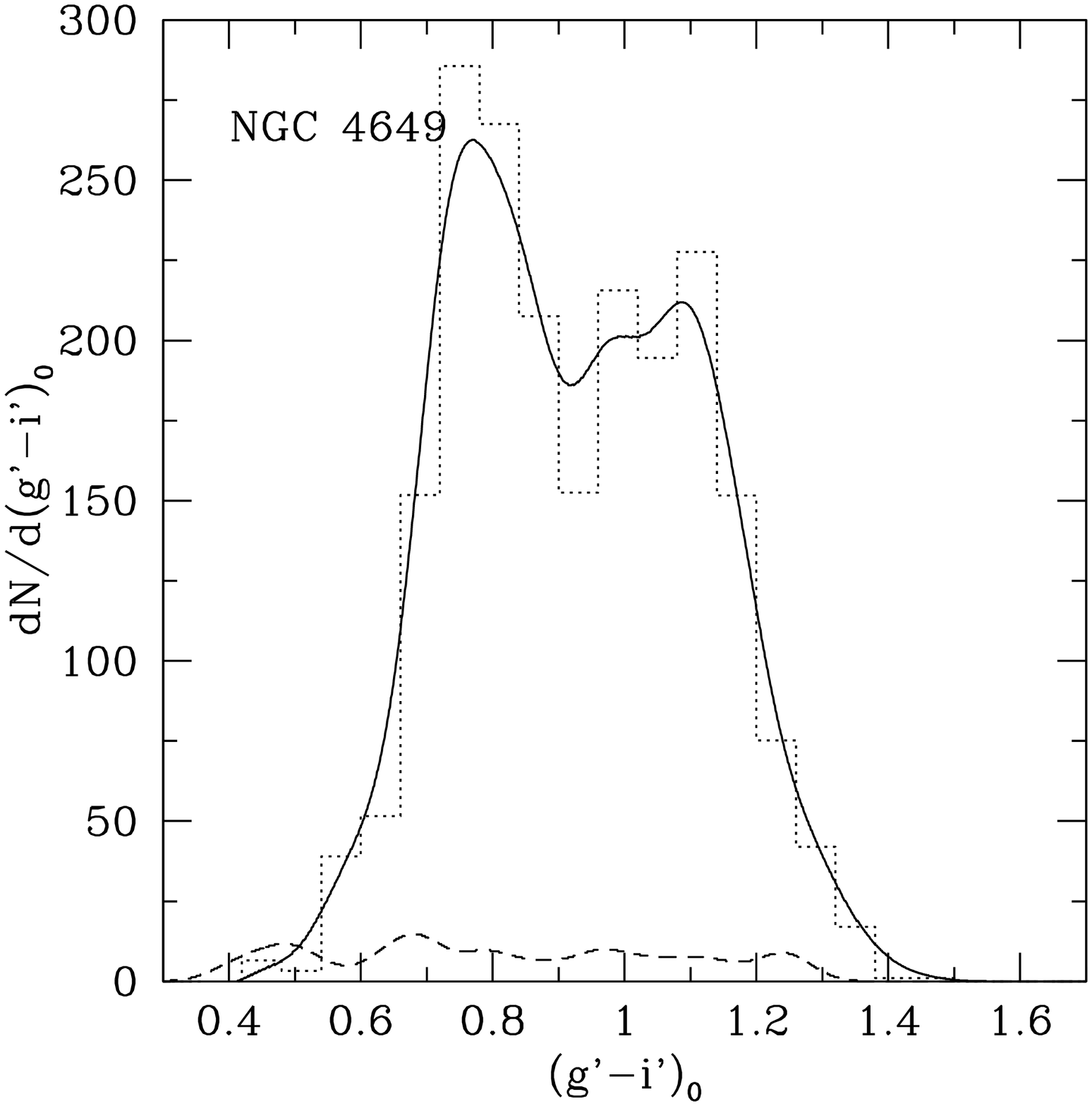}}\\
\resizebox{0.5\hsize}{!}{\includegraphics{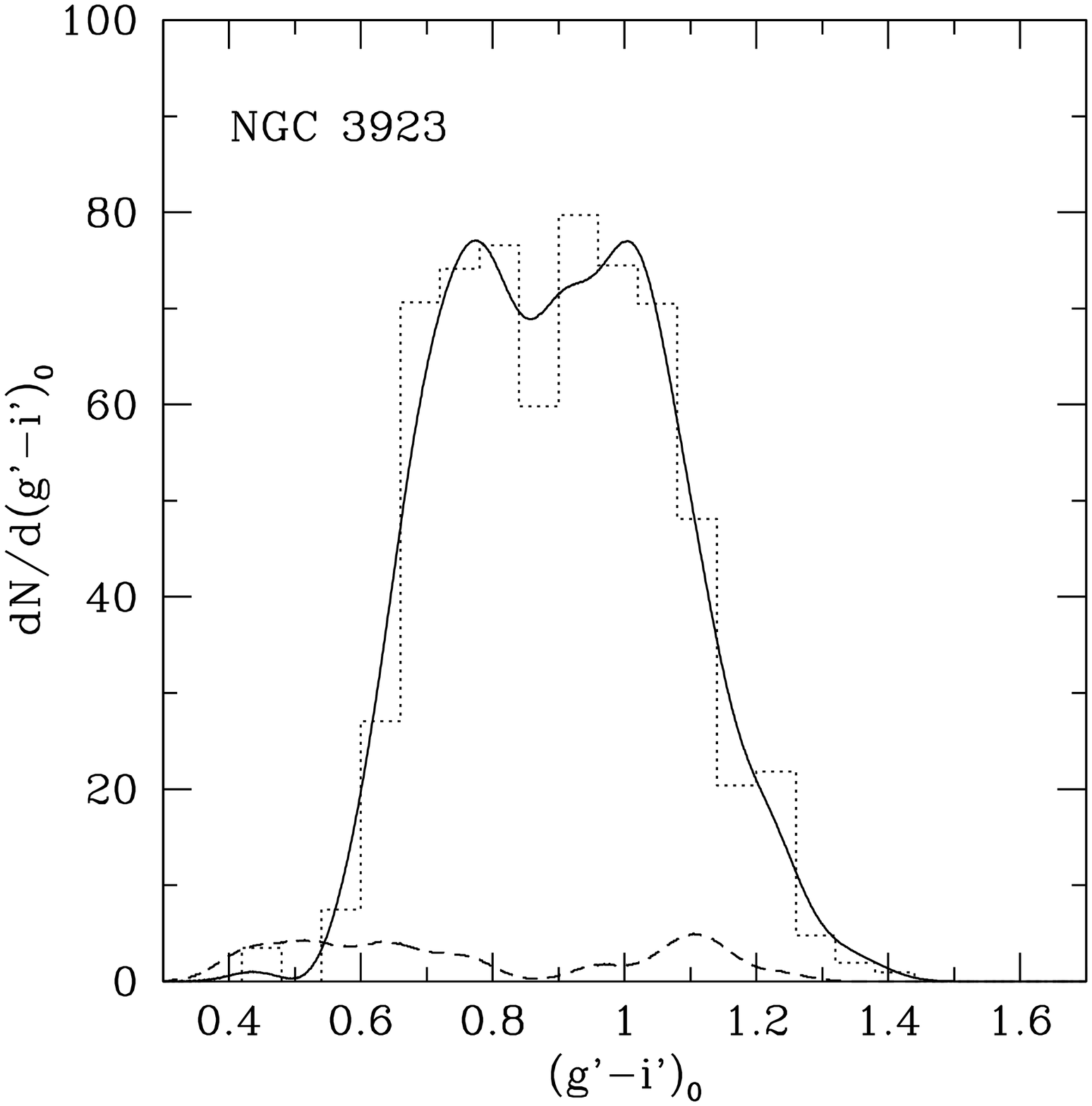}}
\resizebox{0.5\hsize}{!}{\includegraphics{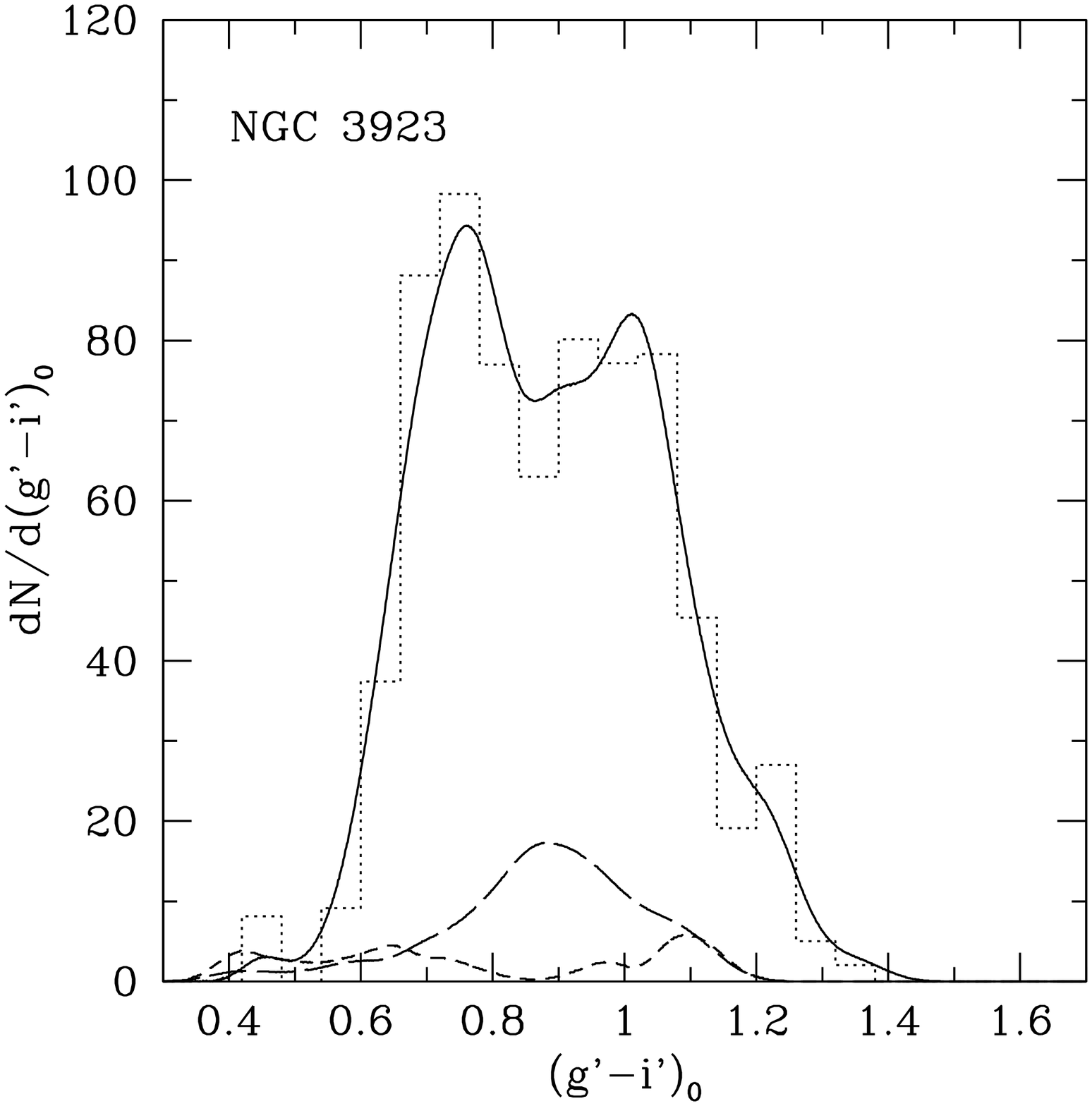}}\\
\resizebox{0.5\hsize}{!}{\includegraphics{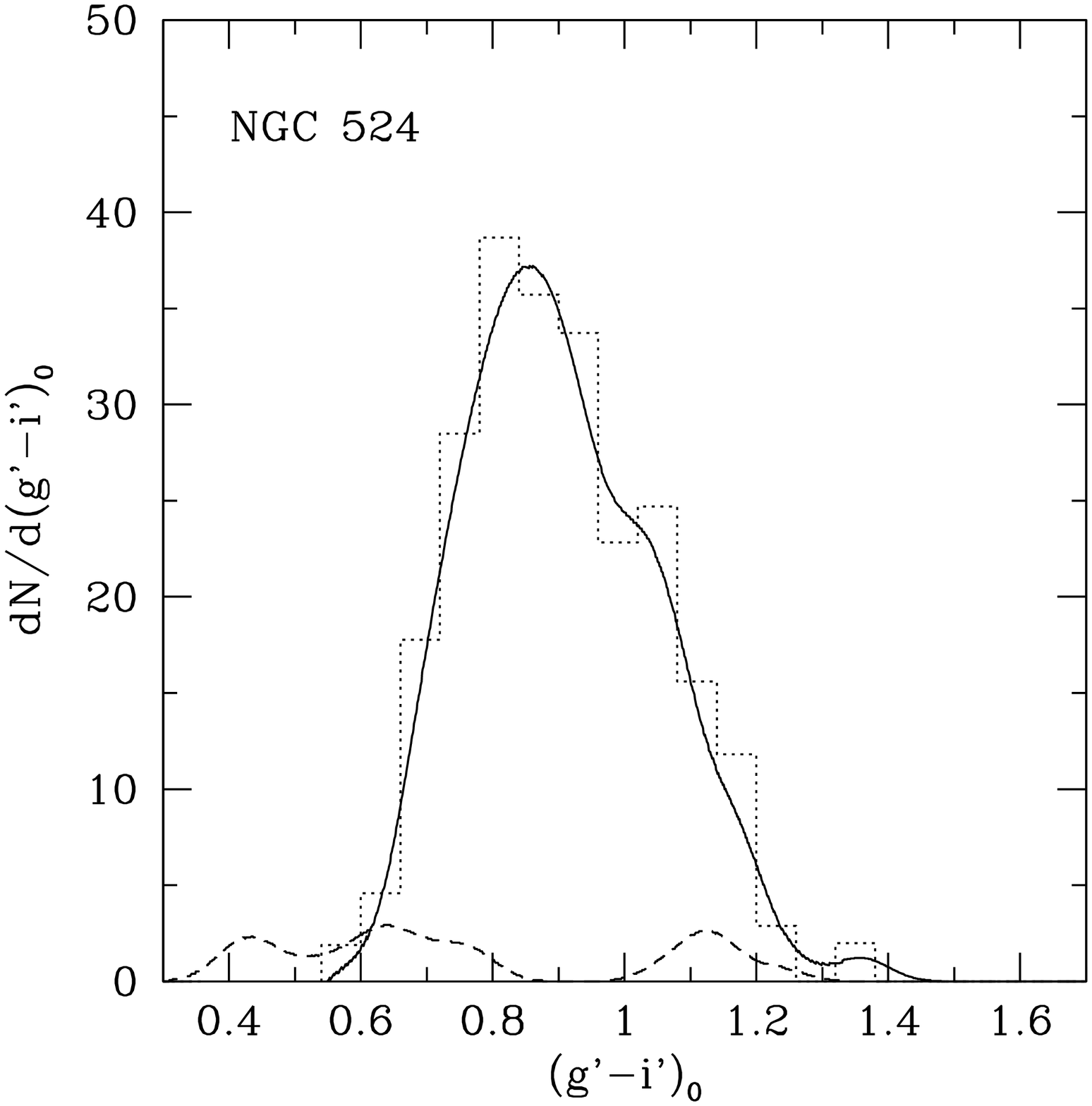}}
\resizebox{0.5\hsize}{!}{\includegraphics{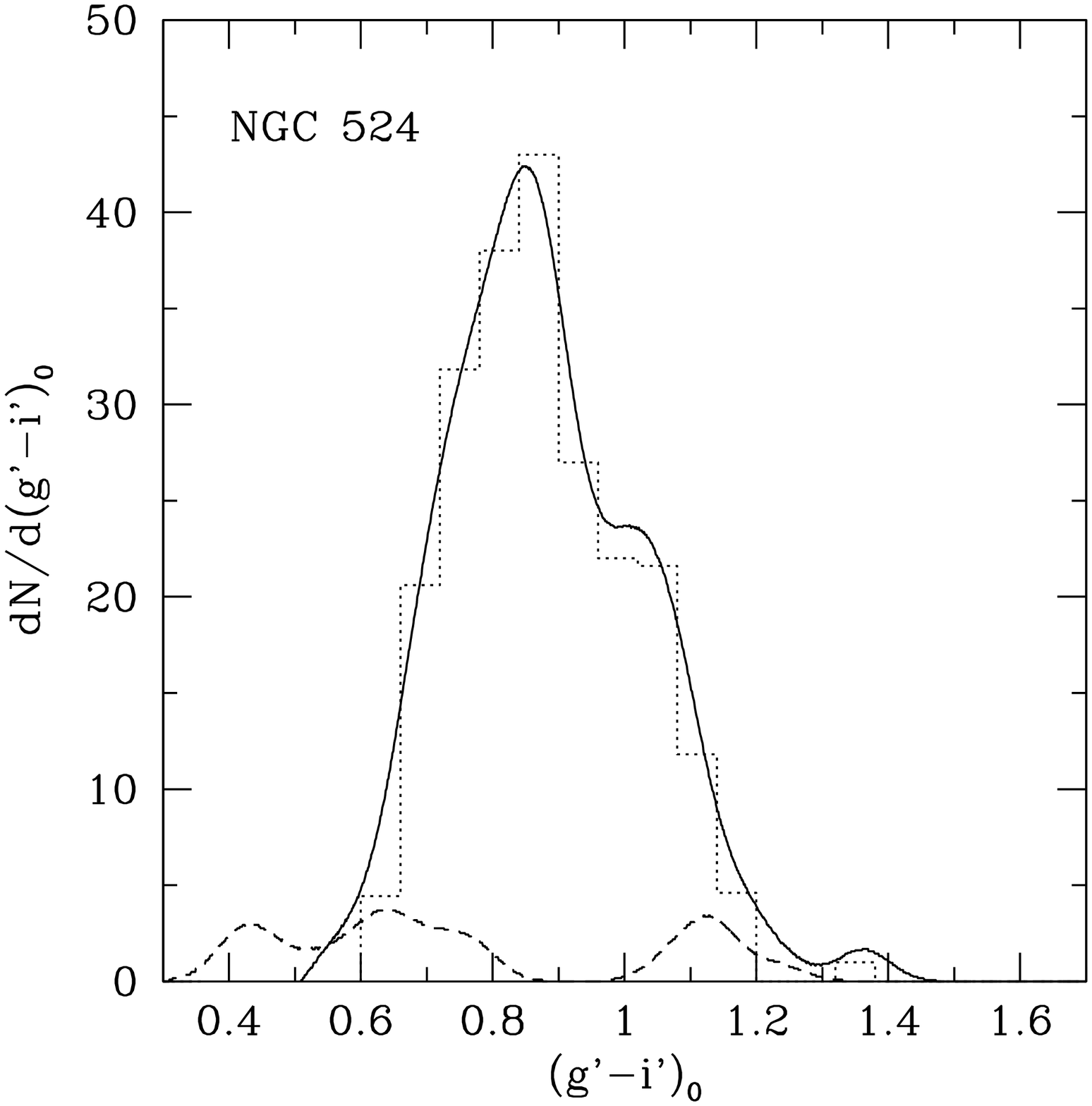}}\\
\resizebox{0.5\hsize}{!}{\includegraphics{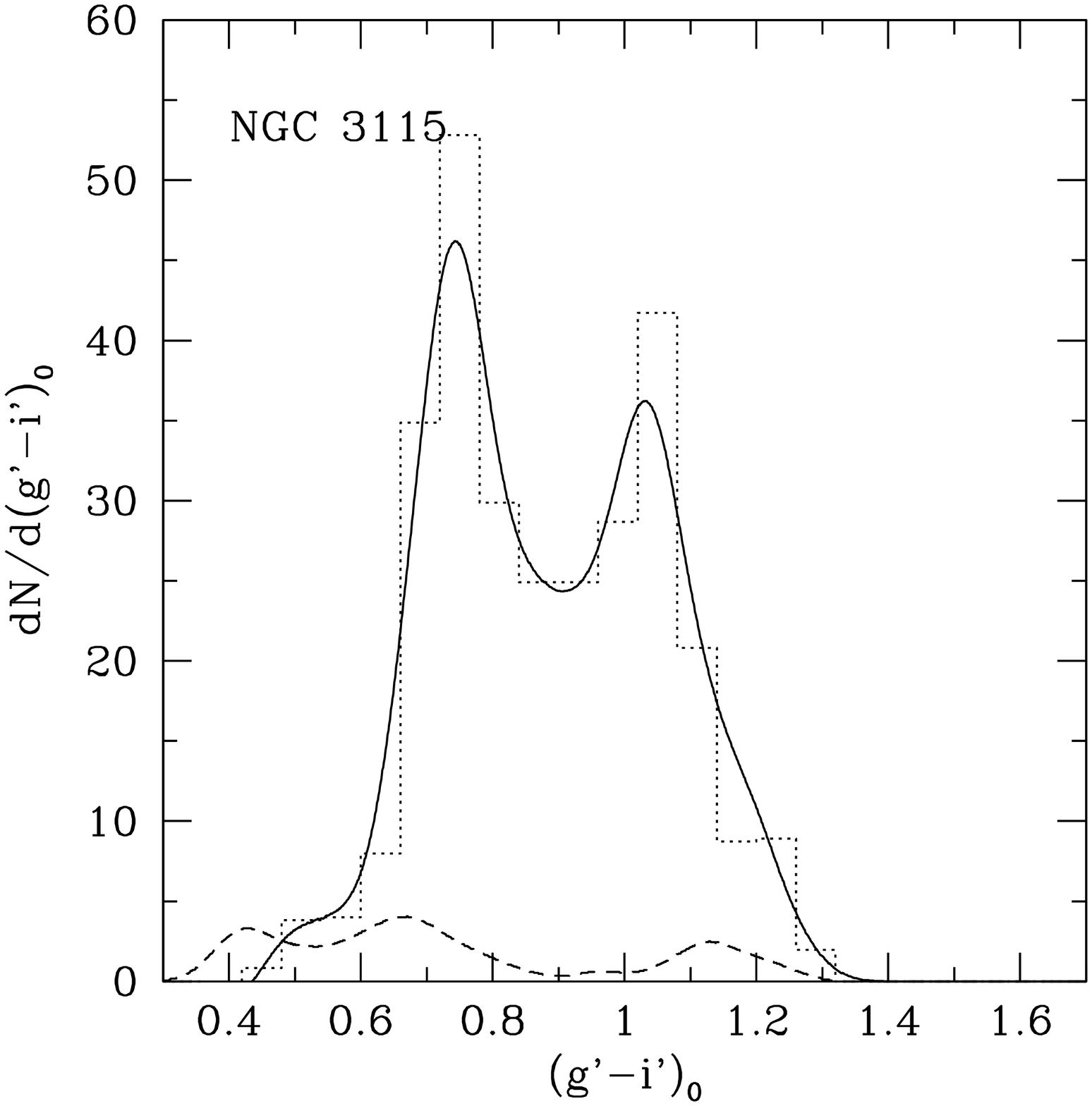}}
\resizebox{0.5\hsize}{!}{\includegraphics{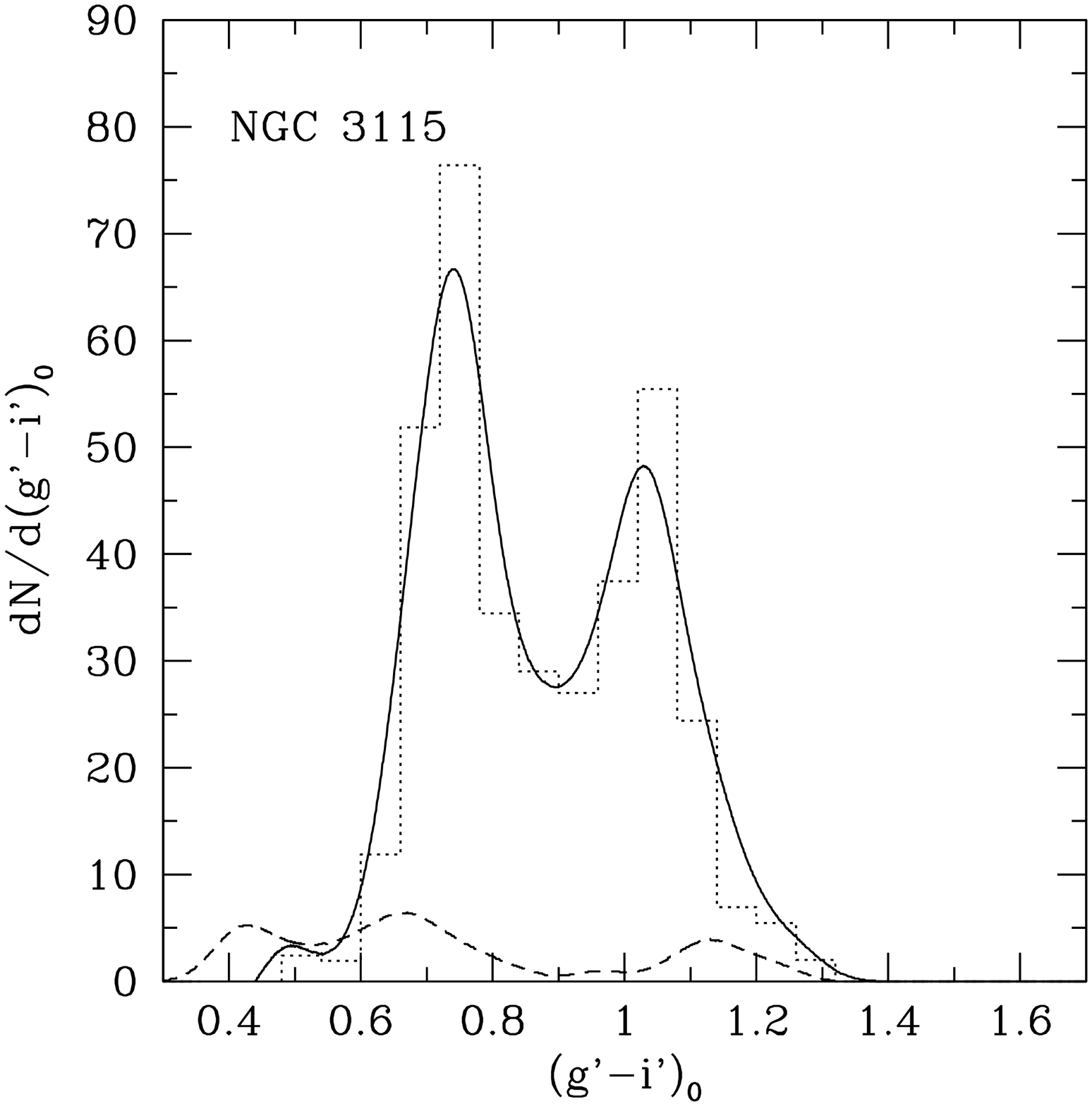}}\\
\resizebox{0.5\hsize}{!}{\includegraphics{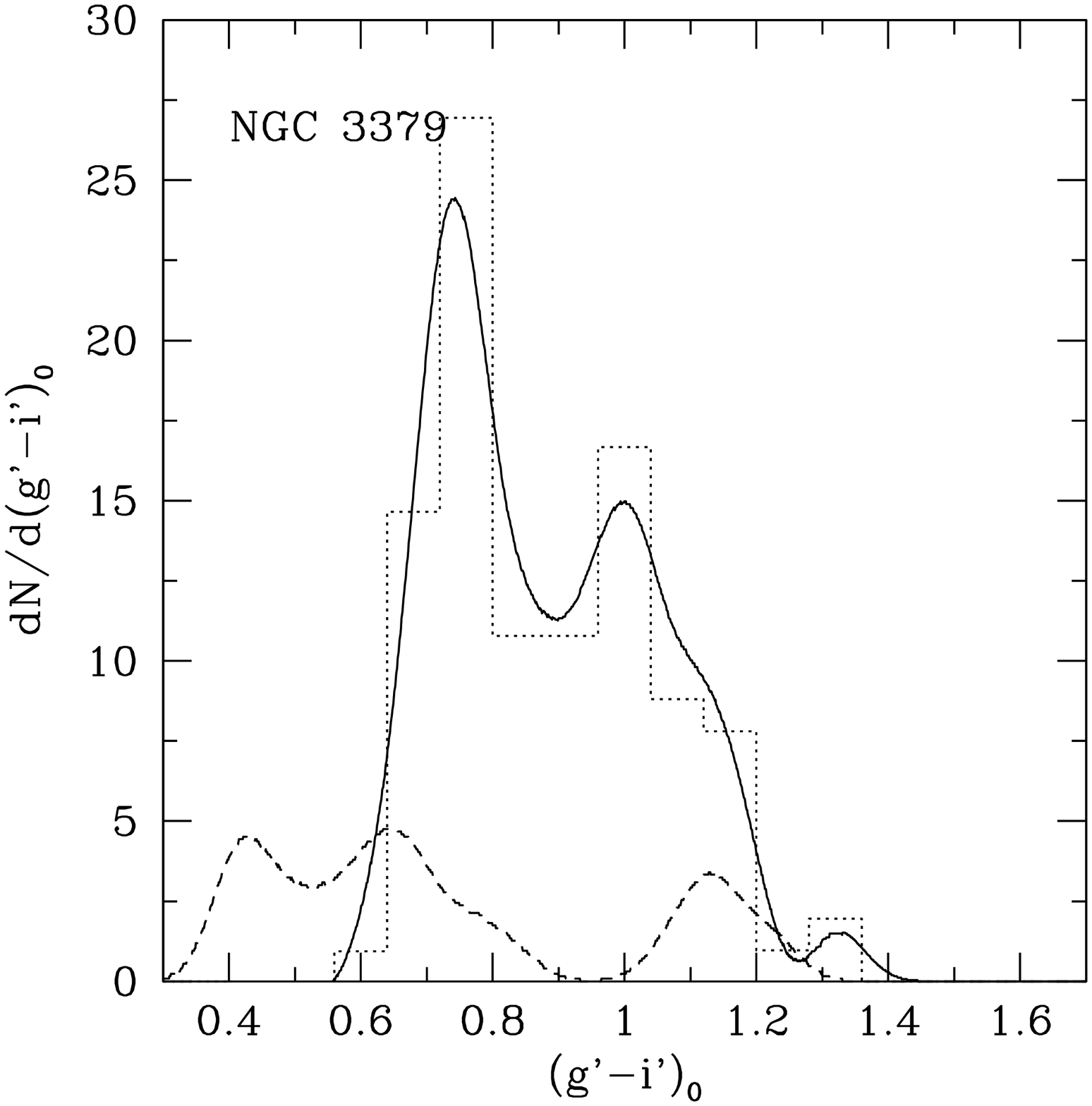}}
\resizebox{0.5\hsize}{!}{\includegraphics{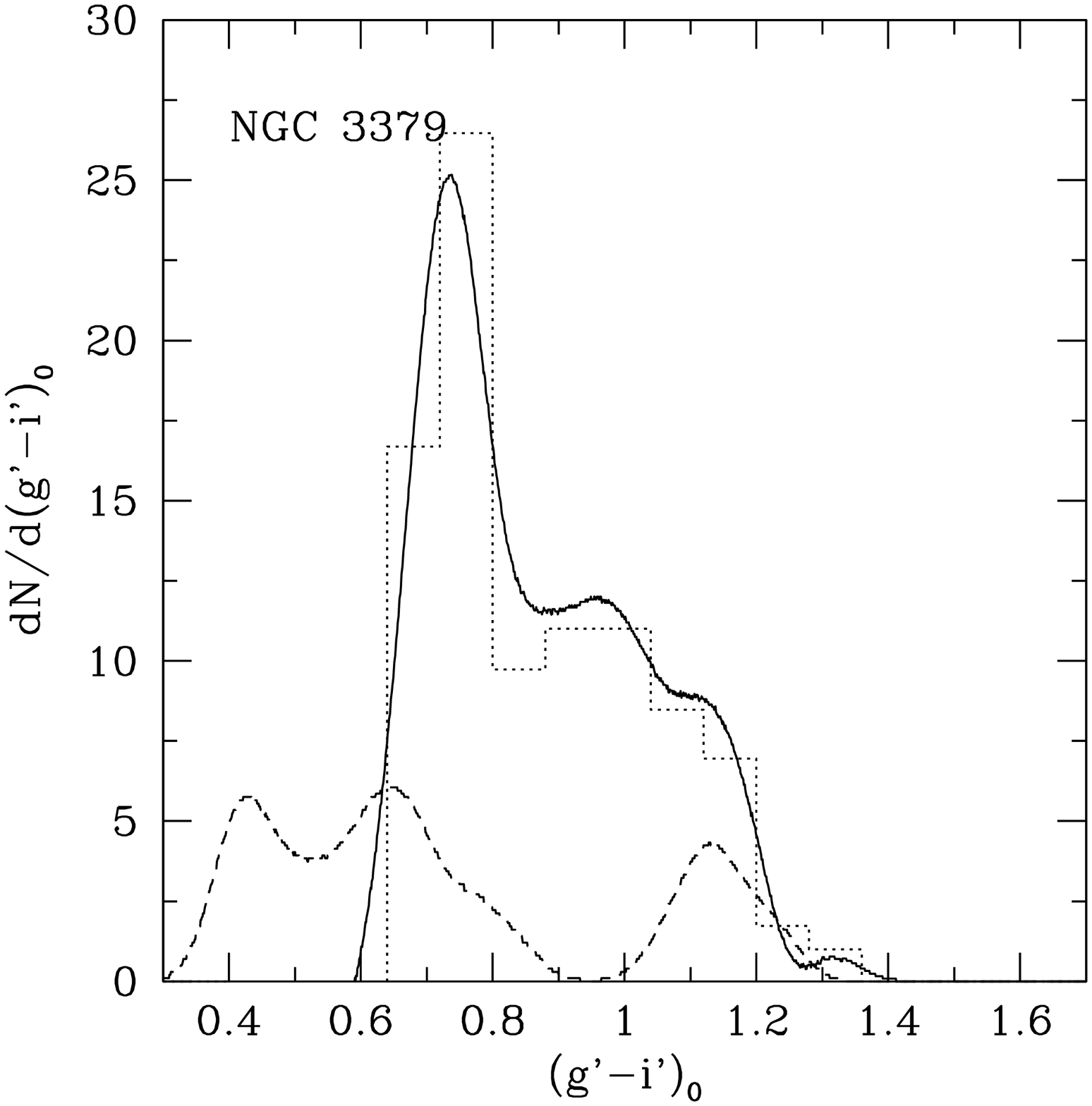}}
\caption{Left panel:  Dotted lines are the {\it (g$'$--i$'$)$_0$} colour 
histograms of all our GC candidates selected from samples 
with 90\% completeness level. The solid lines are smoothed colour 
distributions, obtained as described in the text. The backgrounds used 
to correct the counts are shown as the dashed 
lines. Right panel: The same as left panel, 
but corrected for areal completeness as described in the text. In the case 
of N3923, GCs brighter than {\it M$_I$}=-11 are show as long dashed line 
in this panel.
}
\label{HISTOS_GC_raw}
\end{figure}

Raw {\it (g$'$--i$'$)$_0$} colour distributions for all GC candidates 
selected as indicated in Section \ref{CMD} are presented in the left panel 
of Fig. \ref{HISTOS_GC_raw}, ordered
by decreasing {\it B$_T$} brightness of the host galaxy. These histograms were 
built by counting the objects in bins of 0.06 (NGC 4649, 
NGC 3923, NGC 524 and NGC 3115) and 0.08 mag (NGC 3379), depending 
on the total number of candidates in each sample (dotted lines in 
Fig. \ref{HISTOS_GC_raw}). 
Additionally, a smoothed colour distribution was created by replacing 
each object by a Gaussian of $\sigma=0.04$ mags (solid lines in 
Fig. \ref{HISTOS_GC_raw}). This $\sigma$ value was 
considered to be representative of the mean error in {\it (g$'$--i$'$)} colours 
for fainter candidates. Each histogram is background-corrected, as 
described in the following, and the distribution of 
the contaminant objects are plotted as dashed lines.
 
For the case of NGC 4649, we adopted the WHDF as our control 
field. In all other cases, the average of counts in the WHDF and our  
Comparison Field was taken as representative of the 
background and subtracted.

As noted in Section \ref{CompF}, MW
stars are the most important source of contamination for {\it i$'_0$} $<23.5$, 
while unresolved background galaxies are the main contaminants at fainter 
magnitudes. However, none of the targets listed in Table \ref{Tsample} 
are at very low galactic latitudes and the expected contamination level is
probably less than 10 percent (as demonstrated by our previous spectroscopic 
results).

The left panel of Fig. \ref{HISTOS_GC_raw} shows that, except for 
NGC 524, all the GCSs appear bimodal. NGC 4649 and NGC 3115 show very 
well defined blue and red peaks of {\it (g$'$--i$'$)$_0 \sim 0.75$} and 
{\it (g$'$--i$'$)$_0 \sim 1.0-1.1$}. NGC 3923 appears bimodal, but the
separation between peaks is less clear. NGC 3379 shows a marked blue
 peak around {\it (g$'$--i$'$)$_0 \sim 0.7-0.75$}, and a less conspicuous red 
population.

 As noted before, NGC 524 is the only target where no clear bimodality 
is seen. However, its colour distribution looks rather broad and further
high quality photometry will be required to clarify the situation.

As shown in Fig. \ref{DSS2}, the areal coverages 
are different in each observed GCS. As a consequence of this, we have attempted 
to recover the global colour distribution  of the entire GCS 
by correcting for areal incompleteness.
 In order to do this, we split each GCS into different 
galactocentric radial bins. As seen in Section \ref{clas}, the inner 
regions always have lower photometric completeness. We therefore exclude 
the zones within R$_{gal}<45$ arcsecs. The outer zones have 
very small areal completeness (and hence introduce noise), so 
 the points with R$_{gal} > r_{max}$ were excluded too. Here $r_{max}$ was
 taken as the R$_{gal}$ which shows areal completeness smaller than $20\%$ 
 (in Table \ref{Histo_fit} we listed the values of $r_{max}$ for each GCS). After 
that, we calculated the 
 fraction of each ring ``effectively''  observed. And then, the inverse 
of these fractions were used as a correction factor. The number of 
radial bins was three in the two most massive galaxies (NGC 3923 and 
NGC 4649), and two in the other less populated ones (NGC 3115, NGC 3379 and 
NGC 524). Finally, the total colour distribution was obtained by 
adding the colour distributions of each individual radial bin. 

In the right panel of Fig. \ref{HISTOS_GC_raw} we show the resulting 
background corrected histograms and smoothed colours distributions. The 
areal corrections lead to more prominent blue GCs populations. This 
results from the fact that blue GCs are less concentrated 
towards the  center of the host galaxy than the red ones.

In the case of NGC 3923, 
the colour-magnitude diagram shows a unimodal colour distribution for the
brightest GCs. Then we have split the sample in GCs 
having {\it M$_I$}$<-11$ (long 
dashed line in Fig. \ref{HISTOS_GC_raw}) and {\it M$_I$}$>-11$ (the same codes as the other GCSs). Doing 
that, we see that the bright GC candidates are not 
bimodal. For NGC 3115 and NGC 3379, the small number of bright GCs 
prevents a similar analysis.

In order to characterize  the presence of bimodality in
the integrated GC colours, we have run RMIX on our 
histograms \citep{McD2007}. This program allows the user to 
choose the kind of distribution, number of components, and constraints. 
In this work  we adopted two normal components, allowing  
RMIX to fit the position of the peak, the standard deviation and 
the number of GC candidates in each populations. Using 
reasonable starting values, in all cases RMIX 
converged easily, so we did not
constrain any parameter. The results  are shown in Table \ref{Histo_fit}.
As a comparison we list the modal values for all the GCSs 
showing two clear peaks in their colour distribution. In the case of 
NGC 4649, we list two values for 
each parameter. The first set of values are obtained using the whole 
sample, and the second one taking GC candidates 
with 21.5 $<$ {\it i$'_0$} $<$ 23.9.
In this last sample  both peaks are bluer. In particular the blue peak moves
 0.025 mag as a consequence of the ``blue tilt'' phenomenon. 

In the case of NGC 524 we specifically tested different kinds of fits. We 
found that a double Gaussian distributions fit was strongly preferred 
over only one leading to a $\chi$ value 2.7 times smaller. This 
supports the conclusion that NGC 524 is  
also a bimodal system. However, the blue peak in this 
galaxy ({\it (g$'$--i$'$)$_0 = 0.815 \pm 0.021$})
is the reddest one in our sample. Deeper photometry could help to
obtain a more conclusive result for this galaxy.

\begin{figure}

\resizebox{1\hsize}{!}{\includegraphics{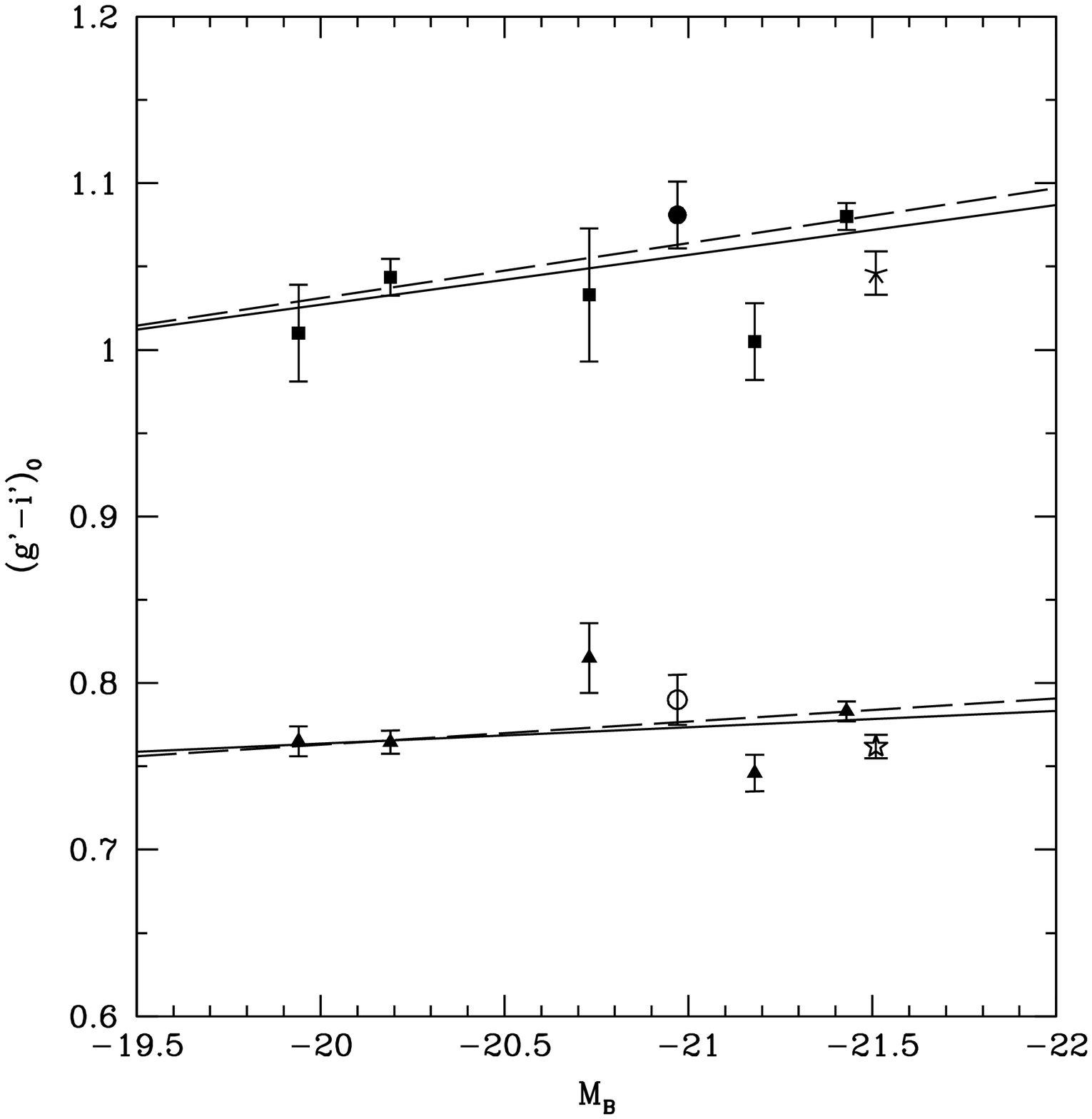}}

\caption{{\it (g$'$--i$'$)$_0$} colour for blue (filled triangles) and 
red peaks (filled squares) obtained by 
fitting two Gaussian to the colours histograms. We have included 
here the NGC 3311 (open and filled circles) 
and M87 (open and filled stars) points from \citet{WHWRW08} and 
Harris (2009), respectively. Solid lines are 
the fits obtained as indicated in the text. Dashed lines are the 
results excluding NGC 3923 ({\it M$_B$}= -21.18) in the fits.
}
\label{COL_peaks}
\end{figure}

Fig. \ref{COL_peaks} shows the positions of the GC colour peaks for 
the different cluster populations as listed in Table \ref{Histo_fit} 
versus the galaxy M$_B$.  Blue and red peaks are depicted
as filled triangles and squares, respectively. One of the most 
striking characteristics of this figure is that the red peak of 
NGC 3923 seems to be too blue according to its absolute magnitude.
To complement this, we have included the values corresponding to 
NGC 3311 as open and filled circles \citep{WHWRW08}.  The 
following (error weighted) linear fits were obtained for the 
red (eq. \ref{Col_L_blue}) and blue GCs peaks (eq. \ref{Col_L_blue}): 

\begin{equation}
(g'-i')_0=-0.03 (\pm 0.01) M_B + 0.42 (\pm0.33)
\label{Col_L_blue}
\end{equation}

\begin{equation}
(g'-i')_0=-0.01 (\pm 0.01) M_B + 0.56 (\pm0.22)
\label{Col_L_red}
\end{equation}

These fits are shown as solid lines in Fig. \ref{COL_peaks} 
where we also include the peaks corresponding to M87 from \citet{H09a} 
for comparison (open and filled stars). The M87 points
are very consistent with our relations. However, it is important 
to note that \citet{H09a} calibrated
his M87 photometry with SDSS stars. As a result, some mismatch 
may be expected.
 
\citet{PACS06} found that the slopes of $(g-z)$ vs. M$_B$ 
relations were, -0.036 and -0.026 for red and blue peaks, respectively. 
Using some common objects between \citet{JPBCEFTW09}, and our 
NGC 4649 photometry, we have found that $d(g-z)/d(g'-i')= 1.36\pm 0.05$. 
With this value, the slopes from Peng et al., translate to 
-0.026 and -0.019. Comparing with Equations (\ref{Col_L_blue}) and 
(\ref{Col_L_red}), we see 
that the red values are in very good agreement. However, our 
results for the blue peaks are consistent with no correlation.

We note that \citet{PACS06} obtain [Fe/H] $\propto L^{0.26 \pm 0.03}$
for the red GCs which is identical to our value after adopting the
colour-metallicity relation given in the next section.

\begin{table*}
\centering
\caption{RMIX fits to the {\it (g$'$--i$'$)$_0$} color distributions. The 
mean colours of the two modes $\mu_b$ (blue) and $\mu_r$ (red), their 
dispersions $\sigma_b$ and $\sigma_r$, and the fraction of object 
assigned by RMIX to the blue GC population are listed. The last 
two columns give the galactocentric radii of the annuli which show areal 
completeness smaller than 20\% and the fraction of area effectively 
observed within those radii.
}
\label{Histo_fit}
\scriptsize
\begin{tabular}{lccccccccc}
\hline
\hline
\multicolumn{10}{c}{}\\
\multicolumn{1}{c}{\textbf{}} &
\multicolumn{2}{c}{\textbf{Modal values}} &
\multicolumn{5}{c}{\textbf{RMIX}} &
\multicolumn{1}{c}{\textbf{r$_{max}$}} &
\multicolumn{1}{c}{\textbf{f$_T$}} \\
\multicolumn{1}{c}{\textbf{Galaxy}} &
\multicolumn{1}{c}{\textbf{$\mu_b$}} &
\multicolumn{1}{c}{\textbf{$\mu_r$}} &
\multicolumn{1}{c}{\textbf{$\mu_b$}} &
\multicolumn{1}{c}{\textbf{$\sigma_b$}} &
\multicolumn{1}{c}{\textbf{$\mu_r$}} &
\multicolumn{1}{c}{\textbf{$\sigma_r$}} &
\multicolumn{1}{c}{\textbf{f$_b$}} &
\multicolumn{1}{c}{\textbf{arcsec}} &
\multicolumn{1}{c}{\textbf{}}\\
\hline \multicolumn{10}{c}{}\\
 NGC 4649    & 0.765 & 1.095 & 0.783 $\pm$ 0.006 &  0.098 $\pm$ 0.004 & 1.080 $\pm$ 0.008 &  0.110 $\pm$ 0.004 &  0.518 $\pm$ 0.024 & 490 & 0.41\\
             &       &       & 0.758 $\pm$ 0.007 &  0.096 $\pm$ 0.004 & 1.067 $\pm$ 0.008 &  0.127 $\pm$ 0.004 &  0.442 $\pm$ 0.025 & '' & ''\\
 NGC 3923    & 0.758 & 1.011 & 0.746 $\pm$ 0.011 &  0.072 $\pm$ 0.008 & 1.005 $\pm$ 0.023 &  0.135 $\pm$ 0.014 &  0.365 $\pm$ 0.069 & 350 & 0.58\\ 
 NGC 524     & -     &  -   & 0.815 $\pm$ 0.021 &  0.086 $\pm$ 0.011 & 1.033 $\pm$ 0.040 &  0.076 $\pm$ 0.021 &  0.713 $\pm$ 0.120 & 315 & 0.70\\ 
 NGC 3115    & 0.764 & 1.053 & 0.765 $\pm$ 0.007 &  0.067 $\pm$ 0.005 & 1.044 $\pm$ 0.011 &  0.092 $\pm$ 0.008 &  0.521 $\pm$ 0.037 & 320 & 0.58\\
 NGC 3379    & 0.769 &  -   & 0.765 $\pm$ 0.009 &  0.036 $\pm$ 0.008 & 1.010 $\pm$ 0.029 &  0.135 $\pm$ 0.018 &  0.408 $\pm$ 0.082 & 290 & 0.72\\ 
\multicolumn{10}{l}{}\\
\hline                                                                             
\multicolumn{10}{l}{}\\ 
\end{tabular}
\end{table*}

\subsection{Blue tilt}

As noted in Section \ref{CMD}, NGC 4649 is the only galaxy in the sample 
where the blue GCs show a clear colour-magnitude trend, a feature called
the ``blue tilt'', and noticed by \citet{SBSB06} 
and \citet{H06}. In contrast, all the 
other galaxies discussed in this paper show two broad color sequences,
without an obvious tilt (for a further study of the NGC 3923, see 
\citealt{NK2011}).
The GC colour distributions in these  
galaxies become unimodal at bright magnitudes, a situation different from
that in NGC 4649, where blue and red clusters are clearly 
separated even at higher luminosities.
 
In particular for NGC 4649, \citet{SBSB06} find a colour trend
$d(g-z)/dz= -0.040\pm 0.005$ identical, within the errors, to $d(g-z)/dz=
 -0.037 \pm 0.004$ derived by \citet{Metal06} by combining the 
colour magnitude diagrams of M49, M60 and M87.

\begin{figure}
\resizebox{1\hsize}{!}{\includegraphics{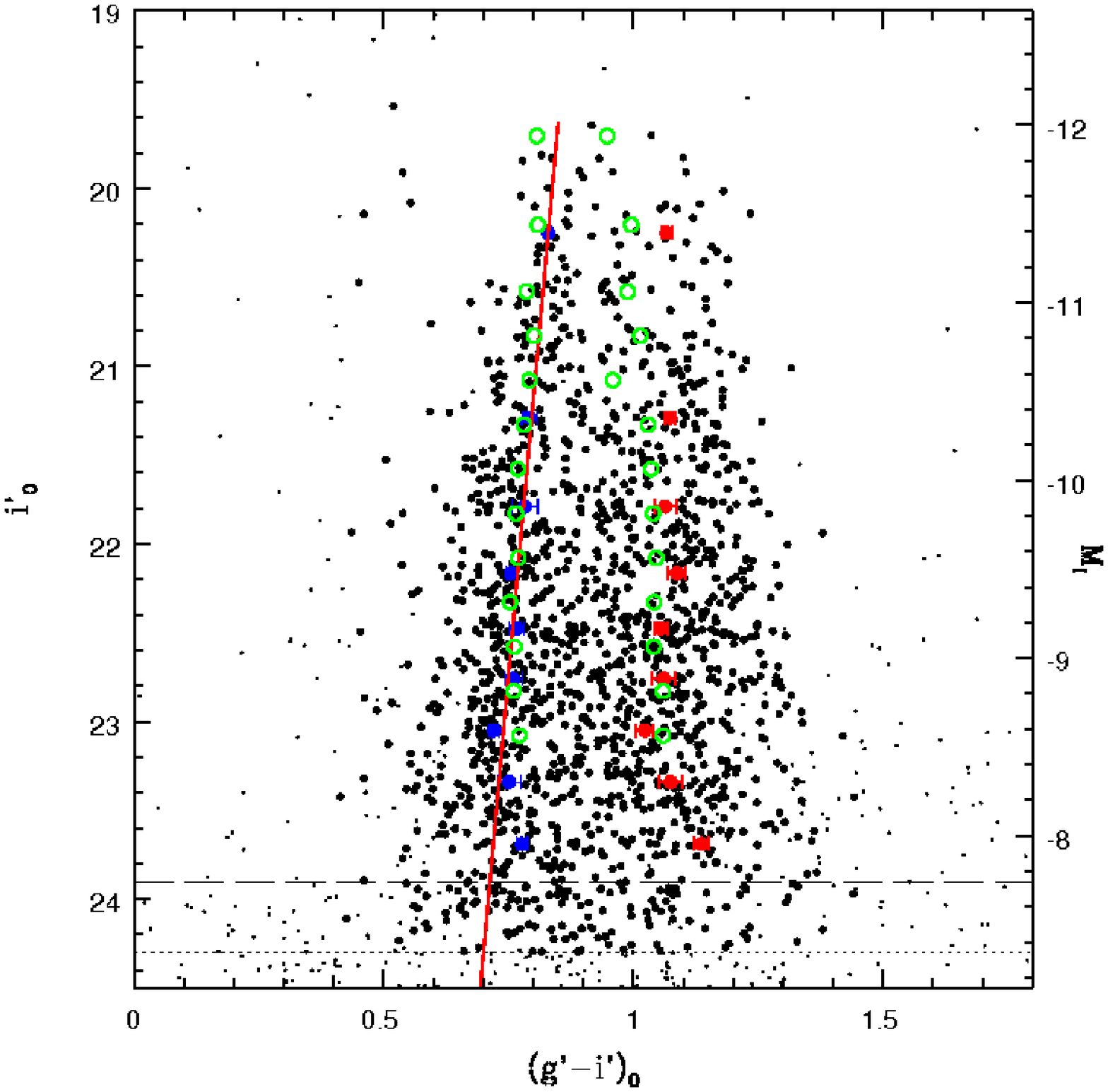}}
\caption{The ``blue tilt'' in NGC 4649. Blue and red points are the positions 
of the peaks of the blue and red GC subpopulations obtained as indicated in 
the text. The green circles are those for the M87 GCS from \citet{H09a}. The 
red line is 
a linear fit to the blue points for NGC 4649. Here we see that the blue 
tilt in this galaxy appears at  {\it i$'_0$ $\sim$ 23} mag, which 
roughly corresponds to $M_I \sim$ -8.5 mag.
}
\label{BT}
\end{figure}

In order to characterize the ``blue tilt'' in NGC 4649, using our
GMOS photometry, we followed an approach similar to that described 
by \citet{WHWRW08}. GCs were grouped in magnitude bins 
containing 150 clusters each and, using RMIX, we obtained the 
positions of the blue and read peaks. This procedure delivered nine
pairs of values depicted as blue and red dots in Fig. \ref{BT}. 
In all the  cases previously studied, the ``blue tilt'' appears to 
have a similar behaviour.  Fig. \ref{BT} shows  that 
the blue sequence seems to become redder  at a magnitude
around {\it M$_I \sim-9$} mag. In turn, and using a combined sample for
 six giant elliptical galaxies, \citet{H09b} found evidence that the 
``blue tilt''  starts to be detectable at absolute magnitudes brighter 
than {\it M$_I=-9.5$} mag and exhibits a nonlinear form. However, in our 
NGC 4649 sample, the three faintest points of the blue and red 
peaks (see Fig. \ref{BT}) were not  well constrained by RMIX as 
they have significant colour errors. So, these points were not 
included and only  those with {\it i$'$}$<23$ were used to perform  
linear fits to the blue and red GCs groups:

\begin{equation}
(g'-i')_0=-0.032(+/-0.003) i'_0+1.478(+/-0.072) 
\end{equation}

\begin{equation}
(g'-i')_0=-0.007(+/-0.006) i'_0 +1.227(+/-0.1356)
\end{equation}

These fits indicate that the red GCs do not show a 
significant trend, as noted for other 
galaxies \citep{H09a,H09b,PJBMCFHMM09,HSFB10}. 

We stress that a second order fit, like that performed by \citet{H09b}, on our data,  
produces very unstable solutions. However, we cannot rule out that
the inclusion of a better sample of faint objects may require a 
higher order fit.

As a comparison Fig. \ref{BT} also includes the position
of the blue and red peaks obtained by \citet{H09a} for the M87
GCS also using g and i photometry. Both galaxies exhibit remarkably
similar blue tilts.

The fact that the tilt is detectable in some but
not all galaxies is an issue that  still deserves clarification. A tentative
interpretation was presented in \citet{FFG07} who showed
that a large spread of the characteristic metallicity scale of the
blue GCs smears the colour-magnitude relation in such a way that
the tilt becomes eventually undetectable. In the specific case 
of the galaxies in our sample, the situation is not so clear. Although the 
colour magnitude diagrams for NGC 3379, NGC 3115 and NGC 524 in 
Fig. \ref{CMD_CCD_GC} show the presence of very massive GCs (those 
with $M_I<-10$ mag) their numbers are
quite small thereby preventing a meaningful detection of the 
blue tilt.
 
It is worth mentioning  that \citet{HSFB10} found a well
defined tilt in the blue GC sequence in M 104, using a 
sample with several hundred GCs. In contrast, NGC 3923 with a comparable
number of clusters ($\sim 670$), exhibits no detectable tilt, suggesting that
the size of the statistical sample cannot be the main reason for
the absence of the tilt. We note, however, that our GCs sample 
does not include a significant number of clusters
brighter than $M_I \approx -9.5$, which conspires against the eventual
detection of the tilt. Rather, our colour magnitude diagram suggests
a "unimodal" distribution for these bright clusters (but see Norris and
Kannappan 2011). Another question that deserves 
a more complete wide field analysis is why NGC 3923 seems to have 
a relatively sparse population of massive blue GCs.

We note that the ``blue tilt'' has been identified as a
possible instrumental artifact specifically affecting the HST-ACS photometry \citep{K08} 
and even dismissed on the basis of {\it (V-I)} photometry of M87
\citep{WZLB09}. However, the observations presented in this 
paper together with other ground based works (see 
also \citealt{FFG07},  \citealt{WHWRW08}, \citealt{H09a}), and new 
HST/ACS based analyses \citep{PJBMCFHMM09,H09b}, 
strongly argue in favour of the existence of the blue tilt.

Previous works have presented an interpretation of the ``blue
tilt'' as the result of increasing metallicity with GC mass, i.e., 
a mass-metallicity relation (MMR). \citet{BH09} and \citet{SS08} 
give models where the physical interpretation of the MMR is based on 
self-enrichment during a cluster's formation stage. According to these 
models, GCs with masses above $10^6 M_{\odot}$ will
hold onto enough of their SN ejecta to enrich the gas from which 
stars are still being formed.
A meaningful comparison of the results from different GCS is 
somewhat precluded by the adoption
of different colour-metallicity relations. The photometry presented  
in this paper, combined with our previous spectroscopic results
(\citealt{PBFBGFFZSHP2006}, \citealt{PBFPBGFFZSH2006} and 
\citealt{NSBGFPFFBZH2008}) allows the 
determination of a new relation that includes 67 globulars in three
different galaxies (NGC 3923, 3379, 4649). Chemical abundances [Z/H]
obtained via Lick indices, are on the scale defined by the synthetic
models by \citet{TMK04} and clusters fainter than
{\it g$'_0$=23} were rejected due to their relatively large associated errors.

\begin{figure}
\resizebox{1\hsize}{!}{\includegraphics{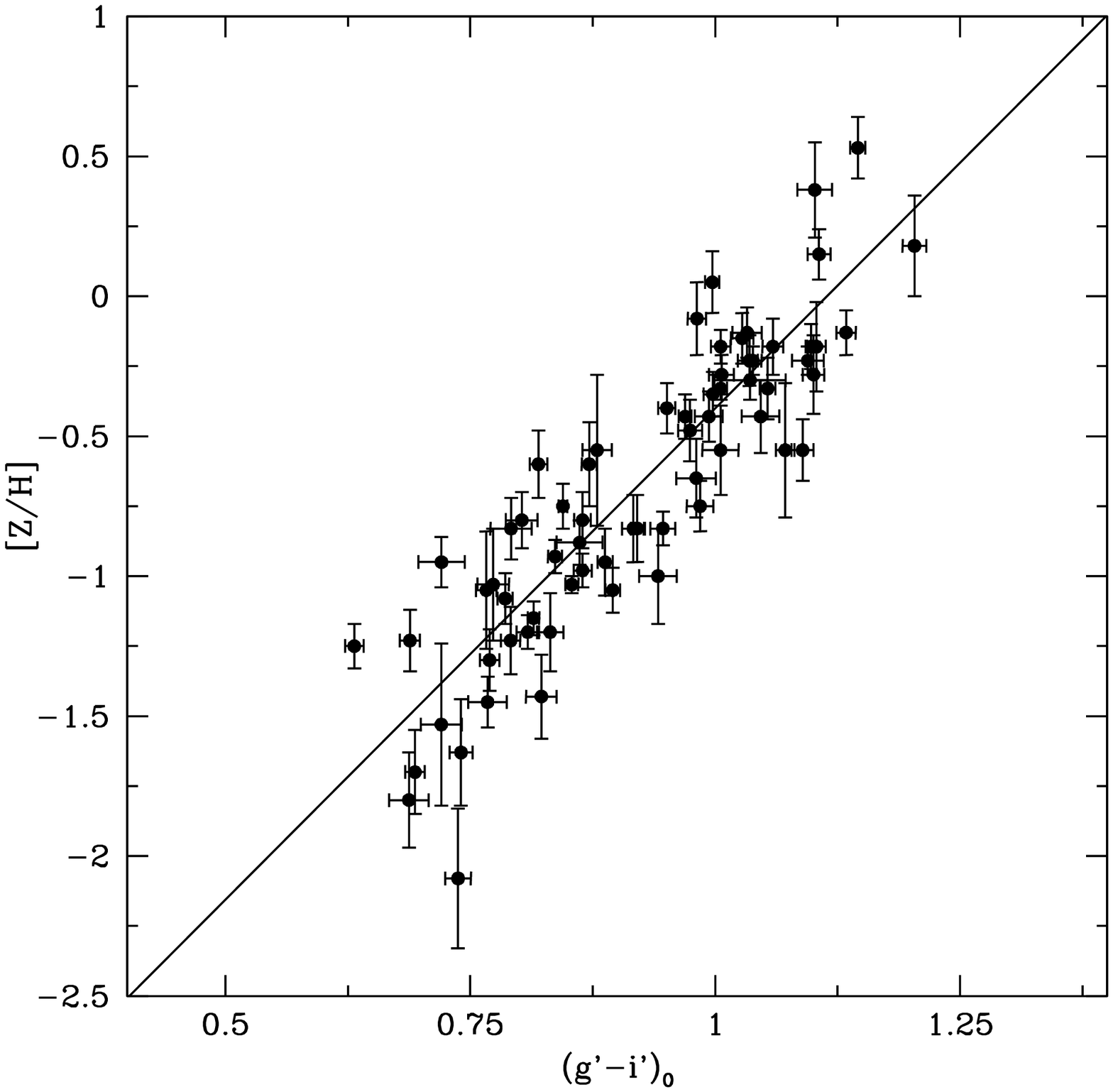}}
\caption{GC colour metallicity relation. The points are GCs 
from the photometry presented in this work and metallicities 
from  \citealt{PBFBGFFZSHP2006}, \citealt{PBFPBGFFZSH2006} and \citealt{NSBGFPFFBZH2008}. The solid line is a linear fit (see the text).
}
\label{colMet}
\end{figure}

The  [Z/H] vs. {\it (g$'$-i$'$)$_0$} values for the calibrating GCs are 
plotted in Fig. \ref{colMet} where a linear fit:

\begin{equation}
[Z/H]=3.51(\pm 0.21) (g'-i')_0  - 3.91 (\pm0.20)
\label{Z_col}
\end{equation}

\noindent is also shown. The overall trend is in agreement with a 
rediscussion of the {\it (C-T1)$_0$} vs [Fe/H] relation presented 
in \citet{FFG07} and with other works where more complex functions 
were fitted
(for example, \citealt{PACS06}, \citealt{LPKHKG2008} and \citealt{BCP10}). However,
it is worth emphasizing that all these last calibrations include higher order colour terms, mostly set by low metallicity MW globulars, whose counterparts 
are scarce in Fig. \ref{colMet}. A more direct comparison is possible 
using the recently published relation obtained by \citet{SHAHW10}. Those 
authors have used integrated colours in the USNO photometric system \citep{S2002} 
and spectroscopically measured abundances of GCs from NGC 5128 and the MW. 
Their linear {\it(g$'$-i$'$)$_0$} vs [Fe/H] relation can be easily translated
to the form of Equation (\ref{Z_col}) using the [Fe/H]=[Z/H]-0.131 relation  
from \citet{MPF07}, leading to [Z/H]$=3.65 \times (g'-i')_0-4.18$, which 
compares very well with our results.

The adoption of a constant (M/L) ratio for the blue clusters, and the above
calibration, leads to

\begin{equation}
Z \propto M^{0.28\pm0.03}
\end{equation}

\noindent in agreement with  \citealt{Metal06}  who, in the 
particular case of NGC 4649, derive an exponent of $0.36 \pm 0.11$.

Our colour-abundance relation was
also applied to the {\it g$'$} and {\it i$'$} photometry presented 
by \citet{CHWWCR09}
 for NGC 5193, IC 4329 and NGC 3311, leading to exponents of
$0.23 \pm 0.14$, $0.37 \pm 0.05$, and $0.47 \pm 0.03$ respectively. These 
values are, in the average, 30 percent smaller than the exponents
obtained by Cockcroft et al. These differences are typical of
the uncertainties connected with the adoption of different metallicity
scales.

\subsection{Spatial distributions}
\label{spatial}

Fig. \ref{DSS2} shows some already known systematics regarding the
spatial distribution of both GC families, i.e., blue clusters usually 
exhibit relatively shallow distributions in contrast with those of
the red ones, which are frequently more concentrated toward the galaxy centers.

In this section we aim to quantify the spatial distributions 
through an analysis of the projected cluster density as a function
of galactocentric distance using both a de Vaucouleurs law (r$^{1/4}$)
and a power law. We also present S\'ersic law fits in the case of the
two brightest galaxies for which we have a good areal coverage.

Blue and red GCs were separated according to a common ``colour
valley'' at {\it (g$'$-i$'$)$_0 \sim 0.95$} and the limiting {\it i$'_0$} 
magnitude was set at the
90 percent completeness level. Due to their relatively low 
flattenings, we used circular annuli with {\it $\Delta$ log r} = 0.1 
to 0.15 for all galaxies except NGC 3115, for which we adopted an 
ellipticity of 0.55.

\begin{figure*}
\resizebox{0.4\hsize}{!}{\includegraphics{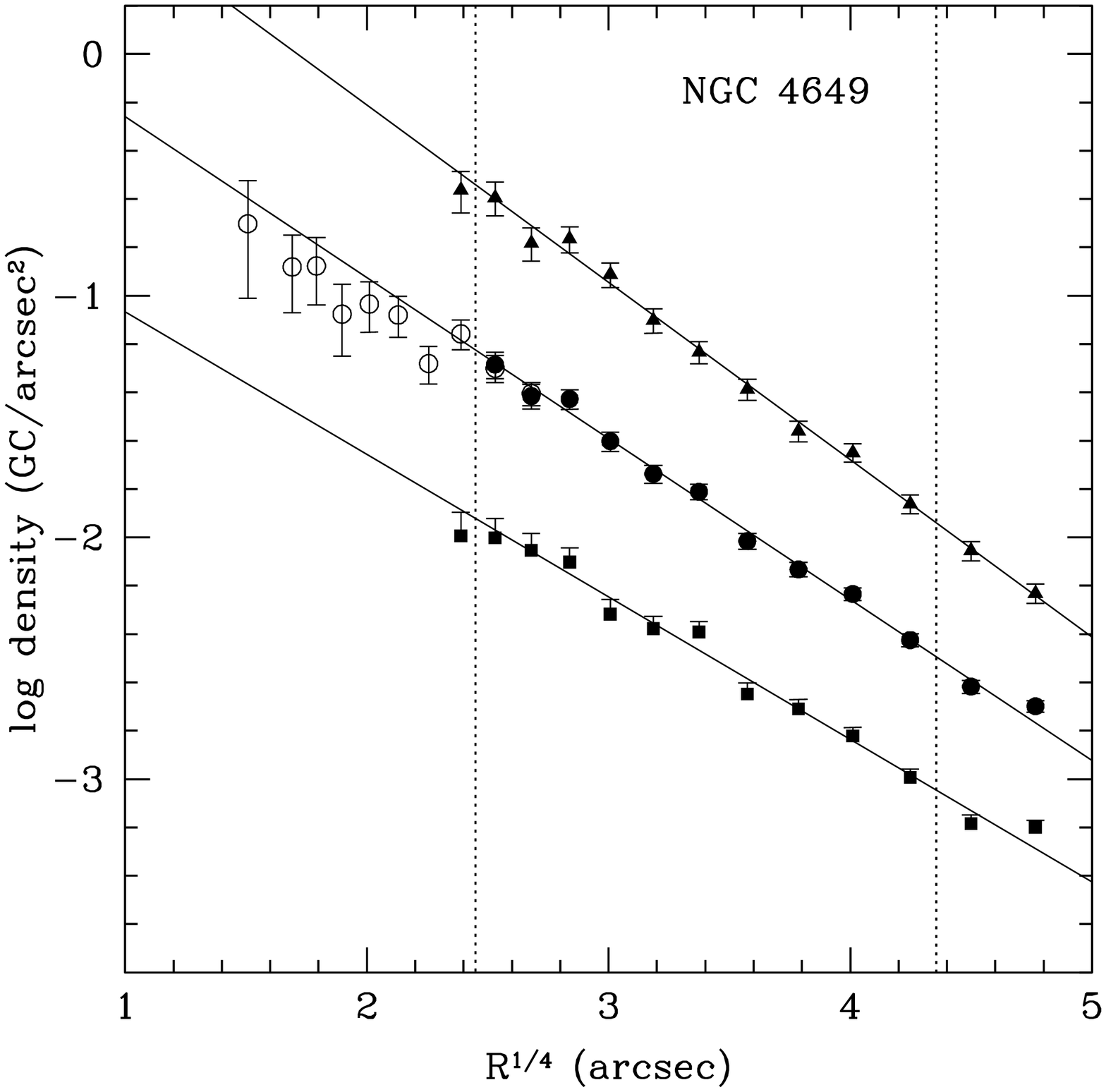}}
\resizebox{0.4\hsize}{!}{\includegraphics{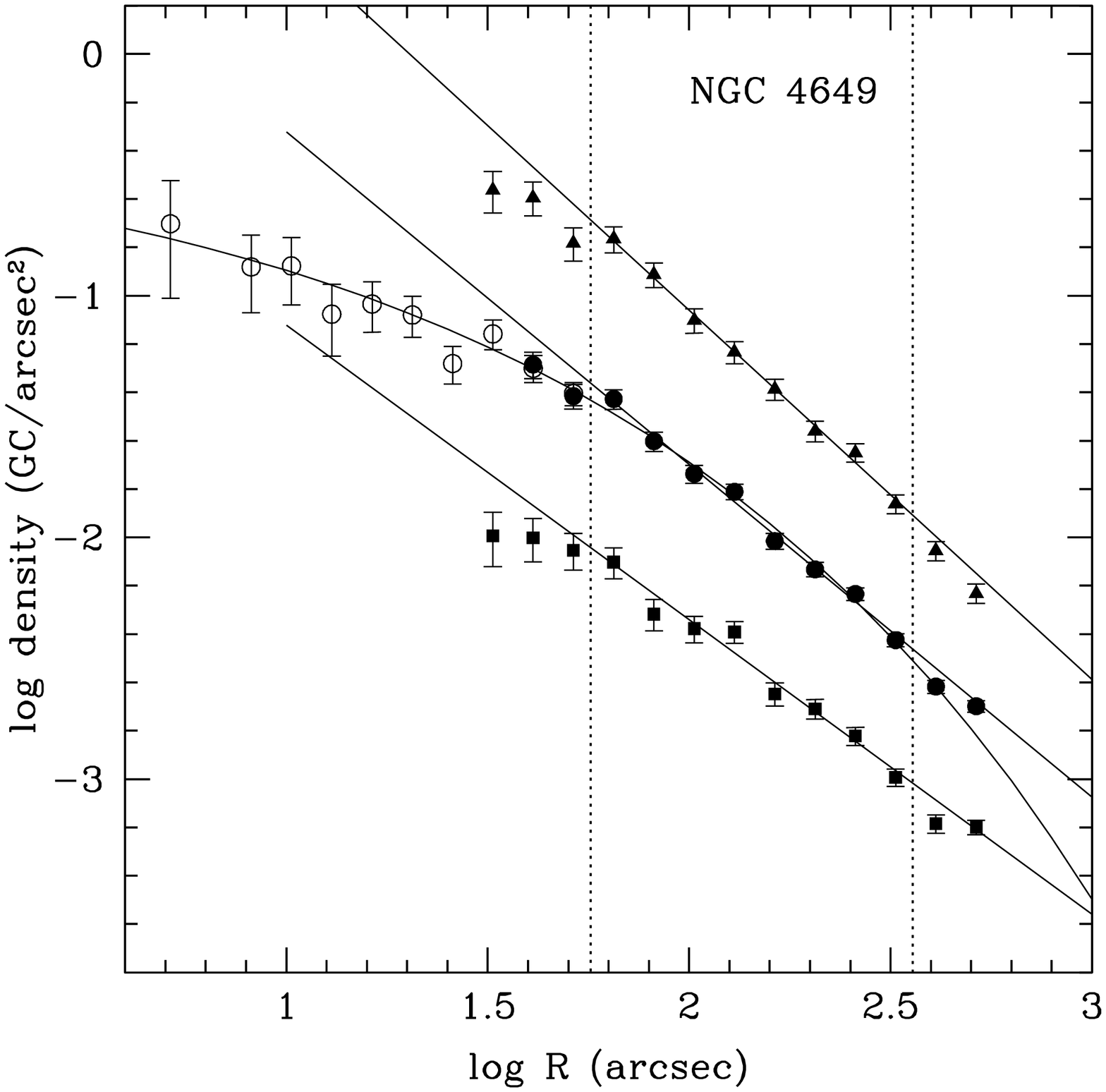}}\\
\resizebox{0.4\hsize}{!}{\includegraphics{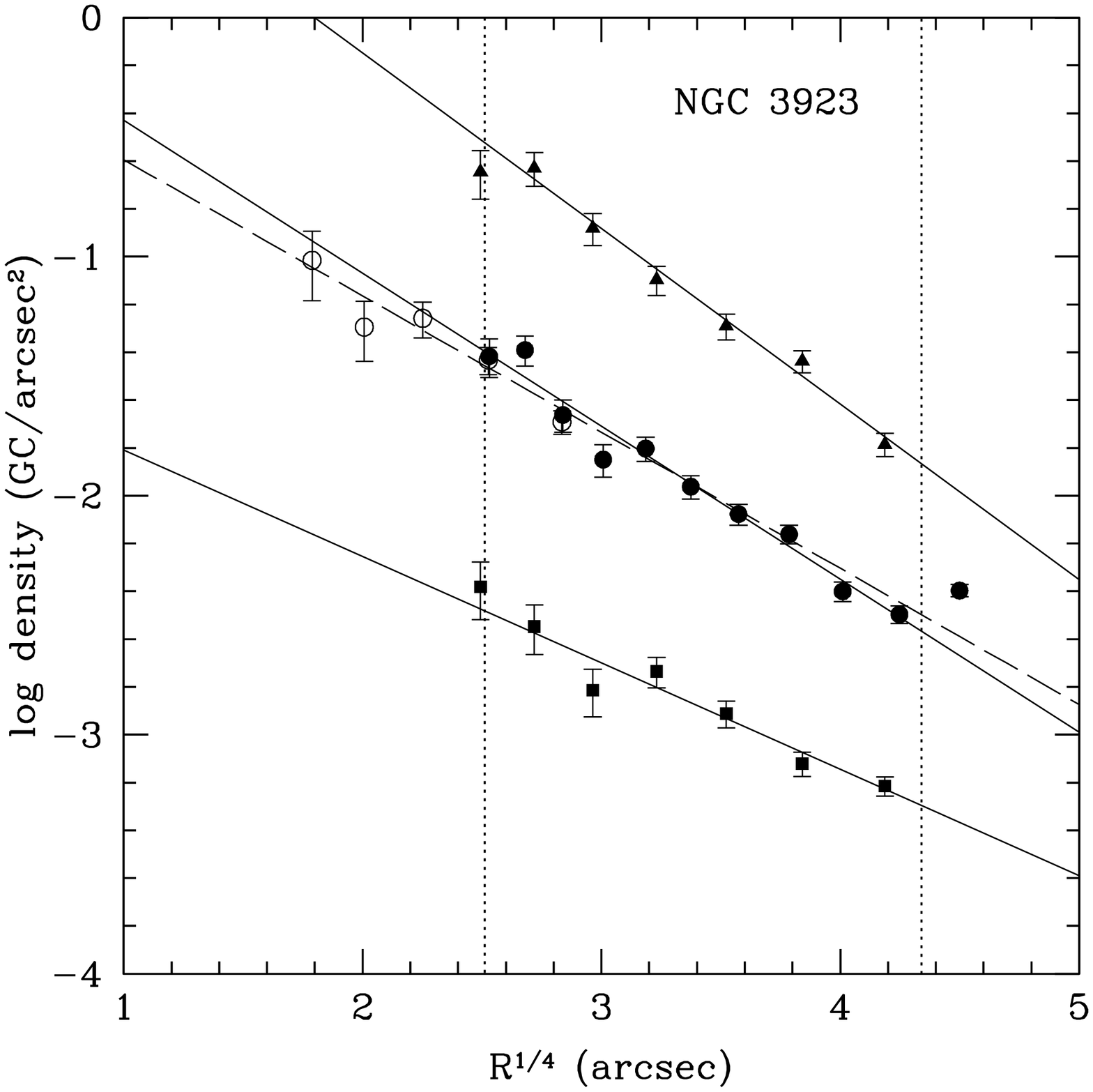}}
\resizebox{0.4\hsize}{!}{\includegraphics{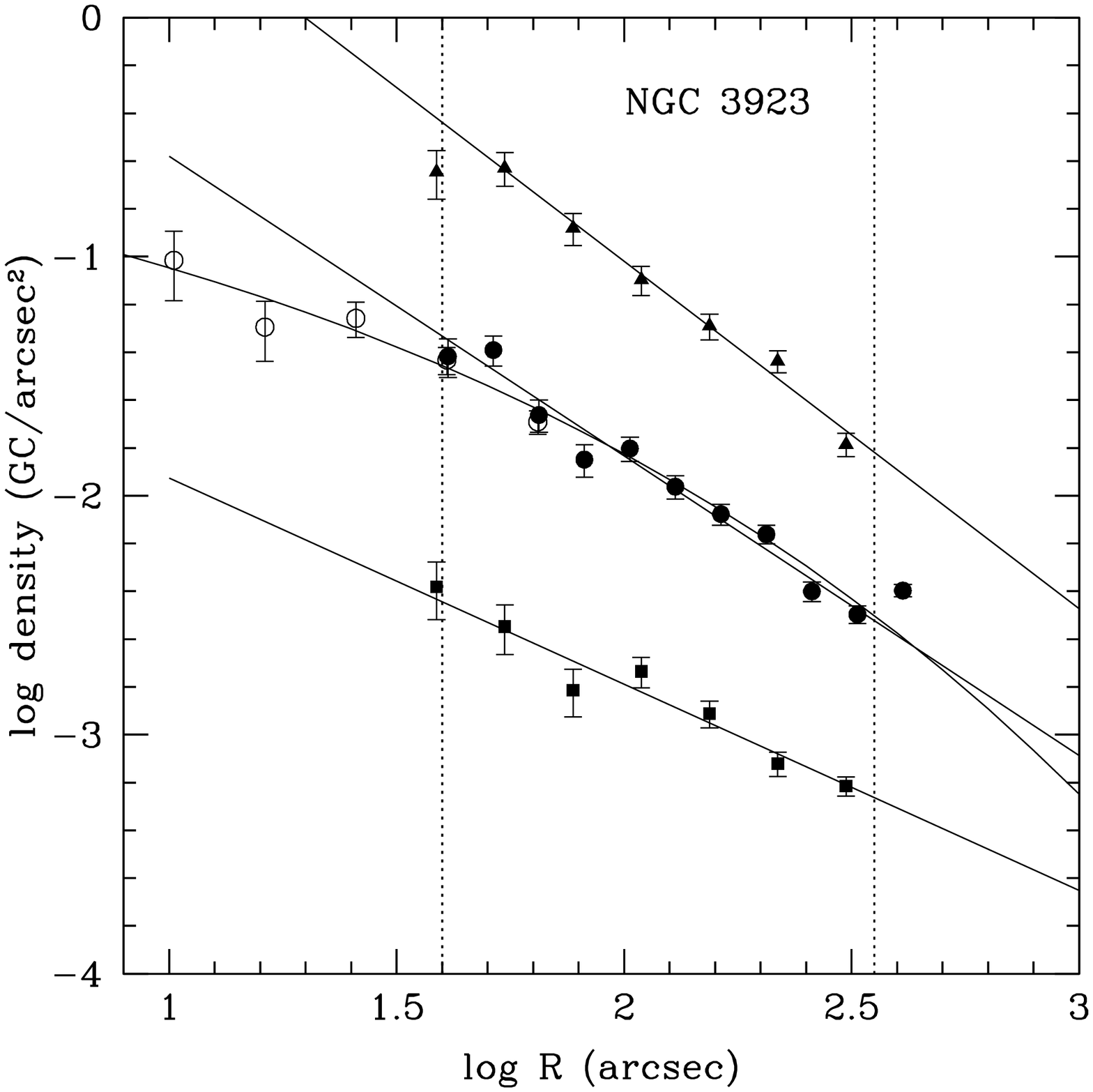}}\\
\resizebox{0.4\hsize}{!}{\includegraphics{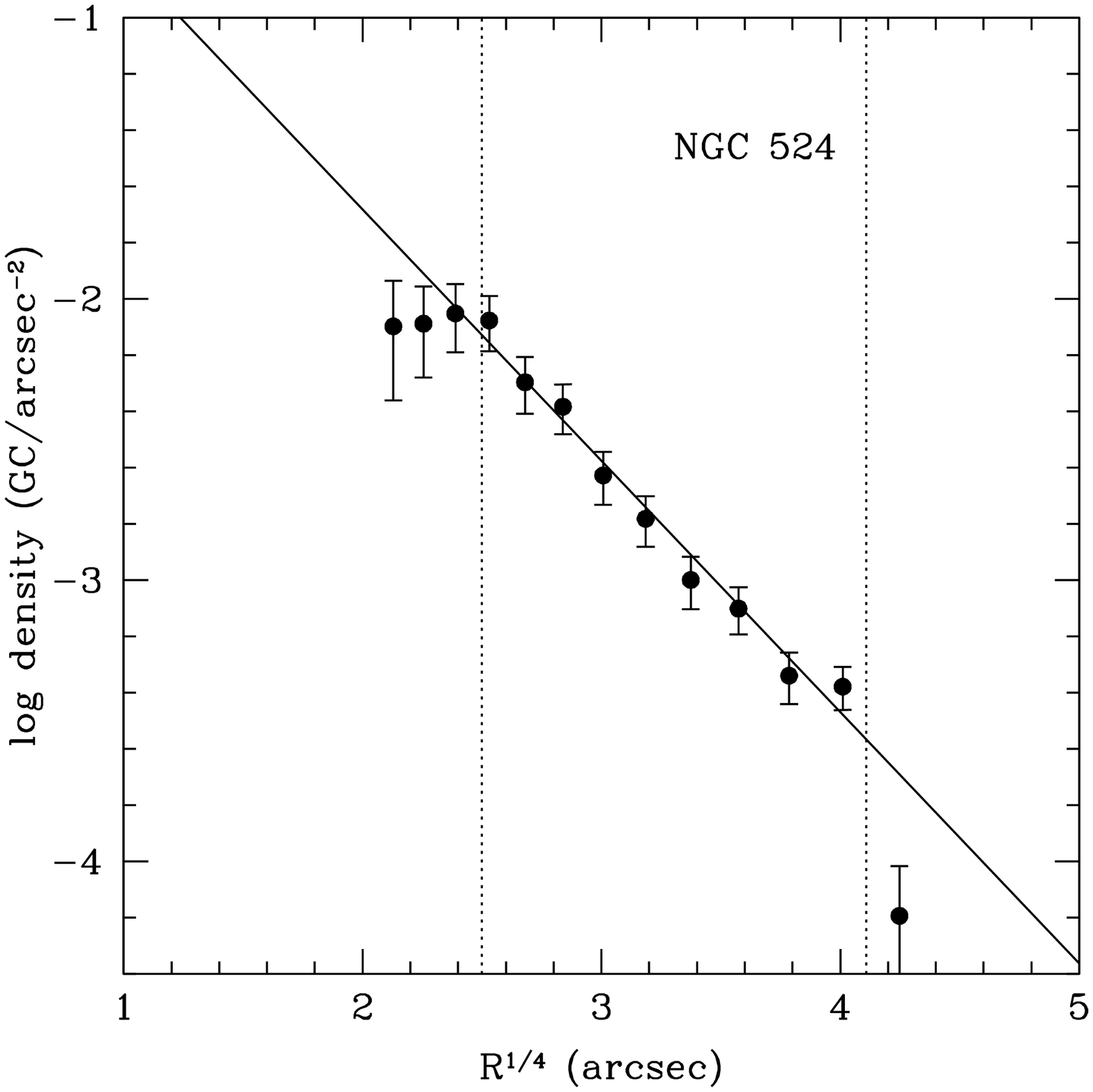}}
\resizebox{0.4\hsize}{!}{\includegraphics{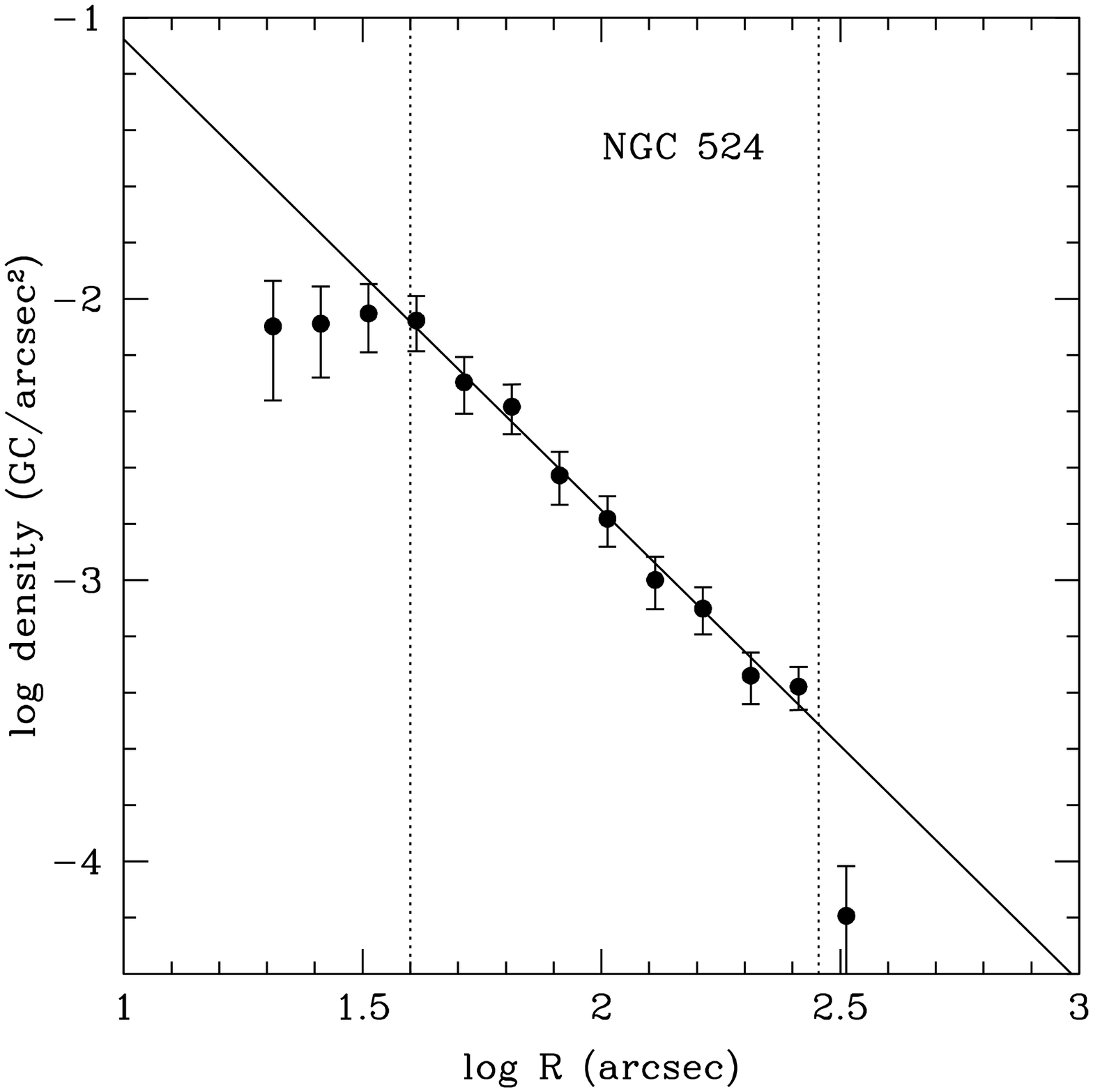}}\\
\caption{Projected radial profiles for all (filled circles), red (triangles) 
and blue GC candidates (squares) fit with a
de Vaucouleurs-law (left panel) and with a power-law 
(right panel). The profiles were arbitrarily shifted in order to avoid 
 overlapping. Since of NGC 3379 shows a very low number of red GCs 
and NGC 524 doesn't show a clear ``valley'', their GCs samples were not 
split into red and blue subpopulations. In the case of NGC 3115 the 
 projected density profile was obtained in elliptical rings. Dashed
lines indicate the radial ranges included in the fits. Open circles 
in the NGC 4649 and NGC 3923 figures are from ACS photometry (see text).
}
\end{figure*}

\setcounter{figure}{8}
\begin{figure*}
\resizebox{0.4\hsize}{!}{\includegraphics{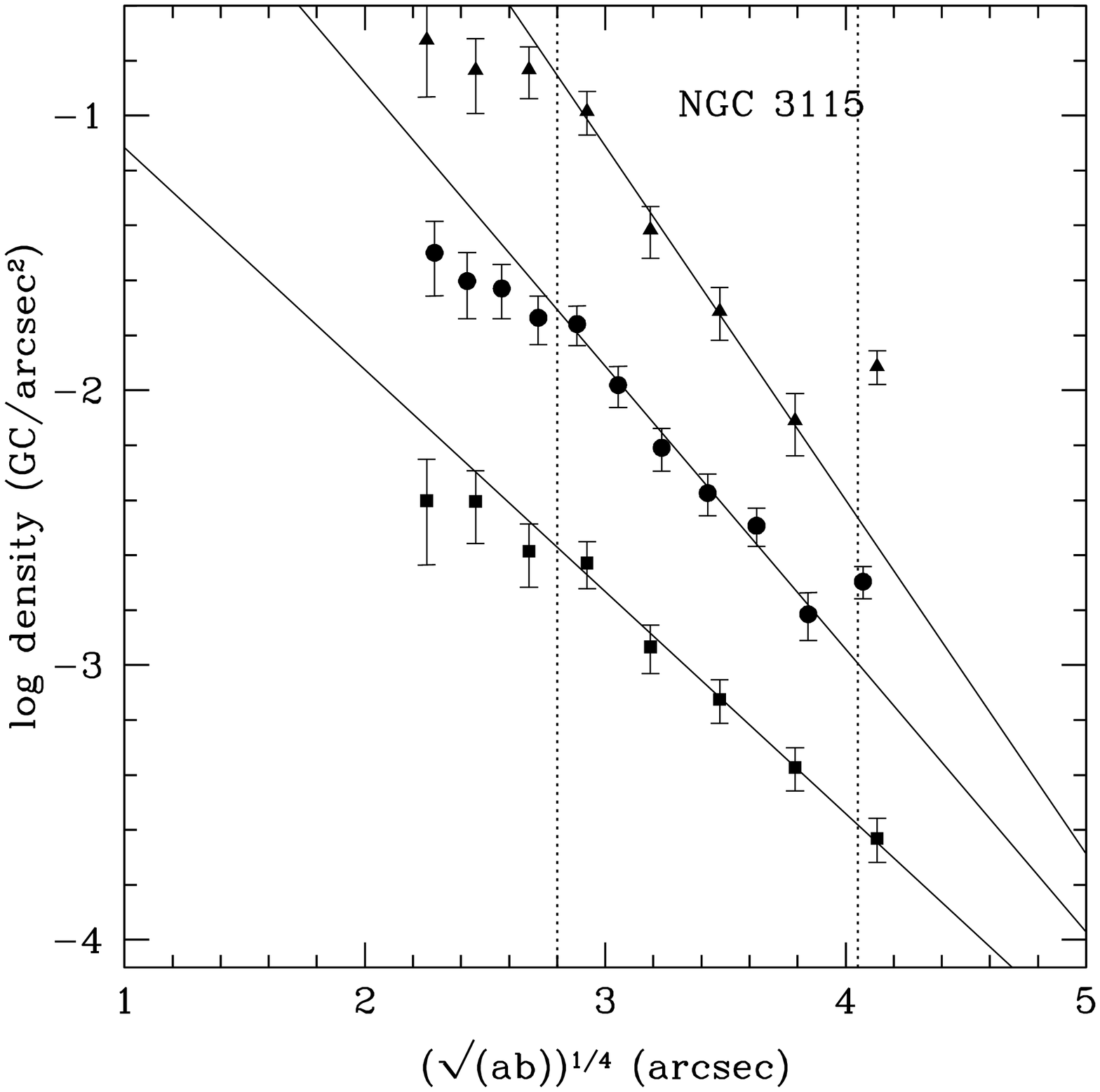}}
\resizebox{0.4\hsize}{!}{\includegraphics{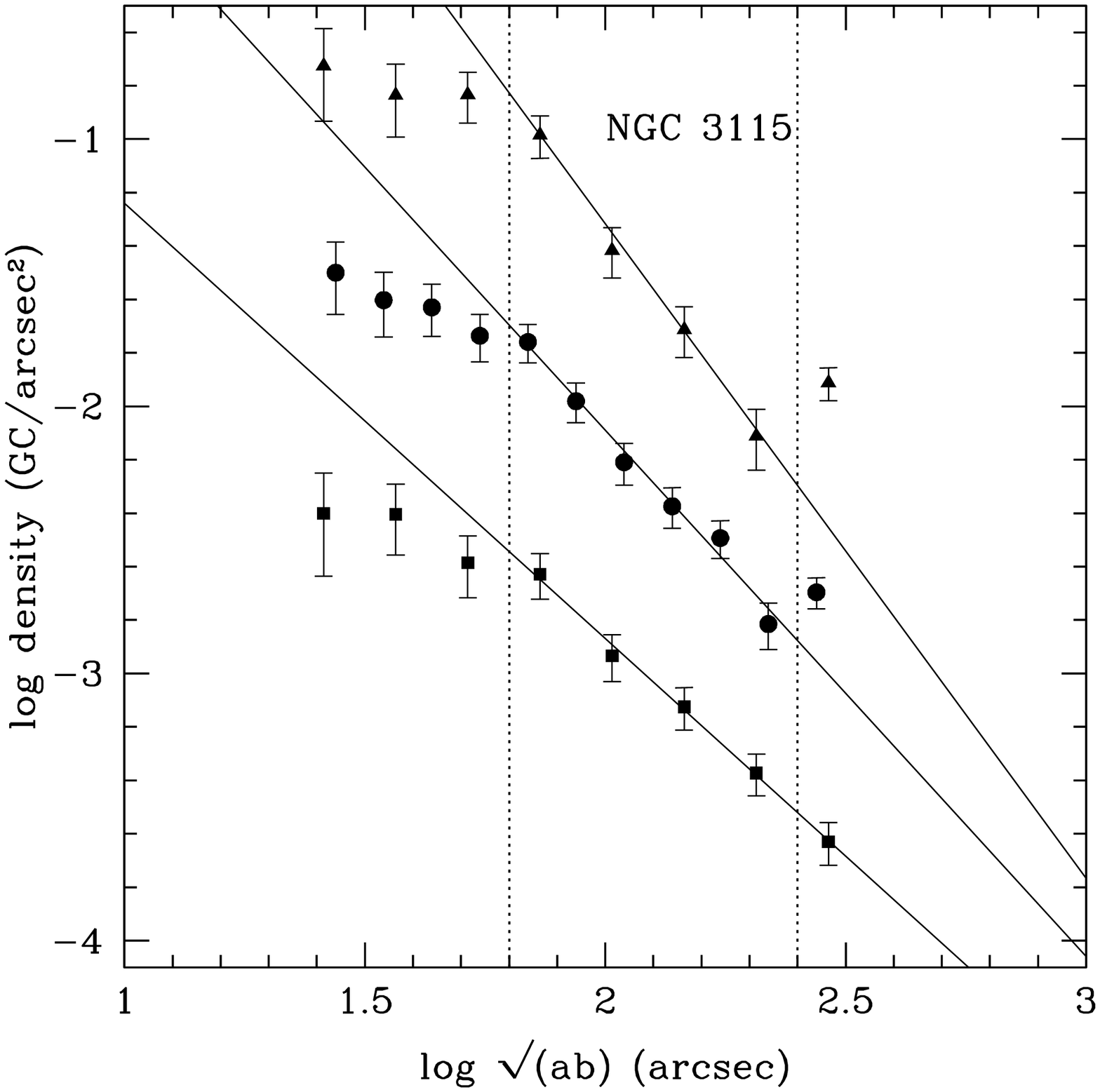}}\\
\resizebox{0.4\hsize}{!}{\includegraphics{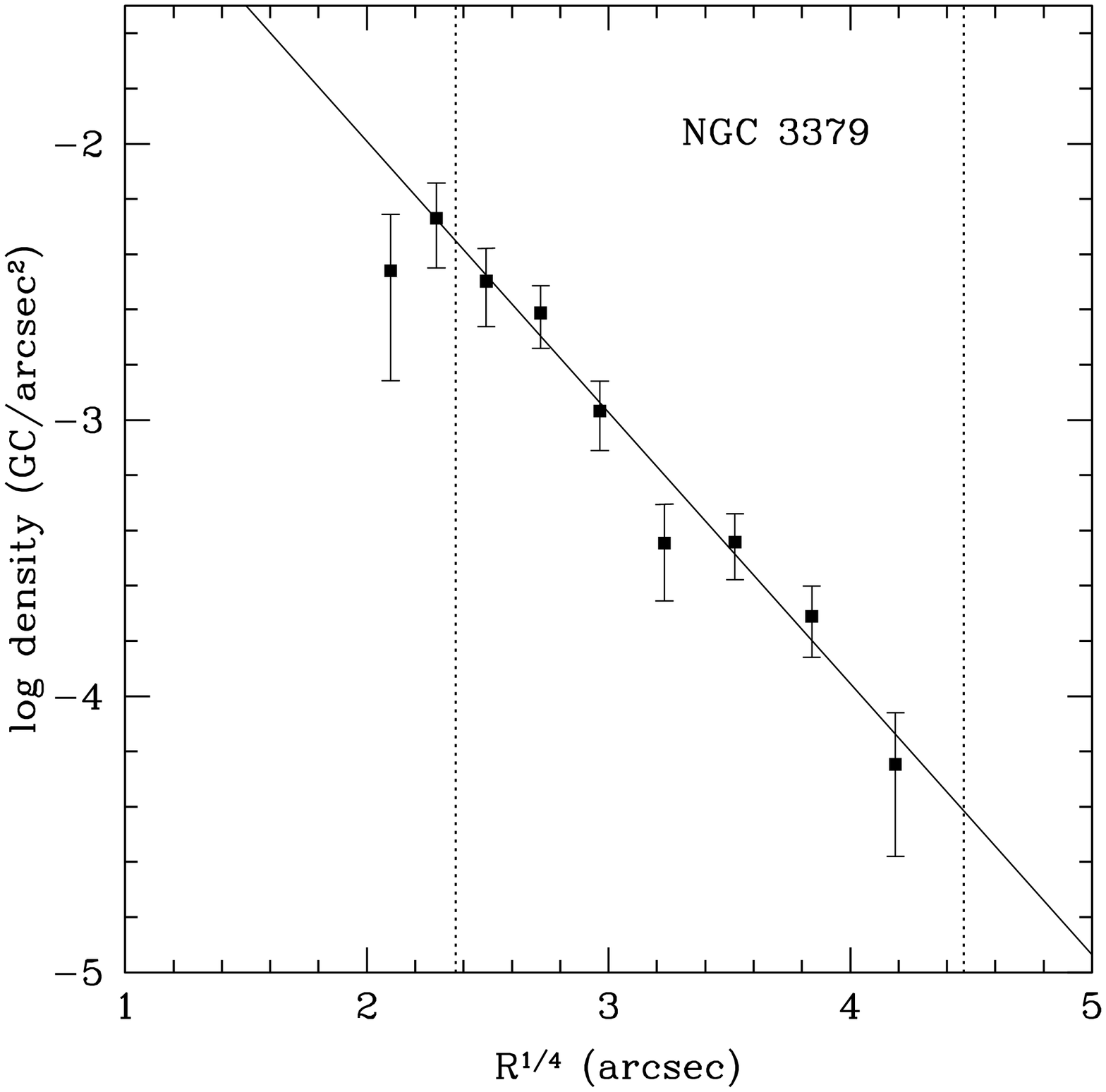}}
\resizebox{0.4\hsize}{!}{\includegraphics{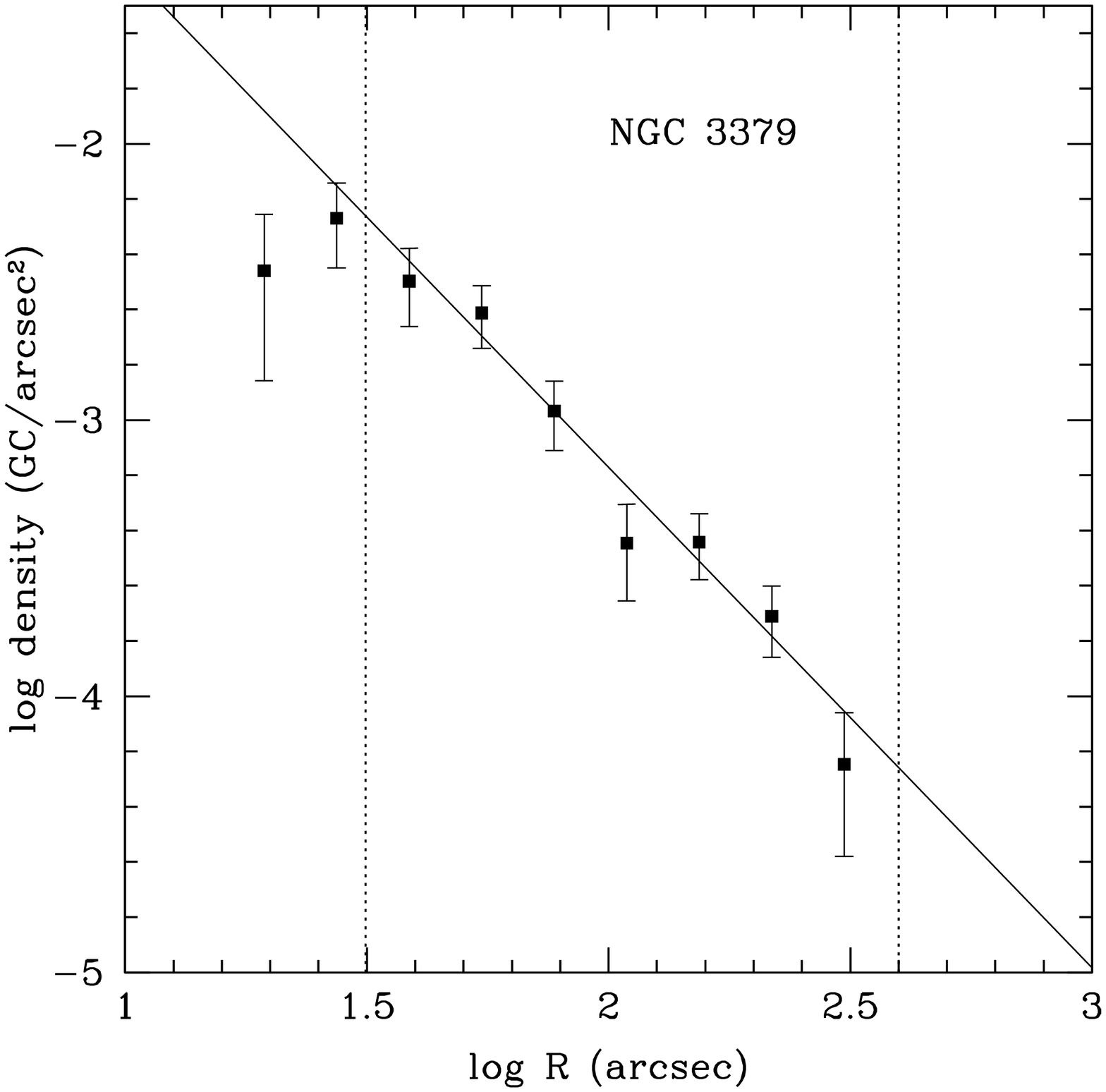}}
\caption{Continued.
}
\label{Perf_D}
\end{figure*}

The resulting areal density distributions are displayed in 
Fig. \ref{Perf_D}, and the fit parameters listed in 
Table \ref{Dens_fit}. As a general
comment, we point out that both the de Vaucouleurs and the power law
give very similar results in terms of the residual errors, although
the former yields marginally better fits in the innermost regions.

In the cases of NGC 4649 and NGC 3923, our observations were combined
with the HST-ACS data allowing a fit to the innermost regions. The GC 
candidates from \citet{JPBCEFTW09} were used for NGC 4649, and 
in NGC 3923 the clusters were selected from our own analysis of the same ACS 
archive images used by \citet{SPCVB2006}. The ACS and GMOS profiles were 
matched by normalizing the counts in common selected radial annuli.

For these galaxies, S\'ersic law fits in the form $log~\sigma(r)=s_1+s_2*r^N$
are also shown in Fig. \ref{Perf_D}. The respective $s_1$, $s_2$ and
 $N$ values are: $-0.57 \pm 0.19$, $-0.27\pm0.09$, $0.36\pm0.04$ for 
NGC 4649, and $-0.60 \pm 0.20$, $-0.51\pm0.20$, $0.26\pm0.09$ for NGC 3923.

A core-like distribution
is detected in the case of NGC 4649, similar to that found, for example,
in M87 by \citet{LK86} and \citet{KWSMZA99}. In turn,
the S\'ersic law seems compatible with a De Vacoucoleurs fit  along the 
whole galactocentric range in NGC 3923.

In what follows we give some brief comments about the characteristics
 of each galaxy.  
    
{\bf NGC 4649.} Our density profile improves that already presented in 
\citet{FFFBBGHSZG04} due to a better determination of the level of the
subtracted background. The slope values given in Table \ref{Dens_fit}, in turn, are in excellent agreement with those presented by \citet{LPKHKG2008}.

{\bf NGC 3923.} Both Fig. \ref{DSS2} and Fig. \ref{Perf_D} show that the 
blue globulars have a very shallow distribution. In fact, the 
slope of the density profile
for these clusters is the lowest in our sample. \citet{SPCVB2006}
also found very low values, although their photometry only includes
the central region of the galaxy.

This feature deserves some attention as it may be connected with the
merger history of this galaxy as suggested by the well known ripples
detectable in its brightness profile (see, \citealt{MC1980}).

{\bf NGC 524.} As mentioned previously, the colour histogram of the GCs in this galaxy
do not show a clearly detectable bimodality and, for this reason,
the density profile displayed in Fig.  \ref{Perf_D} corresponds to the whole
cluster population. The derived value is consistent with \citet{HH85}. 

We note, however, that an inspection of the innermost region in this 
galaxy, indicates that the red GCs seem very concentrated (a situation
that resembles the case of NGC 1427, \citealt{FGOPG01}). However, splitting  
the sample  at {\it $(g'-i')_0=0.9$} as shown by Fig. \ref{DSS2}, 
the ``reddest'' GCs exhibit a clump inside $R_{gal}< 1$ arcmin. In
contrast, this feature is  absent in the ``blue'' clusters.
 
{\bf NGC 3115.} This galaxy exhibits a high flattening and its GCs seem to
follow the brightness isophotal contours. However, our uneven spatial
coverage does not allow a quantitative confirmation (e.g., via
azimuthal counts).

{\bf NGC 3379.} Since this galaxy shows a very low 
number of GC candidates, 
the density profile displayed in Fig.  \ref{Perf_D}
corresponds to the whole GC population. The resulting slope is steeper 
than that given by \citet{RZ2004} who found -1.4 in their wide 
field study. This may indicate
a difference in the background level adopted in these two studies.

Fig. \ref{DSS2} shows that the blue and red candidates in this galaxy 
are not homogeneously distributed around the center. The red GCs show 
a clump to the NE, while the blue ones are clearly less numerous 
in that region. This inhomogeneous distribution is also seen 
in \citet{WFB2003} (their Fig. 3).

\begin{table*}
\centering
\caption{Slopes and errors of the surface density profiles of 
all, red and blue GC candidates obtained using a weighted least 
squares method. We list the results for  power  and  de 
Vaucouleurs laws.
}
\label{Dens_fit}
\scriptsize
\begin{tabular}{lcccccc}
\hline
\hline
\multicolumn{1}{c}{\textbf{Galaxy}} &
\multicolumn{2}{c}{\textbf{all}} &
\multicolumn{2}{c}{\textbf{red}} &
\multicolumn{2}{c}{\textbf{blue}} \\
\multicolumn{1}{c}{} & 
\multicolumn{1}{c}{power law} &
\multicolumn{1}{c}{de Vaucouleurs} &
\multicolumn{1}{c} {power law} &
\multicolumn{1}{c}{de Vaucouleurs} &
\multicolumn{1}{c}{power law} &
\multicolumn{1}{c}{de Vaucouleurs} \\
\hline \multicolumn{7}{c}{}\\
 NGC 4649    & -1.37 $\pm$ 0.04 & -0.67 $\pm$ 0.02 &  -1.52 $\pm$ 0.04 & -0.73 $\pm$ 0.02 & -1.21 $\pm$ 0.08 & -0.58 $\pm$ 0.03\\
 NGC 3923    & -1.25 $\pm$ 0.07 & -0.64 $\pm$ 0.03 &  -1.45 $\pm$ 0.09 & -0.73 $\pm$ 0.04 & -0.86 $\pm$ 0.08 & -0.44 $\pm$ 0.05\\ 
 NGC 524     & -1.67 $\pm$ 0.06 & -0.89 $\pm$ 0.05 &   - & -  &  - & -\\ 
 NGC 3115           & -1.59 $\pm$ 0.13 & -0.78 $\pm$ 0.07 &  -1.74 $\pm$ 0.17 & -1.08 $\pm$ 0.10 & -1.45 $\pm$ 0.15 &- 0.73 $\pm$ 0.07\\
 NGC 3115(ellip)    & -1.97 $\pm$ 0.12 & -1.02 $\pm$ 0.05 &  -2.45 $\pm$ 0.11 & -1.28 $\pm$ 0.08 & -1.62 $\pm$ 0.05 & -0.81 $\pm$ 0.04\\
 NGC 3379    & -1.81 $\pm$ 0.16 & -0.98 $\pm$ 0.08 &   - & -  &  - & -\\ 
\multicolumn{7}{l}{}\\
\hline                                                                             
\multicolumn{7}{l}{}\\ 
\end{tabular}
\end{table*}

\subsection {Galactocentric colour trends}

\begin{figure}

\resizebox{1\hsize}{!}{\includegraphics{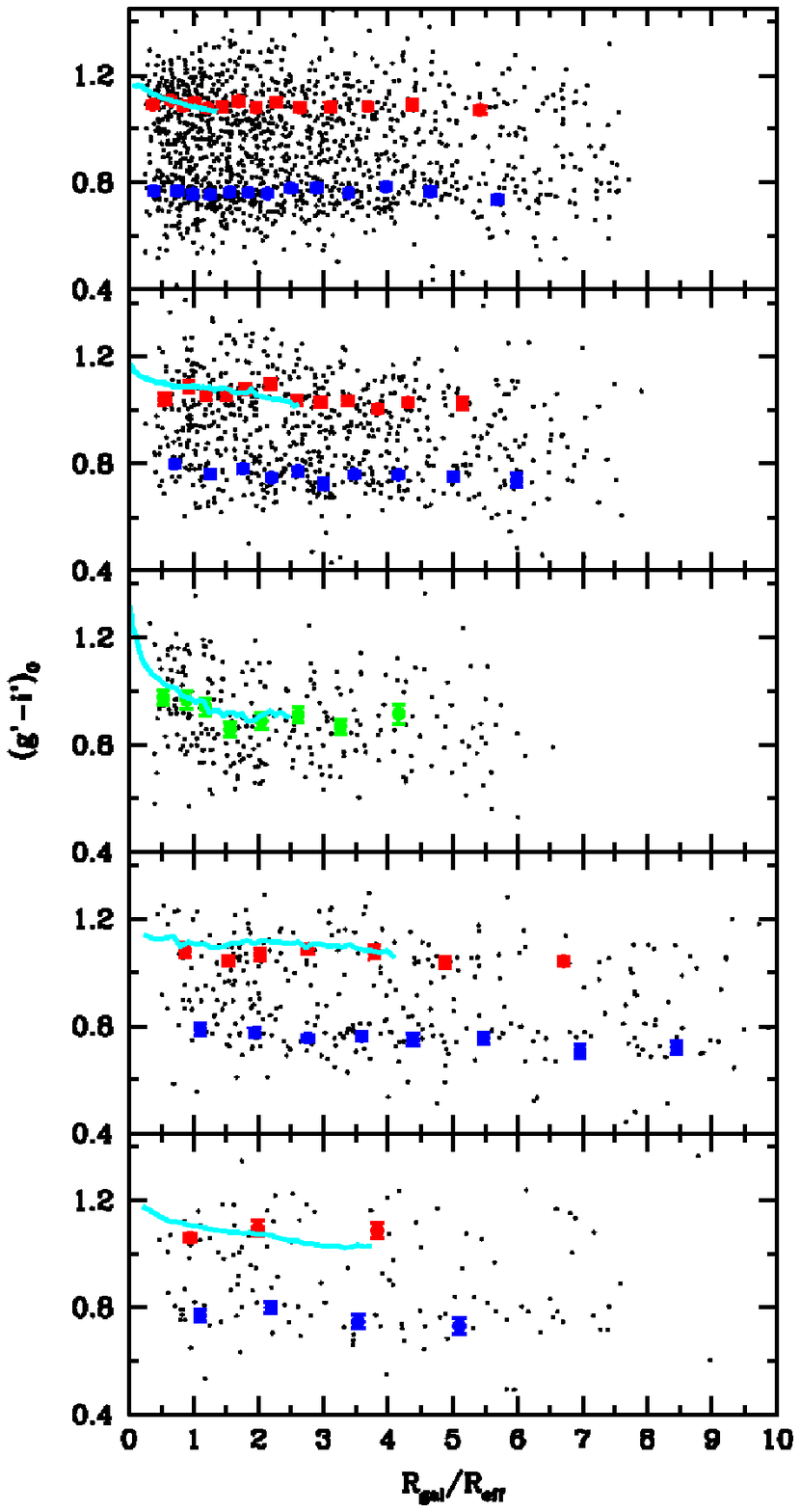}}

\caption{{\it (g$'$-i$'$)$_0$} colour index versus projected galactocentric 
distance normalized by $R_{eff}$( in the B band). The colour profile for each galaxy 
halo, within a radius of 2 arcmin, is shown as the solid line in each 
panel. From top to bottom: NGC 4649, NGC 3923, NGC 524, NGC 3115 and NGC 3379.
}
\label{COL_Rgal}
\end{figure}

Fig. \ref{COL_Rgal} shows the {\it (g$'$-i$'$)$_0$} colours of all 
the globular cluster candidates brighter than the 90 percent 
completeness level as a function of the normalized galactocentric 
radius {\it R$_{gal}$/R$_{eff}$}, where {\it R$_{eff}$} is the 
effective radius in the {\it B} band taken from RC3 data in NED.
  
Mean colours for each GC family were obtained within galactocentric
bins of variable size (containing the same number of clusters) and
fit with a logarithmic law

\begin{equation}
(g'-i')_0= a~log (R/R_{eff}) + b
\end{equation}

The {\it a} and {\it b} parameters are listed in Table \ref{Col_fit}, 
along with the associated uncertainties obtained through a least 
square fit, for each GC subpopulation except for NGC 3379 where, 
due to the low number of clusters, blue and red GCs were 
grouped together. The listed colour gradients were combined with 
Equation \ref{Z_col} to obtain values of $\Delta [Z/H]/ log (R/R_{eff})$ 
and were included in Table \ref{Col_fit}.

The overall appearance of the Fig. \ref{COL_Rgal} suggests very mild gradients
over a galactocentric range larger than $\approx$ 1 to 2 R$_{eff}$. A similar 
trend has been noted by \cite{H09a,H09b} in  M87 and six
other GCS belonging to giant elliptical galaxies. Beyond this, each 
galaxy has its own and distinct behavior. 
In particular, NGC 524 and NGC 3115 (both S0 galaxies) exhibit the 
highest colour gradients in their innermost regions \citep{TNCCJM10}. 

Fig. \ref{COL_Rgal}  also includes the {\it (g$'$-i$'$)$_0$} halo 
colours derived from our images. They were obtained within the central 2 arcmin 
of the galaxies because this zone is not strongly affected by the sky 
level adopted. These colours are remarkably similar to 
the mean colours
of the red GCs, a similarity already noted in the inner regions of 
several galaxies (see, for example, \citealt{FF01}).

\begin{table*}
\centering
\caption{Slopes and errors of the mean {\it (g$'$-i$'$)$_0$} colours 
in function of R$_{gal}$. These values were obtained by fitting the relation: 
$(g'-i')_0 = a \times log (R_{gal}/R_{eff}) + b$. The [Z/H] gradients, obtained
using Equation \ref{Z_col}, are listed in columns four and five. In the cases 
of NGC 4649, NGC 3923 and NGC 3115, two values are listed. The first one show 
the slopes obtained by using the entire sample of points showed in the 
figure, and the second one,  corresponds to a fit excluding the inner point.
}
\label{Col_fit}
\scriptsize
\begin{tabular}{lcccccc}
\hline
\hline
\multicolumn{1}{c}{\textbf{Galaxy}} &
\multicolumn{2}{c}{\textbf{slopes {\it(g$'$-i$'$)$_0$}}}  &
\multicolumn{2}{c}{\textbf{slopes {\it[Z/H]}}}  &
\multicolumn{2}{c}{\textbf{N$_{bin}$}} \\
\multicolumn{1}{c}{} &
\multicolumn{1}{c}{red} & 
\multicolumn{1}{c}{blue} &
\multicolumn{1}{c}{red} & 
\multicolumn{1}{c}{blue} &
\multicolumn{1}{c}{}\\
\hline \multicolumn{6}{c}{}\\
 NGC 4649    & -0.015$\pm$0.007  & -0.001$\pm$0.011   &  -0.052 $\pm$0.024 & -0.004 $\pm$ 0.038 &50   \\
             & -0.021$\pm$0.009  &  0.004$\pm$0.015   &  -0.073 $\pm$0.032 &  0.014 $\pm$ 0.052 &50   \\
 NGC 3923    & -0.050$\pm$0.024  & -0.051$\pm$0.019   &  -0.175 $\pm$0.084 & -0.179 $\pm$ 0.067 &30   \\
             & -0.095$\pm$0.024  & -0.033$\pm$0.025   &  -0.333 $\pm$0.086 & -0.115 $\pm$ 0.088 &30    \\
 NGC 524     & \multicolumn{2}{c}{-0.099$\pm$0.043  / -0.194 $\pm$ 0.069} & \multicolumn{2}{c}{-0.347 $\pm$0.152/ -0.681 $\pm$ 0.245} &30\\
 NGC 3115    & -0.030$\pm$ 0.028 & -0.076$\pm$0.016   & -0.105$\pm$0.098  &  -0.266 $\pm$0.058 & 20\\
             & -0.022$\pm$ 0.037 & -0.081$\pm$0.024   & -0.077$\pm$0.130  &  -0.284 $\pm$0.085 & 20\\      
 NGC 3379    &            -      & -0.066 $\pm$ 0.055   & -  &  -0.232 $\pm$0.193     & 15       \\ 
\multicolumn{3}{l}{}\\
\hline                                                                             
\multicolumn{2}{l}{}\\ 
\end{tabular}
\end{table*}

Our data adds to the
increasing evidence in the literature for colour, and hence
metallicity, gradients in the individual GC subpopulations
(e.g. \citealt{H09b}; \citealt{FSSRBF11}). Such gradients are
indicative of a dissipative formation process for both GC
subpopulations. In a detailed study of the NGC 1407 GC system,
\citet{FSSRBF11} suggested that a transition occurred between
the inner region with strong GC gradients and an outer region
with no colour gradient. This was interpreted as support for the
two phase formation model of early-type galaxies \citep{OONJB10} 
in which such galaxies are built up from a dissipative
core that experiences later accretion. Our data show shallower
gradients and fewer GCs than detected in NGC 1407 by Forbes et
al. Further data on a rich GC system like NGC 4649 would be
needed to search for such a transition.

The GC colour gradients are also compared  to the colour
gradient of the underlying starlight in the galaxy. In each case
where the GC system can be clearly separated into two
subpopulations we find that the galaxy gradient matches the 
red GC subpopulation gradient in both absolute colour and in
slope over a common radial range. This provides strong support
for the idea that the red GCs are associated with the
bulge/spheroid stars of early-type galaxies, and hence may have
shared a common formation epoch \citep{FF01,FVF09,S2010}.

\subsection{GC Integrated Luminosity Functions and GC Specific Frequencies
S$_N$}

\subsubsection{Integrated luminosity functions}

\begin{figure*}
\begin{center}
\resizebox{0.43\hsize}{!}{\includegraphics{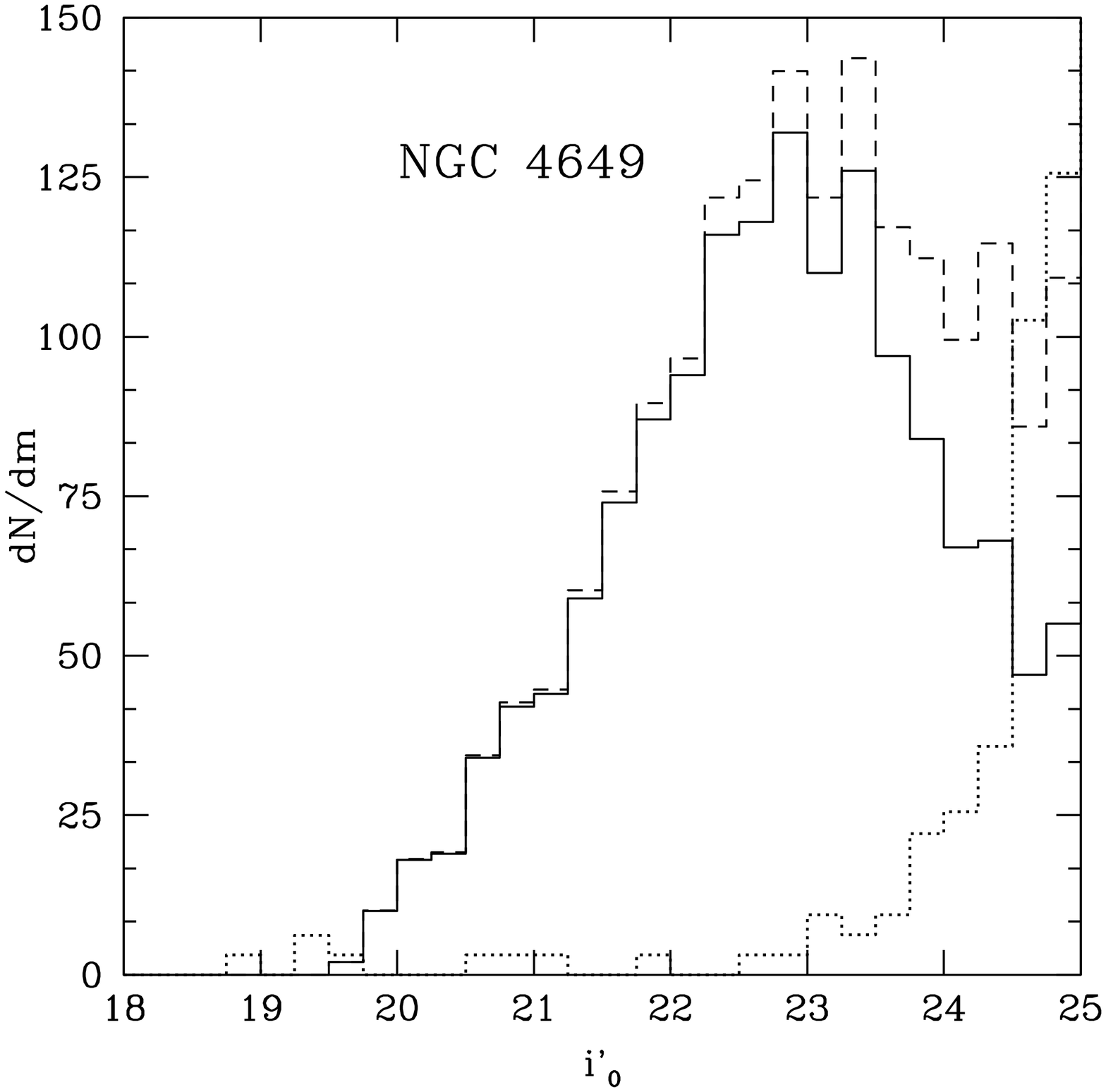}}
\resizebox{0.43\hsize}{!}{\includegraphics{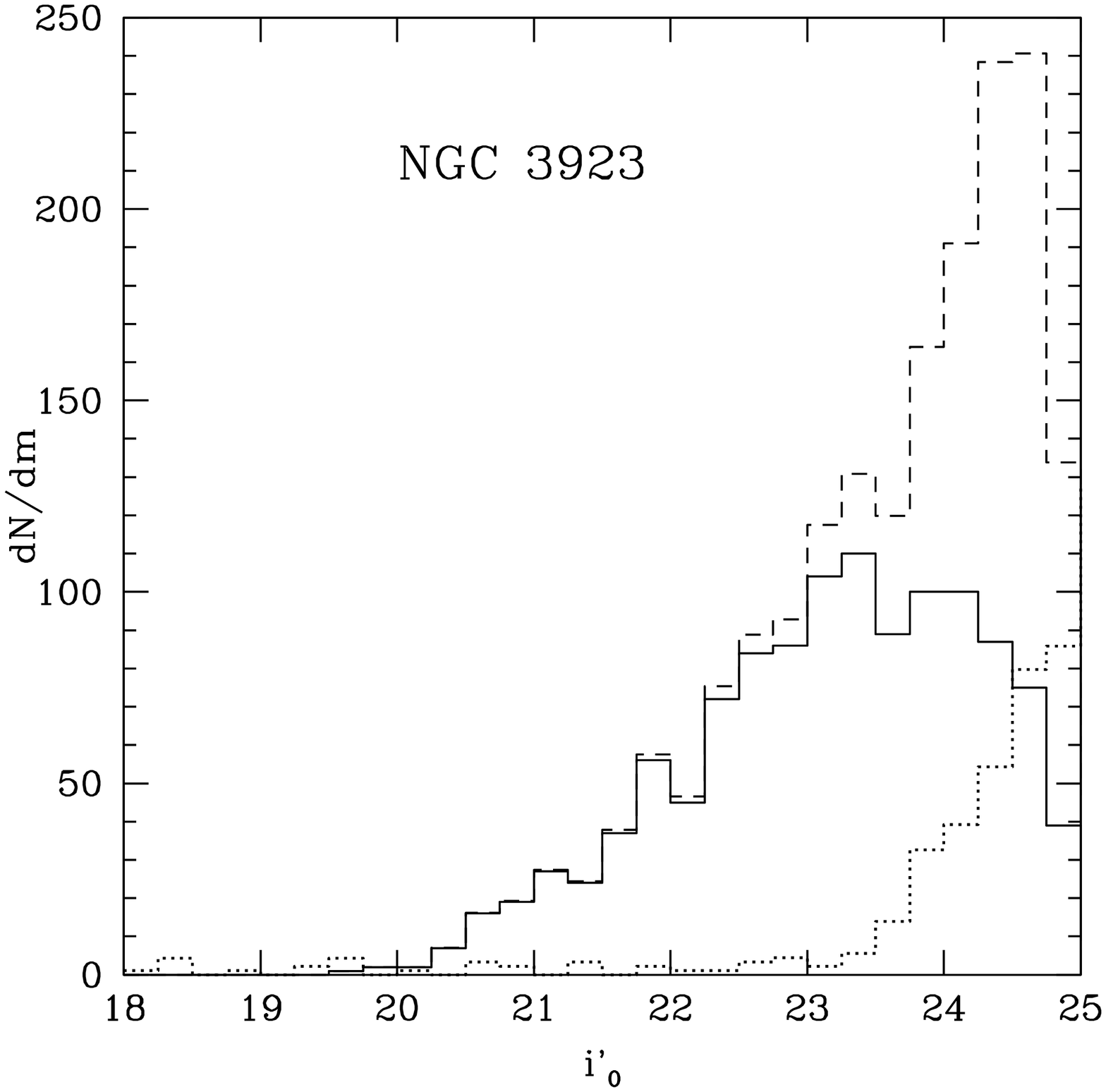}}\\
\resizebox{0.43\hsize}{!}{\includegraphics{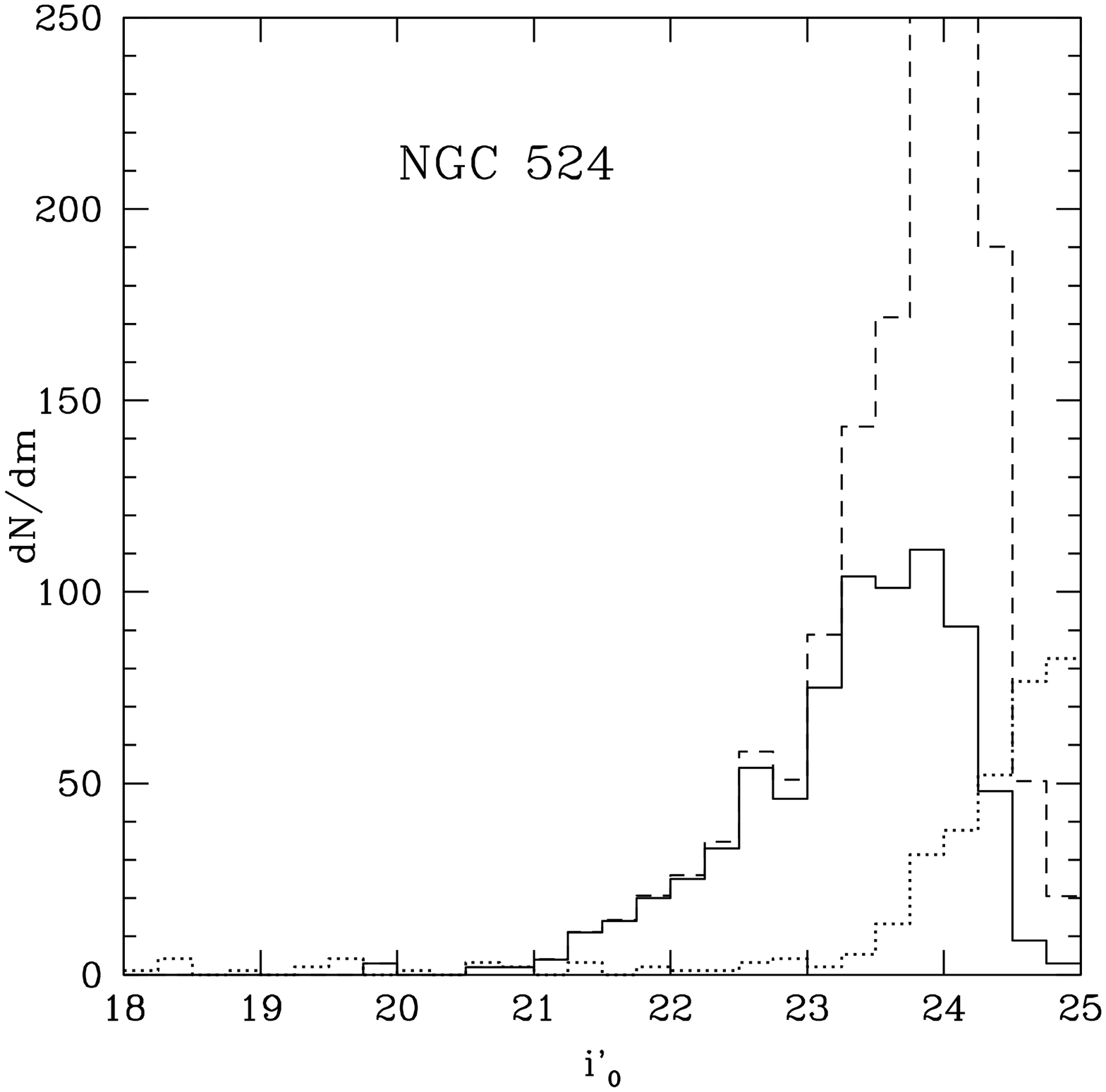}}
\resizebox{0.43\hsize}{!}{\includegraphics{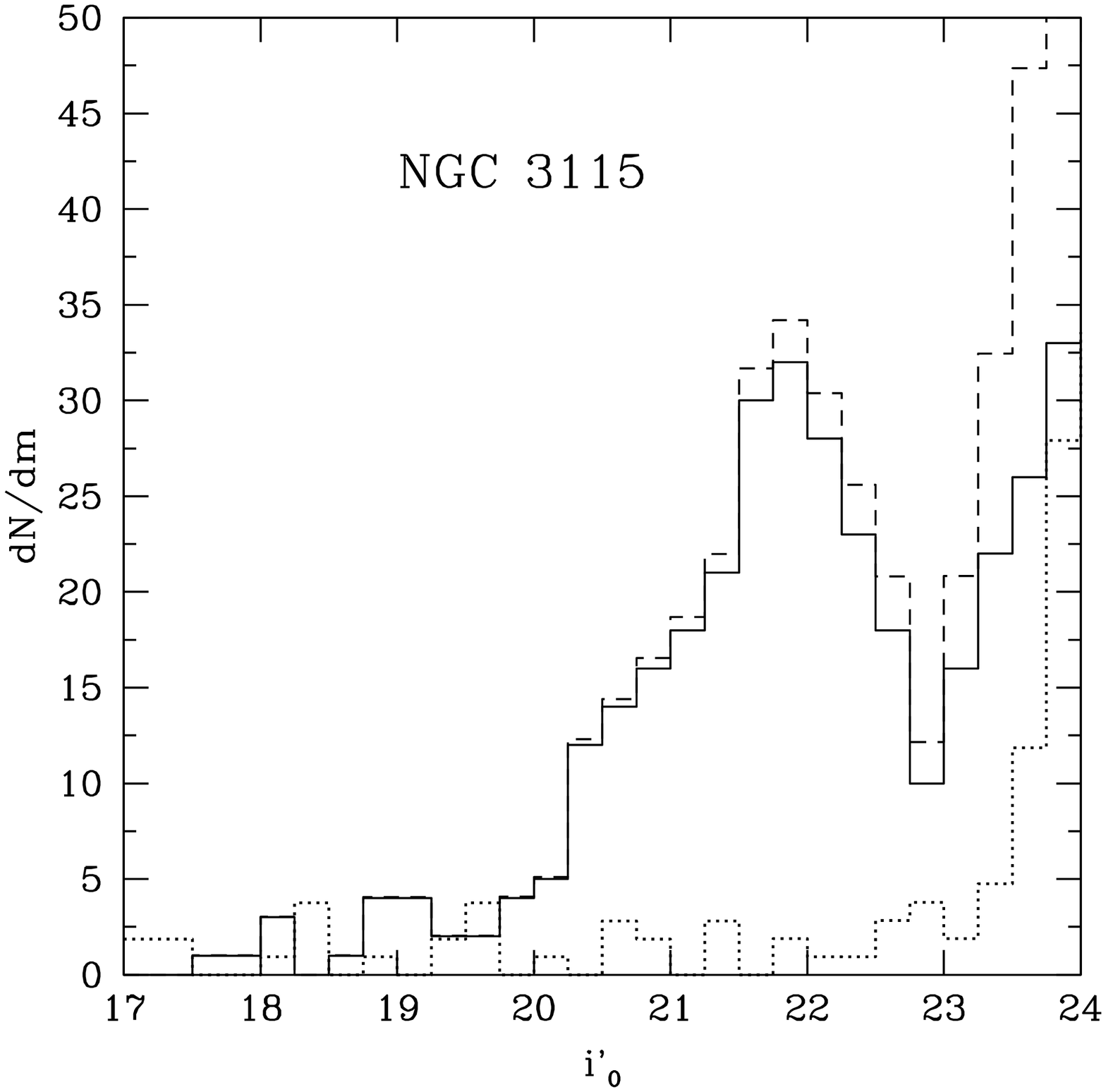}}\\
\resizebox{0.43\hsize}{!}{\includegraphics{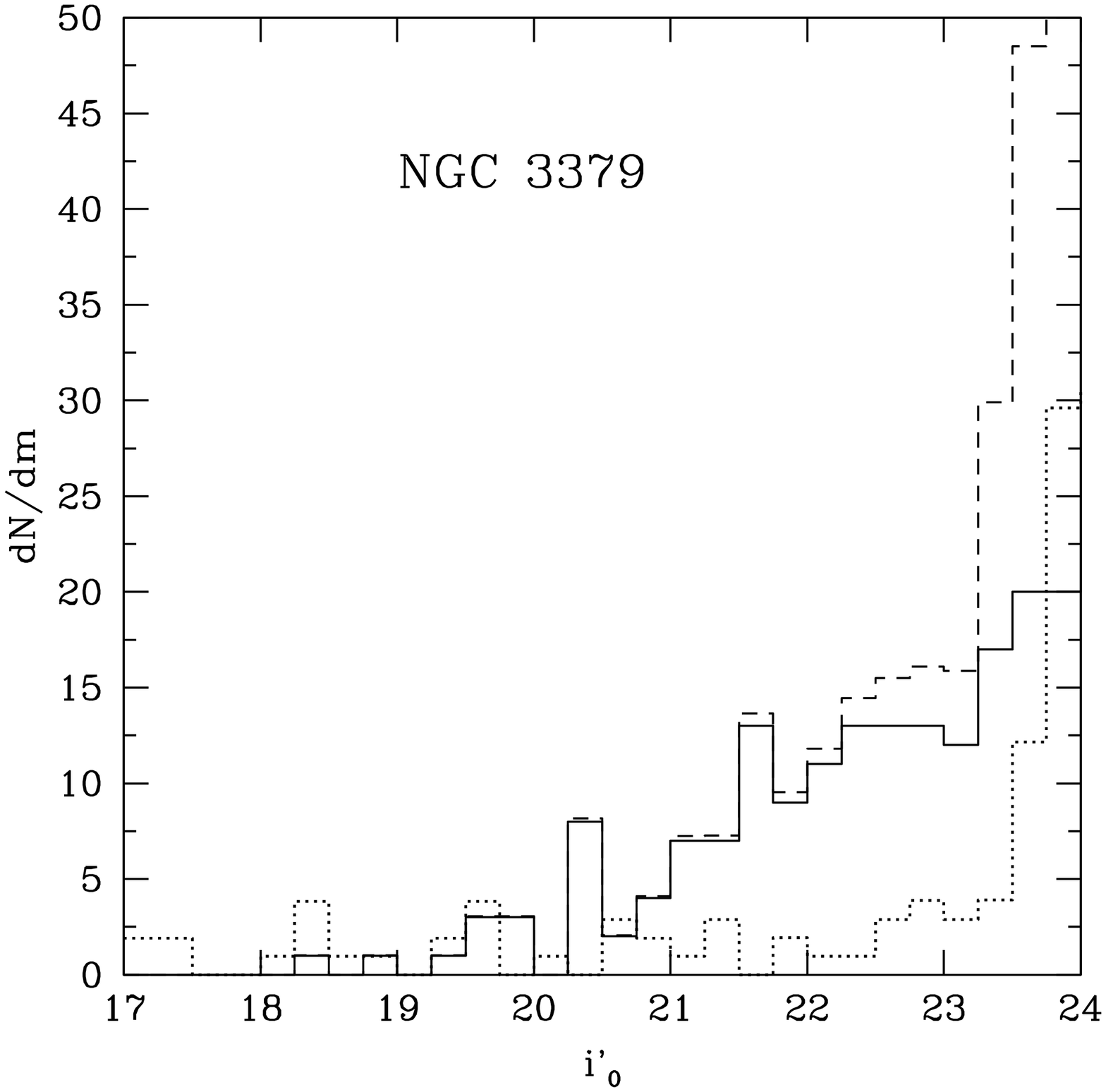}}\\
\end{center}
\caption{Raw counts (solid histograms), and completeness corrected counts (dashed line) 
for the five galaxies in the sample as a function of {\it i$'_0$} magnitudes. The normalized 
backgrounds adopted in each case are shown as dotted lines. 
}
\label{FL_raw}
\end{figure*}

In this section we aim to determine the integrated GC luminosity
function (GCLF) for each galaxy. Fig. \ref{FL_raw} shows the raw  
 and completeness corrected counts of
cluster candidates as well as those of the adopted comparison field
as a function of {\it i$'_0$} magnitude.  NGC 4649 and NGC 3115 
exhibit a good contrast at the expected TOM magnitude
between clusters and contaminant objects allowing this kind of analysis. 
In NGC 3923, we don't have such a good contrast. However, the completeness 
and the background counts are good enough to try to recover some 
useful information. In NGC 3379 the low number of 
candidates, and the strong dependence of the background adopted 
make it difficult to obtain reliable results and the parameters of the 
GCLF are not very well constrained. For that 
reason the points with {\it i$'>$ 23.2} were rejected from 
the fit. In the case of NGC 524, we expect to have the 
TOM at $i'_0\sim 23.9$, where our photometry has a very low 
photometric completeness and therefore this 
galaxy was not included in what follows.

\begin{figure*}
\begin{center}
\resizebox{0.43\hsize}{!}{\includegraphics{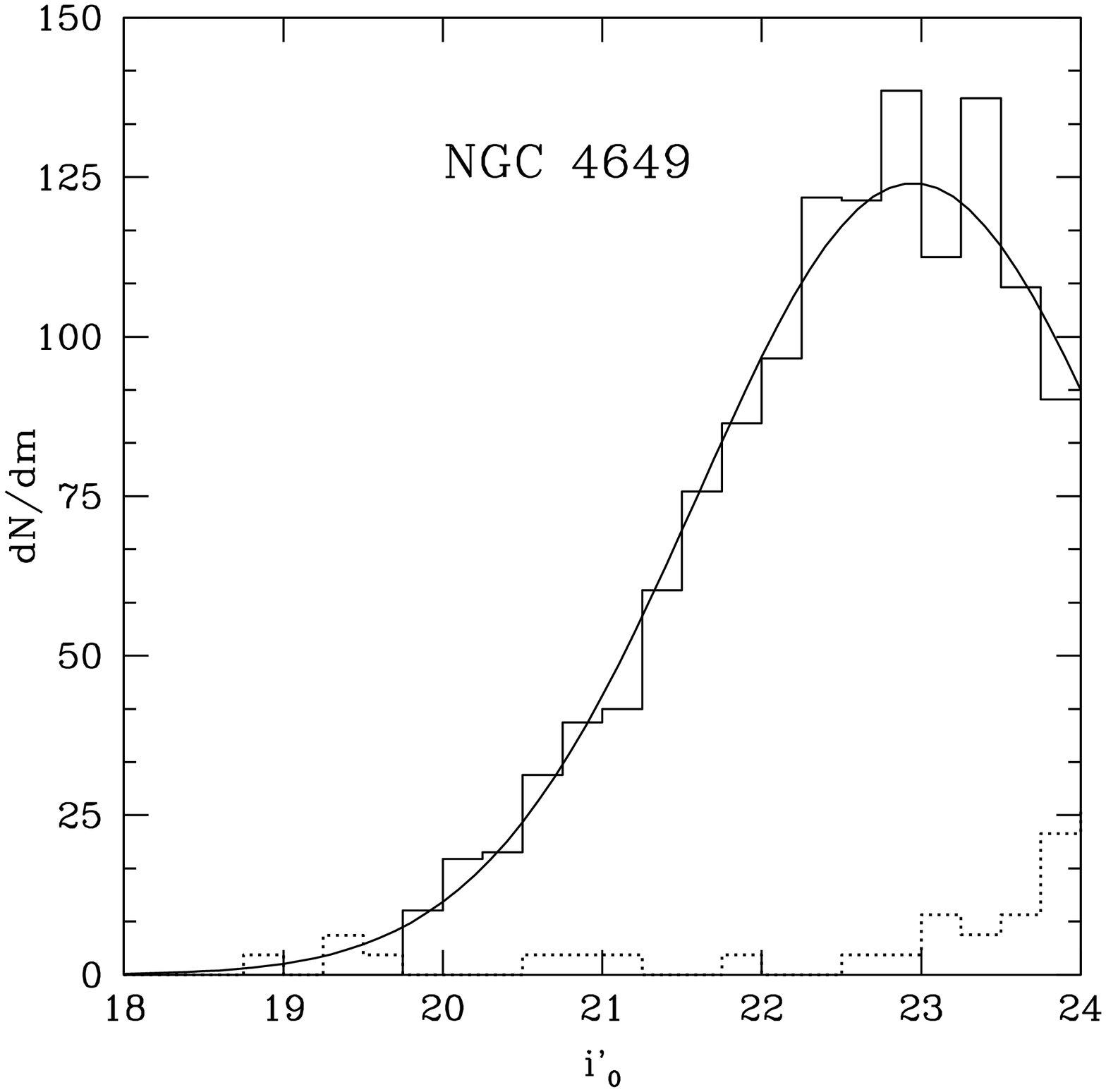}}
\resizebox{0.43\hsize}{!}{\includegraphics{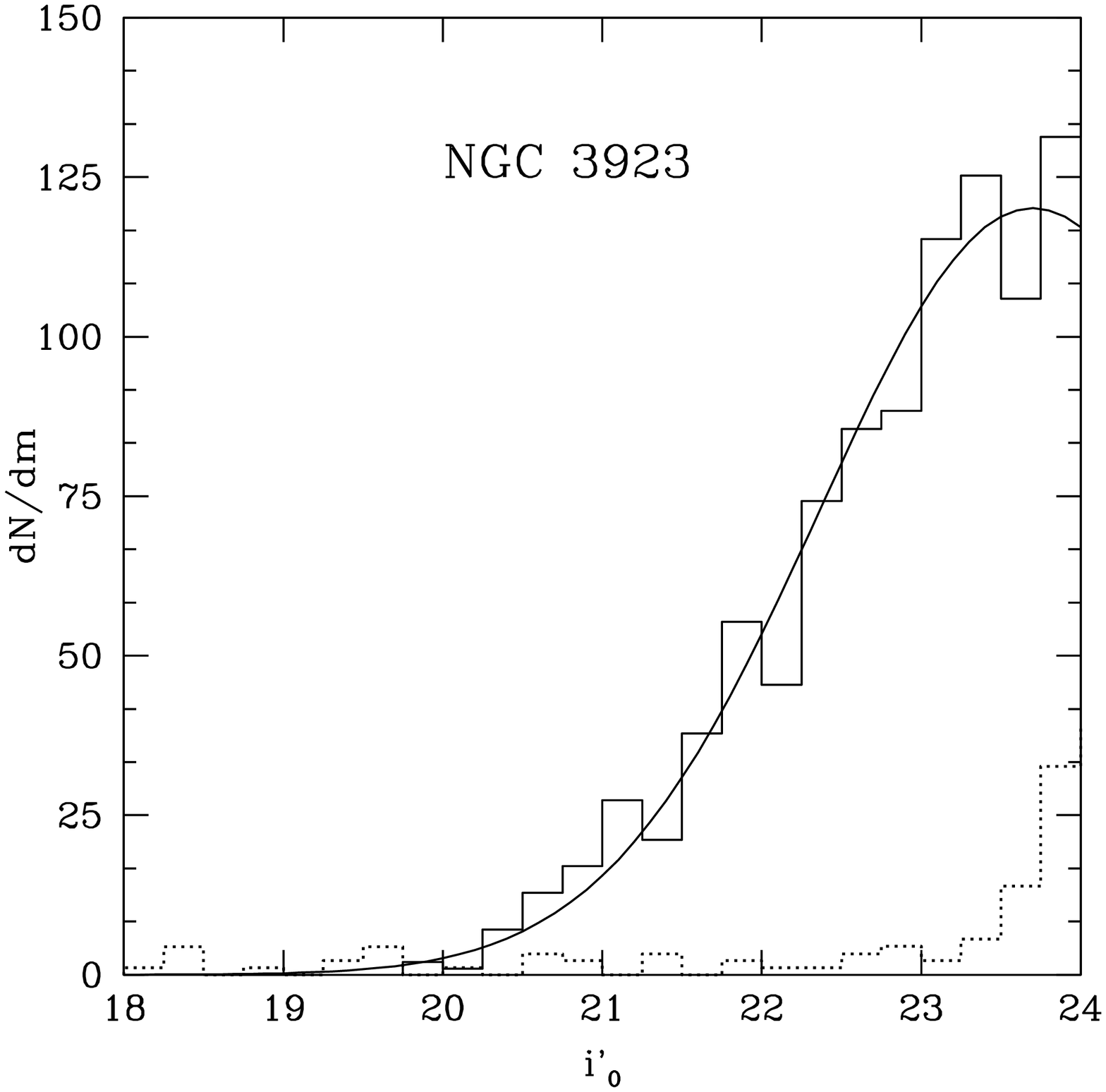}}\\
\resizebox{0.43\hsize}{!}{\includegraphics{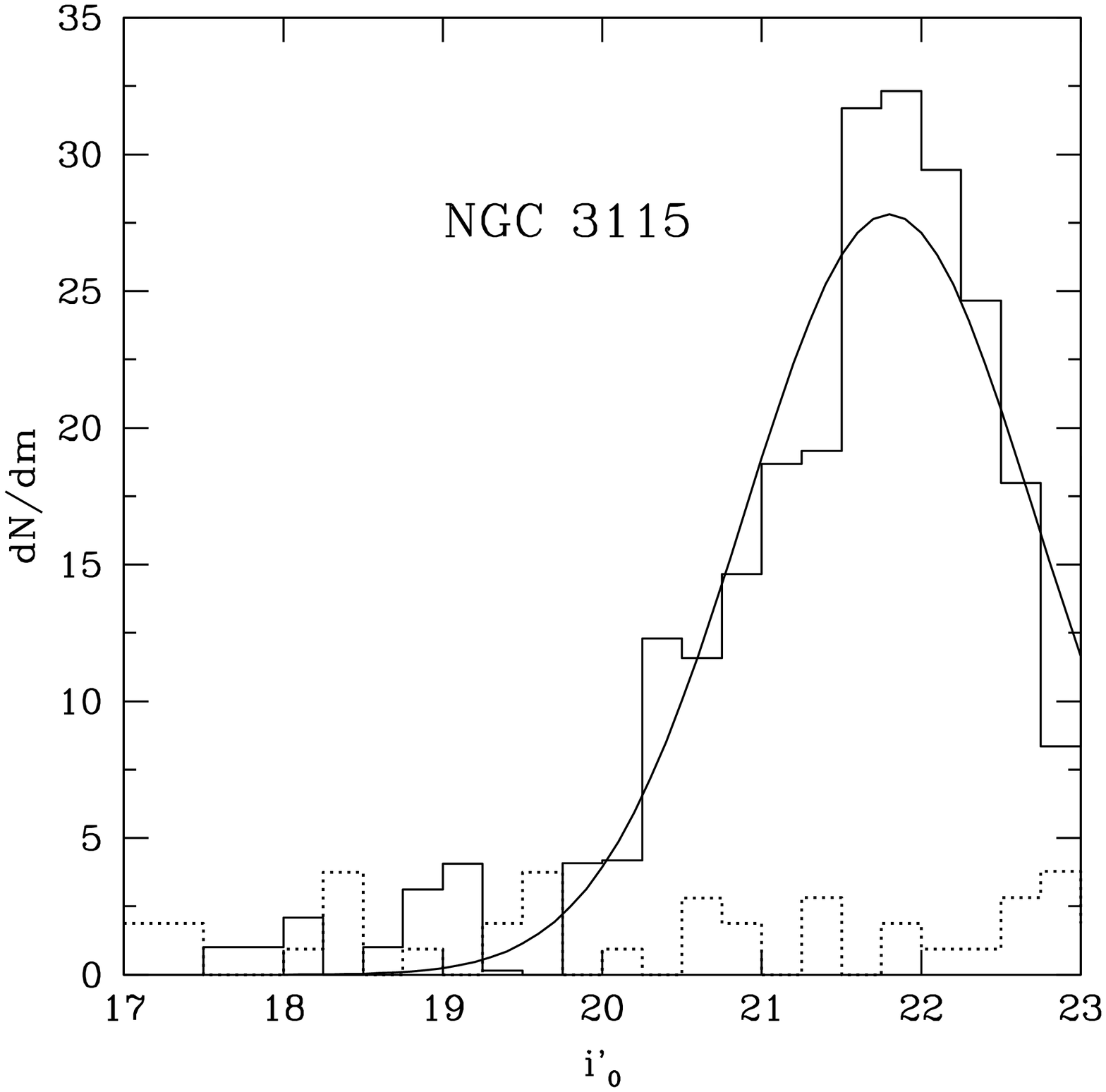}}
\resizebox{0.43\hsize}{!}{\includegraphics{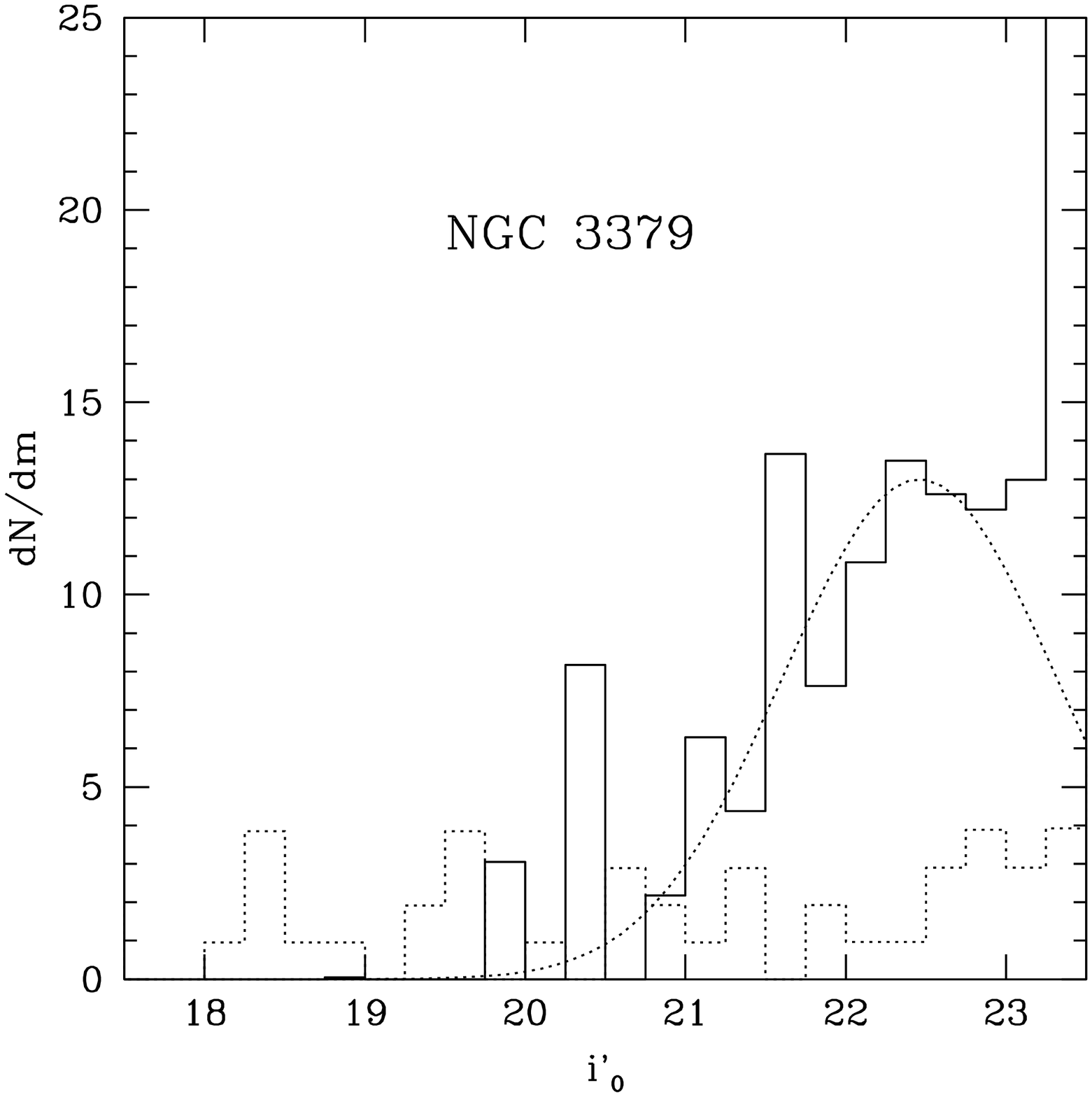}}
\end{center}
\caption{Globular cluster luminosity functions in the {\it i$'_0$} band. We 
show with solid lines the background and completeness corrected histograms of 
the GC candidates. The fitted Gaussian GCLF are shown as  solid lines 
too. The corresponding backgrounds are 
displayed as dotted histograms.
}
\label{FL}
\end{figure*}

The resulting GCLF, determined on the basis 
of the net counts within 0.25 mag bins, are displayed in 
Fig. \ref{FL} together with the normalized counts of the subtracted
background.

Both a Gaussian and a t5 function fit were performed on each
histogram in order to determine the position of the turn-over
magnitude (TOM) and the dispersion parameter. As the results do
not show clear differences between both approaches, we only give
the parameters of the Gaussian fit in Table \ref{GCLF_fit}.

Reliable estimates of the TOMs in NGC 4649 
and NGC 3115 are derived since our adopted limiting magnitude
is fainter than the TOMs by approximately one magnitude.

\begin{figure*}
\begin{center}
\resizebox{0.43\hsize}{!}{\includegraphics{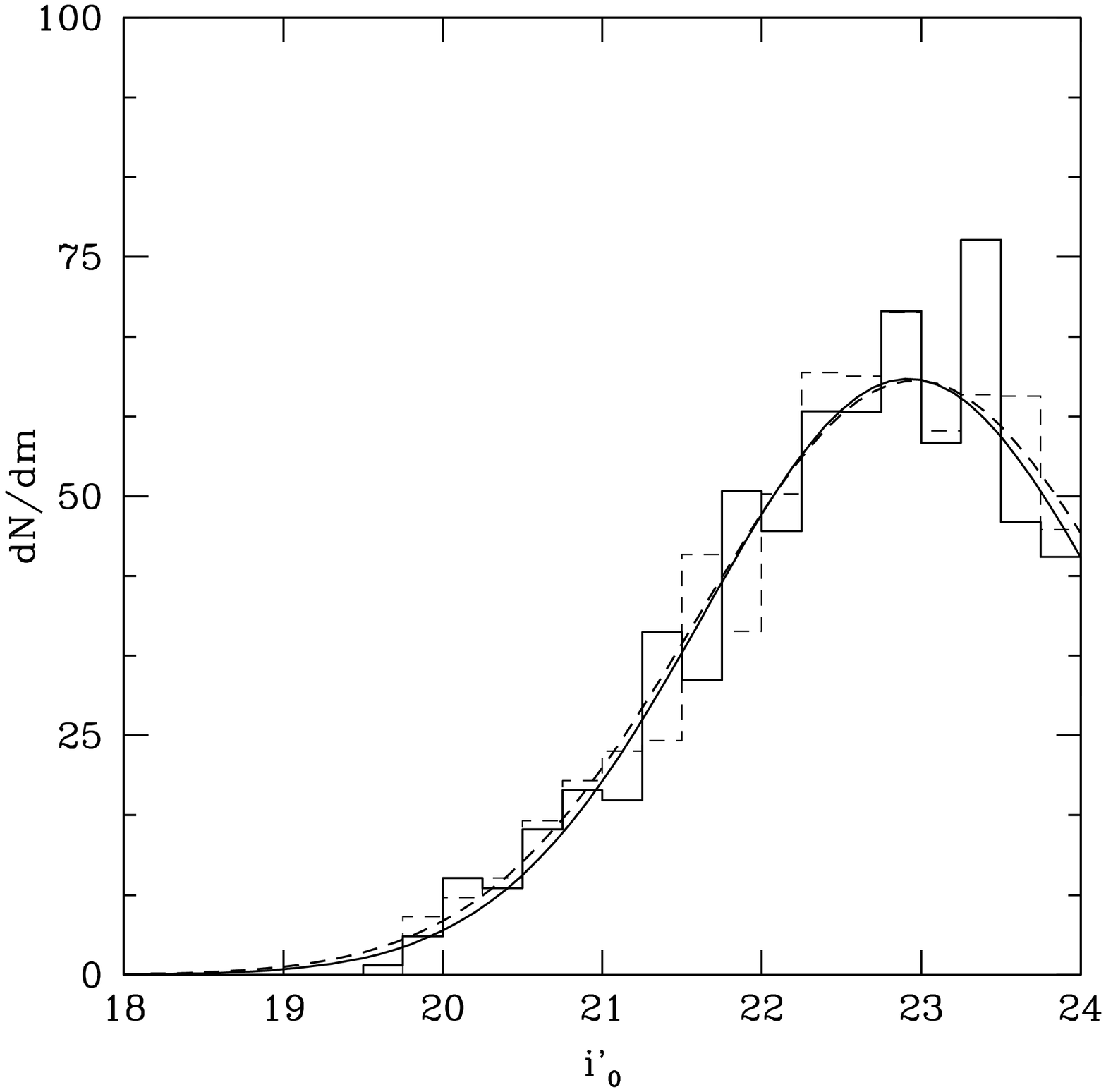}}
\resizebox{0.43\hsize}{!}{\includegraphics{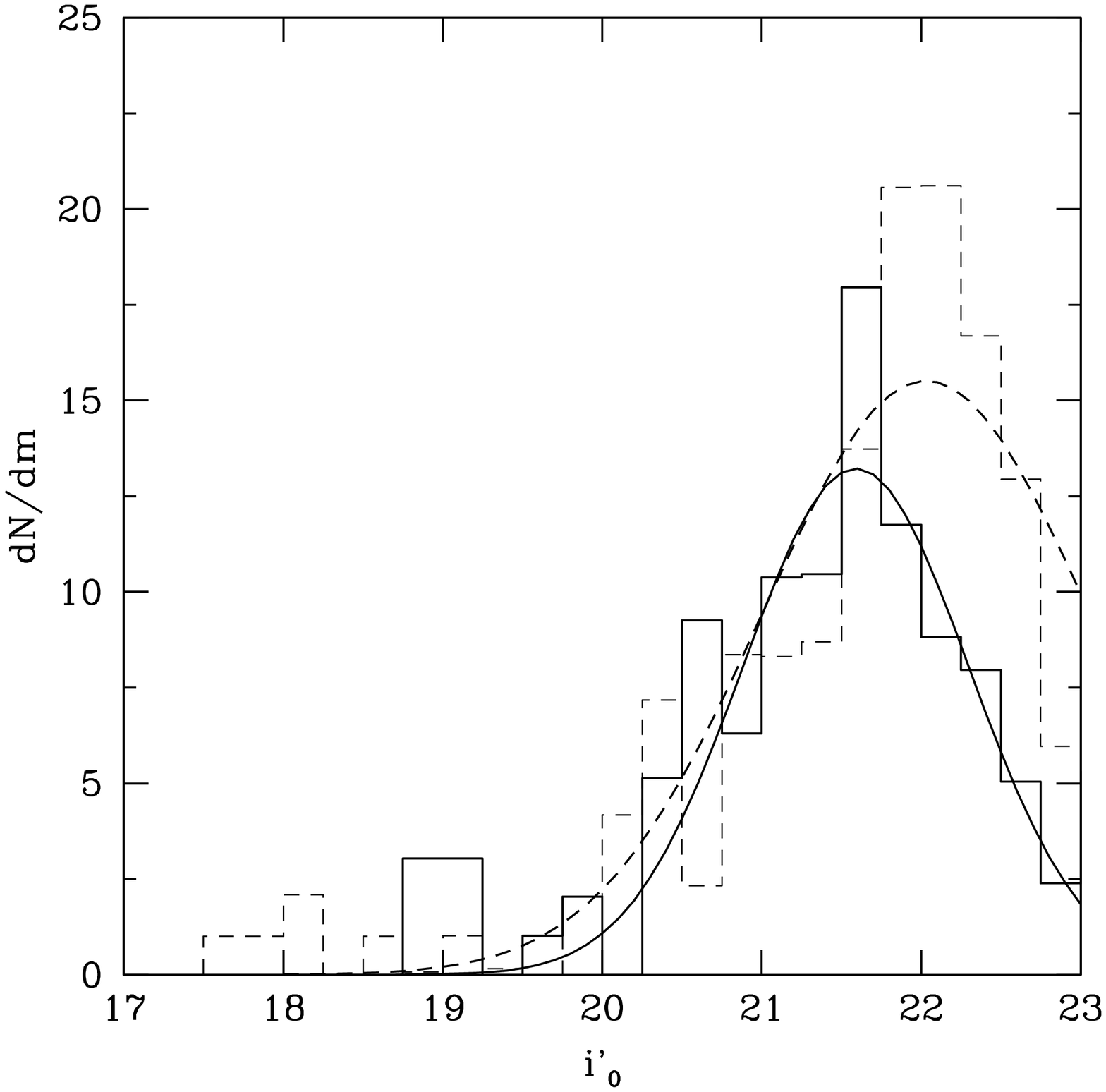}}\\
\end{center}
\caption{Corrected luminosity functions for each GCs subpopulation 
in galaxies NGC 4649 (left)
and NGC 3115 (right). The LF for red GCs are shown as solid 
lines and those for blue GCs as dashed lines.
}
\label{FL_BR}
\end{figure*}

For these galaxies we also obtained the TOMs of the blue and
red GC populations defined in terms of the {\it (g$'$-i$'$)$_0$} ``valley''
in the colour statistics. The corresponding corrected luminosity
functions for each GCs subpopulation are shown in Fig. \ref{FL_BR}. A comparison
shows that, in NGC 4649, the blue and red TOMs are coincident
within the errors. This agreement is expected as the {\it i$'_0$} magnitudes
are not strongly affected by metallicity effects (see \citealt{ACZ95}). 
The situation is not clear in NGC 3115 where the 
blue TOM seems fainter than the red one, contrary to the
expectations for two coeval GC populations with different
metallicities. An inspection of Fig. \ref{DSS2} shows that our areal
coverage of the central region of this galaxy is rather poor,
resulting in the loss of a fraction of the red GCs population
(because red GCs are more concentrated towards the center of 
the galaxy than the blue one) and possibly producing an spurious 
effect on the magnitude statistics of these clusters.   

For NGC 3923, and as the TOM seems very close to the magnitude
cut off (at a 50 percent completeness level), we only derive the
dispersion. In this case we adopted the I band TOM given by \citet{KW2001}
 which, at the adopted distance modulus becomes
{\it i$'_0$}=$23.8  \pm 0.28$.

The adoption of the absolute TOM magnitudes given by \citet{KW2001}, 
and using the relation given in Equation \ref{eq1}, lead to
distance moduli {\it (m-M)}$_I$= $30.92 \pm 0.06$ and 
{\it {(m-M)$_I$}= $29.75 \pm 0.09$} for NGC 4649, NGC 3115 respectively.

A comparison with \citet{LBHFG2001}, \citet{KW1998} 
and  \citet{KW2001} (previous correction by the TOM for the
MW GCs and the {\it M$_V$} adopted by these authors) shows an excellent
 agreement between their distance moduli and ours.

\subsubsection{Specific frequency and total GC population}

Adopting Gaussian integrated luminosity functions defined by
the parameters given in Table \ref{GCLF_fit} in combination with the projected
areal density profiles previously discussed, we derive the cumulative
specific globular frequency $S_N=N_{tot} 10^{0.4(M_V+15)}$ for each galaxy within 
maximum galactocentric radii of 50 and 100 kpc. The first radius seems
a reasonable value in order to include most (if not all) the CGs
associated with a given low mass galaxy. The largest radius seem more
appropriate for massive galaxies  that have very extended GCS (see,
for example, \citealt{BFFDR06}, \citealt{H09a}).

In order to estimate $N_{tot}$ for NGC 4649 and 
NGC 3923, we have integrated the de Vaucouleurs law profiles obtained 
by fitting the GMOS+ACS profile presented in 
Fig. \ref{Perf_D}. In the particular case of NGC 3115, we have integrated
our de Vaucouleurs GMOS profile and we adopted the
WFPC2 photometry given by \citealt{LBHFG2001}  
to complete the counts in the innermost regions of the galaxy not 
covered by our frames. As those density profiles were obtained 
considering samples with 90 percent completeness magnitude cuts, these 
three values were corrected by calculating the fraction of 
the GCLF not included beyond these magnitude limits.

Absolute galaxy luminosities were obtained using the photometric 
values given in the NED database and the distance moduli listed 
in Table \ref{Tsample}. The total number of GCs, galaxy visual 
luminosity and inferred $S_N$ values are included in Table \ref{GCLF_fit}.
As an indicative of the uncertainty in our estimation of $N_{tot}$ and $S_N$, we list 
as error the difference between the results obtained by 
integrating a de Vaucouleurs and power law profiles.

For NGC 4649, there are two previously published values of $S_N$ in the 
literature. Both of them were calculated for 50 Kpc and they are 
from \cite{LPKHKG2008}, $S_N=3.8\pm0.4$, and that from  
\citealt{FFFBBGHSZG04}, $S_N=4.1\pm1$. A comparison with the numbers
listed in Table \ref{GCLF_fit} shows that our new $S_N$ value is very 
good agreement with those values.

For NGC 3923 \citealt{SPCVB2006} have obtained $S_N=5.6\pm1.3$ 
and $S_N=8.3\pm3.6$, for the same two radial limits listed in 
Table \ref{GCLF_fit}. However, these authors have used 
different distance and {\it M$_V$} for this target. Taking this 
into account, we have corrected 
their values and obtained $S_N=3.4$ and $S_N=5.05$, somewhat lower 
than those listed in Table \ref{GCLF_fit}.

For NGC 3115, the only previous estimation of the cumulative 
specific frequency is that from \citet{HH10}, $S_N=2.0$, which is 
very similar to our results for his galaxy.

\citet{RZ2004} and \citet{HH10} determine
a value of $S_N = 1.2 \pm 0.3$ for NGC 3379, in good agreement with 
our result.

\begin{table*}
\centering
\caption{Parameter for fitted Gaussian GCLF. The last four columns give 
the total number of GC and the Specific Frequency obtained considering 
limits of 50 and 100 Kpc for the GCS.
}
\label{GCLF_fit}
\scriptsize
\begin{tabular}{lccccccccccc}
\hline
\hline
\multicolumn{12}{c}{} \\
\multicolumn{1}{c}{\textbf{Galaxy}} &
\multicolumn{1}{c}{\textbf{TOM}} &
\multicolumn{1}{c}{\textbf{$\sigma$}} &
\multicolumn{1}{c}{\textbf{TOM$_{red}$}} &
\multicolumn{1}{c}{\textbf{$\sigma_{red}$}} &
\multicolumn{1}{c}{\textbf{TOM$_{blue}$}} &
\multicolumn{1}{c}{\textbf{$\sigma_{blue}$}} &
\multicolumn{1}{c}{\textbf{M$_V$}} &
\multicolumn{1}{c}{\textbf{Nt$_{50}$}} &
\multicolumn{1}{c}{\textbf{$S_{N50}$}} &
\multicolumn{1}{c}{\textbf{Nt$_{100}$}} &
\multicolumn{1}{c}{\textbf{$S_{N100}$}} \\
\hline \multicolumn{12}{c}{}\\
 NGC 4649    & 22.95 $\pm$ 0.05 &  1.35 $\pm$ 0.05  &  22.92 $\pm$ 0.07 & 1.28 $\pm$ 0.09 & 22.96 $\pm$ 0.07 & 1.28 $\pm$ 0.09 & -22.44 & 3310$\pm$130 & 3.5$\pm$0.1  & 4690$\pm$980 & 5.0$\pm$ 1 \\
 NGC 3923    & 23.84 (fixed)    &  1.42 $\pm$ 0.05  &       ...         &        ...      &        ...       &       ...       & -22.03 & 3019$\pm$70 & 4.1$\pm$0.1  & 4580$\pm$820 & 6.3$\pm$1.1\\ 
 NGC 3115    & 21.78 $\pm$ 0.08 &  0.91 $\pm$ 0.07  &  21.59 $\pm$ 0.07 & 0.71 $\pm$ 0.05 & 22.93 $\pm$ 0.15 & 1.04 $\pm$ 0.12 & -21.13 & 546$\pm$80  & 1.9$\pm$ 0.2  & 571$\pm$190  & 2.0$\pm$0.7\\
 NGC 3379    & 22.46 $\pm$ 0.37 &  0.85 $\pm$ 0.23  &       ...         &        ...      &        ...       &       ...       & -20.88 & 216$\pm$42  & 0.9$\pm$ 0.2  & 226$\pm$100  & 1.0$\pm$0.4\\
\multicolumn{12}{l}{}\\
\hline                                                                             
\multicolumn{12}{l}{}\\ 
\end{tabular}
\end{table*}

\section{Conclusions}
\label{resumen_conclusiones}

Deep GMOS multi-colour photometry  has been obtained for the GCS of
five early type galaxies. The primary results of this study, that
will be combined with the spectroscopic results for a further discussion in
a following paper, are:

\begin{itemize}

\item All the studied GCS show bimodal integrated
colour-distributions. Even the less clear case, NGC 524, appears 
bimodal when analyzed with the RMIX software. The adoption of a 
mass to luminosity ratio $(M/L)_B$= 8 indicates a minimum 
stellar mass close to $1.2 10^{11}M_{\odot}$ for
our galaxy sample, i.e., all these systems are above 
the minimum mass where bimodality becomes detectable as a 
common feature (\citealt{PACS06}).
 
\item The mean colour of the peaks of red GCs follow a well defined
luminosity-colour relation in the sense that these peaks become redder
with increasing galaxy luminosity. This is indicative of a larger GCs
chemical abundance scale for the most luminous galaxies which, being 
more massive, are also able of a more efficient enrichment. We note 
that the most massive system in this work (NGC 4649) lies outside 
the plane defined by GC systems in  the logarithmic 3-D space determined 
by galaxy stellar mass, projected stellar mass
density and GCs formation efficiency, possibly indicating a merger past
history as suggested by \citet{FVF09}.
The GC blue peaks, in turn, show no significant correlation or,
possibly, only a very mild one, with galaxy luminosity a situation already
noted in the literature \citep{PACS06}.

\item An empirical  {\it(g$'$-i$'$)} vs [Z/H]  relation was obtained 
based solely on GMOS data. A linear fit 
yields: $[Z/H]=3.51(\pm 0.21) (g'-i')_0  - 3.91 (\pm0.20)$, which provides
a good approximation, although the eventual presence of a second order
term at the low metallicity end will require the inclusion of more GCs.

\item A comparison of the colour spreads of both blue and red GCs, 
once transformed to metallicity, clearly indicates that both 
populations are widely different in terms of chemical abundance, a 
fact that when associated with the different spatial
distributions of these subpopulations, strongly indicates a distinct nature
for the two populations.

\item A number of resolved/marginally resolved GC candidates were detected
in all five galaxies. Some of them are spectroscopically confirmed as GCs in
the literature. In principle the GCs upper mass shows a dependence with
galaxy luminosity (e.g. \citealt{V2010}) and then the detection of
resolved clusters could be a more frequent situation in these 
systems. However, further analysis will require the inclusion of 
distance effects.

\item A population of Ultra Compact Dwarf (UCD) candidates
(resolved and unresolved ones with --15 $>$ {\it M$_I$} $>$ --12 ) were also
detected in all the galaxies. NGC 3923 shows the highest number of
associated objects, followed by NGC4649 and NGC 3115. This result 
indicates that UCDs are indeed a common feature
although the small size of our sample does not allow a 
definite conclusion in statistical terms.

\item This paper reports the first detection of the so called ``blue tilt''
in NGC 4649 using ground based observations. As a first approach we 
adopted a linear relation between GC colours and magnitudes although, as 
shown by \citet{H09b} on the basis of a more numerous GCs sample, 
the tilt may exhibit some degree of curvature that becomes more 
evident for the brightest clusters.
The tilt  translates into an approximate mass-metallicity relation, given
by {\it Z $\propto$ $M^{0.28\pm0.03}$}. This result is in excellent
agreement withe the mean trend obtained by \citet{CHWWCR09}.

\item All the GCS density profiles are reasonably well fitted by a
power law, although, in the case of NGC 4649, an R$^{1/4}$ dependence
produces a better fit in terms of the residuals. As found in other 
galaxies, the red GCs show a more concentrated spatial distribution
than the blue ones, again lending support to the idea that bimodality 
is a real and distinctive feature and not a mere result of a 
particular colour-metallicity relation.

\item We confirm the very low spatial concentration of the NGC 3923 GCS
found by \citet{SPCVB2006}. In contrast to these authors, we find a
significant difference of the areal density slopes of the red and blue GCs.

\item We have measured radial colour gradients for the blue and red GC
subpopulations separately, and find evidence for statistically
significant gradients in several galaxies out to several
effective radii (tens of kpc). The steeper ones are those present 
in the two S0 galaxies
included in the sample. These galaxies have smaller stellar masses than
the brightest ellipticals in this study showing shallower gradients. This
is in agreement with \citet{TNCCJM10} who find a non monotonic
behaviour of colour gradients with galaxy mass. A possible 
interpretation of this trend assumes that more massive galaxies 
may have experienced mergers that lead to a dilution of the chemical 
gradients.

\item Like in other studied systems, the red GC subpopulation has similar
colours to the galaxy halo stars in their inner region. This is consistent
with the idea that most of the galaxy luminosity in these regions comes from
a diffuse stellar population associated with these clusters (e.g., FFG07).

\item The TOM and $\sigma$ were determined for the GCLF of NGC 4649
and NGC 3115. In the case of NGC 3923, only $\sigma$ was fitted. This
  parameter and the density profiles were used to obtain the total GCs
populations, and the Specific Frequency $S_N$ of these galaxies. The
highest $S_N=4.5$ value was obtained for NGC 3923, followed by NGC 4649,
$S_N=3.6$. We have obtained new estimations of the NGC 3115 
and NGC 3379 specific frequencies; $S_N=2.0\pm07$ and $S_N=1.0\pm0.4$
respectively.
\end{itemize}

\vspace{1.cm}
Acknowledgments:
FF acknowledges financial support from the
Agencia de Promoci\'on Cient\'ifica y Tecnol\'ogica (BID AR
PICT 885). This work was partially supported by CONICET
funds through a PIP 2009/712 grant. SEZ acknowledges support for this 
work in part from the NSF grant AST-0406891
Data were based on observations obtained at
the Gemini Observatory, which is operated by the Association of 
Universities for Research in Astronomy, Inc., under a 
cooperative agreement with the NSF on behalf of
the Gemini partnership: the National Science Foundation
(United States), the Particle Physics and Astronomy Re-
search Council (United Kingdom), the National Research
Council (Canada), CONICYT (Chile), the Australian Re-
search Council (Australia), CNPq (Brazil), and 
Ministerio de Ciencia, Tecnolog\'ia e Innovaci\'on 
Tecnol\'ogica (Argentina). The Gemini program ID are 
GN-2007A-Q-37, GS-2007A-Q-49, GN-2007A-Q-37, GN-2002B-Q-25,   
GN-2003A-Q-22, GS-2004A-Q-9, GS-2004A-Q-9, GN-2003A-Q-22, 
GN-2001B-SV-104. This
research has made use of the NASA/IPAC Extragalactic
Database (NED), which is operated by the Jet Propulsion
Laboratory, Caltech, under contract with the National Aero-
nautics and Space Administration.

\label{lastpage}

\appendix
\section{Photometric data}

\begin{table*}
\centering
\begin{minipage}{140mm}
\caption{Photometric catalog of all the GC candidates in the five GCS studied in this work. The magnitudes are from psf fitting for the unresolved sources and from aperture photometry for the resolved ones. These magnitudes are extintion corrected as indicated in the text.
}
\label{GCphoT}
\scriptsize
\begin{tabular}{cccccccccccc}
\hline
\hline
\multicolumn{12}{c}{} \\
\multicolumn{1}{c}{\textbf{Object ID}} &
\multicolumn{1}{c}{\textbf{RA}} &
\multicolumn{1}{c}{\textbf{DEC}} &
\multicolumn{1}{c}{\textbf{$x_{ccd}$}} &
\multicolumn{1}{c}{\textbf{$y_{ccd}$}} &
\multicolumn{1}{c}{\textbf{$g'_0$}} &
\multicolumn{1}{c}{\textbf{$\sigma_{g'}$}} &
\multicolumn{1}{c}{\textbf{$r'_0$}} &
\multicolumn{1}{c}{\textbf{$\sigma_{r'}$}} &
\multicolumn{1}{c}{\textbf{$i'_0$}} &
\multicolumn{1}{c}{\textbf{$\sigma_{i'}$}} &
\multicolumn{1}{c}{\textbf{notes}}  \\
\hline \multicolumn{12}{c}{}\\
N4649GC2	&	12:43:33.74	&	11:34:02.5	&	1708.009	&	53.761	&	23.8009	&	0.023	&	23.1895	&	0.023	&	22.8163	&	0.024 &\\
N4649GC3	&	12:43:33.97	&	11:34:01.3	&	1685.015	&	62.249	&	22.7809	&	0.016	&	22.0355	&	0.008	&	21.6613	&	0.013 & \\
N4649GC4	&	12:43:29.01	&	11:34:01.2	&	2185.490	&	62.391	&	24.0299	&	0.013	&	23.5035	&	0.017	&	23.1393	&	0.025 & \\
N4649GC5	&	12:43:38.04	&	11:34:02.4	&	1273.952	&	55.273	&	23.7579	&	0.028	&	22.9325	&	0.016	&	22.5213	&	0.019 & \\
N4649GC6	&	12:43:41.82	&	11:34:02.4	&	 892.493	&	55.250	&	23.6949	&	0.015	&	23.2035	&	0.015	&	22.9523	&	0.025 & \\
N4649GC7	&	12:43:41.24	&	11:34:03.2	&	 950.987	&	50.004	&	23.9169	&	0.020	&	23.3875	&	0.019	&	23.3253	&	0.032 & \\
N4649GC8	&	12:43:34.64	&	11:33:58.9	&	1617.590	&	78.958	&	21.2469	&	0.008	&	20.7345	&	0.011	&	20.5343	&	0.013 & \\
N4649GC9	&	12:43:37.67	&	11:34:00.1	&	1310.816	&	70.643	&	22.5349	&	0.010	&	22.0455	&	0.017	&	21.7833	&	0.022 & \\
... & ... & ... & ... & ... & ... & ... & ... & ... & ... & ... & ...{\footnote{The complete photometry for 
GC candidates is available in the on-line version.
}}\\

\multicolumn{12}{l}{}\\
\hline                                                                             
\multicolumn{12}{l}{}\\ 
\end{tabular}
\end{minipage}
\end{table*}

\begin{table*}
\centering
\begin{minipage}{140mm}
\caption{The photometric data for all the UCD candidates in the five GCS studied in this work are given in Table \ref{UCDphotT}. The magnitudes are from psf fitting for the unresolved sources and from aperture photometry for the resolved ones. These magnitudes are extintion corrected as indicated in the text.}
\label{UCDphotT}
\scriptsize
\begin{tabular}{cccccccccccc}

\hline
\hline
\multicolumn{12}{c}{} \\
\multicolumn{1}{c}{\textbf{Object ID}} &
\multicolumn{1}{c}{\textbf{RA}} &
\multicolumn{1}{c}{\textbf{DEC}} &
\multicolumn{1}{c}{\textbf{$x_{ccd}$}} &
\multicolumn{1}{c}{\textbf{$y_{ccd}$}} &
\multicolumn{1}{c}{\textbf{$g'_0$}} &
\multicolumn{1}{c}{\textbf{$\sigma_{g'}$}} &
\multicolumn{1}{c}{\textbf{$r'_0$}} &
\multicolumn{1}{c}{\textbf{$\sigma_{r'}$}} &
\multicolumn{1}{c}{\textbf{$i'_0$}} &
\multicolumn{1}{c}{\textbf{$\sigma_{i'}$}} &
\multicolumn{1}{c}{\textbf{notes}}  \\
\hline \multicolumn{12}{c}{}\\
N4649UCD1	&	12:43:41.23	&	11:30:48.3	&	952.676	&	1388.875	&	18.4779	0.003	&	17.8995	&	0.007	&	17.7183	&	0.006	& ... \\
... & ... &   .... & ... & ... & ... &  ... &  ... & ... &  ... & ... & ...{\footnote{The complete photometry for 
UCD candidate  is available in the on-line version.}} \\
\multicolumn{12}{l}{}\\
\hline                                                                             
\multicolumn{12}{l}{}\\ 
\end{tabular}
\end{minipage}
\end{table*}


\begin{thebibliography}{}

\bibitem[\protect\citeauthoryear{Ajhar et al.}{1994}]{AT94} Ajhar E. A.; Tonry J. L., 1994, ApJ, 429, 557

\bibitem[\protect\citeauthoryear{Ashman \& Zepf}{1992}]{AZ92} Ashman K. M., ZepfS. E., 1992, ApJ, 384, 50

\bibitem[\protect\citeauthoryear{Ashman et al.}{1995}]{ACZ95} Ashman K. M., Conti A., Zepf S. E., 1995, AJ, 110, 1164

\bibitem[\protect\citeauthoryear{Bailin \& Harris}{2009}]{BH09} Bailin J., Harris W., 2009, ApJ, 695, 1082

\bibitem[\protect\citeauthoryear{Bassino et al.}{2006}]{BFFDR06} Bassino L., Faifer F., Forte J., Dirsch B., Richtler T., Geisler D., Schubert Y., 2006, A\&A, 451, 789

\bibitem[\protect\citeauthoryear{Beasley et al.}{2002}]{BBFSF02} Beasley M. A., Baugh C. M., Forbes D. A., Sharples R. M., \& Frenk C. S., 2002, MNRAS, 333, 383

\bibitem[\protect\citeauthoryear{Beasley et al.}{2004}]{BFBK2004} Beasley M., Forbes D., Brodie J., Kissler-Patig M., 2004, MNRAS, 347, 1150

\bibitem[\protect\citeauthoryear{Bergond et al.}{2006}]{BZRSR2006} Bergond G. Zepf S. E., Romanowsky A., Sharples R., Rhode K., 2006, A\&A, 448, 155

\bibitem[\protect\citeauthoryear{Bertin \&  Arnouts}{1996}]{BA96} Bertin E., Arnouts S., 1996, A\&AS, 117, 393

\bibitem[\protect\citeauthoryear{Blakeslee et al.}{2010}]{BCP10}Blakeslee J. P., Cantiello M., Peng E. W., 2010, ApJ, 710

\bibitem[\protect\citeauthoryear{Bridges et al.}{2006}]{BGSFFBZFHP2006} Bridges T., Gebhardt K., Sharples R., Faifer F. R., Forte J. C., Beasley M. A., Zepf S., Forbes D., Hanes D., Pierce M., 2006, MNRAS, 373, 157

\bibitem[\protect\citeauthoryear{Brodie \& Strader}{2006}]{BS06} Brodie, J.P., Strader, J. 2006, AR\&A, 44, 193

\bibitem[\protect\citeauthoryear{Buote \& Canizares}{1999}]{BC1999} Buote D. A., Canizares C. R., MNRAS, 1999, 298, 811

\bibitem[\protect\citeauthoryear{Couture et al.}{Couture, Harris \& Allwright}{1991}]{CHA91} Couture J., Harris W. E., Allwright J. W. B., 1991, ApJ, 372, 97

\bibitem[\protect\citeauthoryear{Cockcroft et al.}{2009}]{CHWWCR09} Cockcroft R., Harris W., Wehner E., Whitmore B., Rothberg B., 2009, AJ, 138, 758

\bibitem[\protect\citeauthoryear{Denicol\'o et al.}{2005}]{DTT2005} Denicol\'o G., Terlevich R., Terlevich E., et al., MNRAS, 2005, 358, 813

\bibitem[\protect\citeauthoryear{Dirsch et al.}{2003}]{DRGFBG03} Dirsch B., Richtler T., Geisler D., Forte J. C., Bassino L., Gieren W, AJ, 125, 1908

\bibitem[\protect\citeauthoryear{de Vaucouleurs et al.}{1991}]{dV91} de Vaucouleurs G., de Vaucouleurs A., Corwin H. G., Buta R. J., Paturel G.,  Fouque P., 1991, Third Reference Catalog of Bright Galaxies (Springer: New York)

\bibitem[\protect\citeauthoryear{Eggen, Lynden Bell \& Sandage}{Eggen et al.}{1962}]{ELBS} Eggen O., Lynden Bell D., Sandage A.R., 1962, ApJ, 136, 748 

\bibitem[\protect\citeauthoryear{Forbes et al.}{1997}]{FBG97} Forbes D. A., Brodie J. P., Grillmair C. J., 1997, AJ, 113, 1652

\bibitem[\protect\citeauthoryear{Forbes \& Forte}{2001}]{FF01} Forbes D. A., Forte J. C., 2001, MNRAS, 322, 257

\bibitem[\protect\citeauthoryear{Forbes et al.}{2004}]{FFFBBGHSZG04} Forbes D.A., Faifer F. R., Forte J. C., Bridges T., Beasley M., Gebhardt K., Hanes D., Sharples R., Zepf S., 2004, MNRAS, 355, 608

\bibitem[\protect\citeauthoryear{Forbes et al.}{2011}]{FSSRBF11} Forbes D., Spitler L., Strader J., Romanowsky A., Brodie J., Foster C., 2011, MNRAS, acepted

\bibitem[\protect\citeauthoryear{Forte et al.}{2001}]{FGOPG01} Forte J. C., Geisler D., Ostrov P. G., Piatti A. E., Gieren W., 2001, AJ, 121, 1992

\bibitem[\protect\citeauthoryear{Forte, Faifer \& Geisler}{Forte et al.}{2007}]{FFG07} Forte J. C., Faifer F., Geisler D., 2007, MNRAS, 382, 1947

\bibitem[\protect\citeauthoryear{Forte, Vega \& Faifer}{Forte et al.}{2009}]{FVF09} Forte J. C., Vega I., Faifer F. R., 2009, MNRAS, 397, 1003

\bibitem[\protect\citeauthoryear{Fukazawa et al.}{2006}]{FBPOK2006} Fukazawa Y., Botoya-Nonesa J. G., Pu J., Ohto A., Kawano N., 2006, ApJ, 636, 698

\bibitem[\protect\citeauthoryear{Fukugita et al.}{1995}]{FSI95} Fukugita M., Shimasaku M., Ichikawa T., 1995, PASP, 107, 945

\bibitem[\protect\citeauthoryear{Fukugita et al.}{1996}]{FIGSS96} Fukugita M., Ichikawa T., Gunn J., Shimasaku M., Schneider D., 1996, AJ, 111, 1748

\bibitem[\protect\citeauthoryear{Geller \& Huchra}{1983}]{GH83} Geller M. J., Huchra J. P., 1983, ApJS, 52, 61

\bibitem[\protect\citeauthoryear{Gregg et al.}{2004}]{GFMTC2004} Gregg M. D., Ferguson H. C., Minniti D., Tanvir N., Catchpole R., 2004, AJ, 127, 1441


\bibitem[\protect\citeauthoryear{Harris \& Hanes}{1985}]{HH85} Harris W.E., Hanes D. A., 1985, ApJ, 291, 147

\bibitem[\protect\citeauthoryear{Harris et al.}{2006}]{H06} Harris, W. E., et al. 2006, ApJ, 636, 90

\bibitem[\protect\citeauthoryear{Harris}{2009a}]{H09a} Harris W., 2009a, ApJ, 703, 939

\bibitem[\protect\citeauthoryear{Harris}{2009b}]{H09b} Harris W., 2009b,ApJ, 699,254

\bibitem[\protect\citeauthoryear{Harris \& Harris}{2011}]{HH10} Harris G., Harris W., 2011, MNRAS, 410, 2347
\bibitem[\protect\citeauthoryear{Harris et al.}{2010}]{HSFB10} Harris W., Spitler L., Forbes D., Bailin J., 2010, MNRAS, 401, 1965

\bibitem[\protect\citeauthoryear{Hook et al.}{2004}]{HJADMMC04} Hook I. M., Jorgensen I., Allington-Smith J. R., Davies R. L., Metcalfe N., Murowinski R. G., Crampton D., 2004, PASP, 116, 425

\bibitem[\protect\citeauthoryear{Jensen et al.}{2003}]{JTBTLRAB03} Jensen J., Tonry J., Barris B., Thompson R., Liu M., Rieke M., Ajhar E., Blakeslee J., 2003, ApJ, 583, 712

\bibitem[\protect\citeauthoryear{Jord\'an et al.}{2009}]{JPBCEFTW09} Jord\'an A., Peng E., Blakeslee J., C\^ot\'e P., Eyheramendy S., Ferrarese L., Mei S., Tonry J., West M., 2009, ApJS, 180, 54

\bibitem[\protect\citeauthoryear{Kissler-Patig}{2009}]{K2009} Kissler-Patig M., in Globular Clusters - Guides to Galaxies, Eso Astrophysics Symposia, Volume . ISBN 978-3-540-76960-6. Springer Berlin Heidelberg, 2009, p. 1

\bibitem[\protect\citeauthoryear{Kundu \& Whitmore}{1998}]{KW1998}  Kundu A., Whitmore B. C., 1998, AJ, 116, 2841

\bibitem[\protect\citeauthoryear{Kundu et al.}{1999}]{KWSMZA99} Kundu A., Whitmore B., Sparks W., Macchetto F., Zepf S., Ashman K., 1999, ApJ, 513, 733

\bibitem[\protect\citeauthoryear{Kundu \& Whitmore}{2001}]{KW2001}  Kundu A., Whitmore B. C., 2001, AJ, 121, 2950
 
\bibitem[\protect\citeauthoryear{Kundu}{2008}]{K08} Kundu A., 2008, AJ, 136, 1013

\bibitem[\protect\citeauthoryear{Kuntschner et al.}{2002}]{KZSWF2002} Kuntschner H., Ziegler B. L., Sharples R. M., Worthey G., Fricke K. J., 2002, A\&A, 395, 761

\bibitem[\protect\citeauthoryear{Larsen et al.}{2001}]{LBHFG2001} Larsen S. S.,Brodie J. P., Huchra J. P., Forbes D., Grillmair C. J., 2001, AJ, 121, 2974

\bibitem[\protect\citeauthoryear{Lauer \& Kormendy}{1986}]{LK86} Lauer T. R., Kormendy J., 1986, ApJ, 303, 1

\bibitem[\protect\citeauthoryear{Lee et al.}{2008}]{LPKHKG2008} Lee M. G., Park H. S., Kim E., Hwang H. S, Kim S. C., Geisler, 2008, ApJ, 682, 135

\bibitem[\protect\citeauthoryear{MacDonald}{2007}]{McD2007} MacDonald, P.D.M. 2007, documentation and code at http://www.math.mcmaster.ca/peter/mix/mix.html, Department of Mathematics and Statistics, McMaster University

\bibitem[\protect\citeauthoryear{Malin \& Carter}{1980}]{MC1980} Malin D. F., Carter D., 1980, Nature, 285, 643

\bibitem[\protect\citeauthoryear{Mendel et al.}{2007}]{MPF07} Mendel J. T., Proctor R. N., Forbes D. A., 2007, MNRAS, 379, 1618

\bibitem[\protect\citeauthoryear{Metcalfe et al.}{2001}]{MSCMcF2001} Metcalfe N., Shanks T., Campos A., McCracken H. J., Fong R., 2001, MNRAS,323, 795

\bibitem[\protect\citeauthoryear{Mieske et al.}{2006a}]{MHIJ2006} Mieske S., Hilker M., Infante L., Jord\'an A., 2006, AJ, 131, 2442

\bibitem[\protect\citeauthoryear{Mieske et al.}{2006b}]{Metal06} Mieske S., Jord\'an A., C\^ot\'e P., Kissler-Patig M., Peng E., Ferrarese L., Blakeslee J., Mei S., Merritt D., Tonry J., West M., 2006, ApJ, 653, 193


\bibitem[\protect\citeauthoryear{Mieske et al.}{2008}]{MHJI2008} Mieske S., Hilker M., Jord\'an A., Infante L., Kissler-Patig M., Rejkuba M., Richtler T., C\^ot\'e P., Baumgardt H., West M., Ferrarese L., Peng E., 2008, A\&A, 487, 921

\bibitem[\protect\citeauthoryear{Meylan et al.}{2001}]{M01}Meylan G., Sarajedini A., Jablonka P., Djorgovski S. G., Bridges T., Rich R. M., 2001, AJ, 122, 830

\bibitem[\protect\citeauthoryear{Mulchaey et al.}{2003}]{MDMB2003} Mulchaey J. S., Davis D. S., Mushotzky R. F., Burstein D., 2003, ApJS, 145, 39

	
\bibitem[\protect\citeauthoryear{Muratov \& Gnedin}{2010}]{MG10} Muratov A., Gnedin O., 2010, ApJ, 718, 1266

\bibitem[\protect\citeauthoryear{Navarro, Frenk \& White}{Navarro et al.}{1996}]{NFW96} Navarro, J.F., Frenk, C. S., White, S.D.M., 1996, ApJ, 462, 563

\bibitem[\protect\citeauthoryear{Norris et al.}{Norris, Sharples \& Kuntschner}{2006}]{NSK2006} Norris M. A., Sharples R. M., Kuntschner H., 2006, MNRAS, 367, 815

\bibitem[\protect\citeauthoryear{Norris et al.}{2008}]{NSBGFPFFBZH2008} Norris M., Sharples R., Bridges T., Gebhardt K., Forbes D., Proctor R., Faifer F. R., Forte J. C., Beasley M. A., Zepf S., Hanes D., 2008, MNRAS, 385, 40

\bibitem[\protect\citeauthoryear{Norris \& Kannappan}{2011}]{NK2011} Norris M., Kannappan S., 2011, MNRAS, acepted
\bibitem[\protect\citeauthoryear{Norris et al.}{2011}]{N2011} Norris M. et al., 2011, MNRAS, submmited

\bibitem[\protect\citeauthoryear{Oser et al.}{2010}]{OONJB10} Oser L., Ostriker J., Naab T., Johansson P., Burkert A., 2010, ApJ, 725, 2312

\bibitem[\protect\citeauthoryear{O'Sullivan, Forbes \& Ponman}{O'Sullivan et al.}{2001}]{OFP2001} O'Sullivan E., Forbes D., Ponman T., 2001, MNRAS, 328, 461

\bibitem[\protect\citeauthoryear{Ostrov et al.}{1998}]{OFG98} Ostrov P., Forte J. C., Geisler D., 1998, AJ, 116, 2854



\bibitem[\protect\citeauthoryear{Peng et al.}{2006}]{PACS06} Peng E. W., Jord\'an A., C\^ot\'e P., Blakeslee J. P., Ferrarese L., Mei S., West M. J., Merritt D., Milosavljevi\'c, M., Tonry J. L., 2006, ApJ, 639, 95

\bibitem[\protect\citeauthoryear{Peng et al.}{2008}]{Pet08} Peng E. W., Jord\'an A., C\^ot\'e P., Takamiya M., West M. J., Blakeslee J. P., Chin-Wei Chen, Ferrarese L., Mei S., Tonry J., West A., 2008, ApJ, 681, 197

\bibitem[\protect\citeauthoryear{Peng et al.}{2008}]{PJBMCFHMM09} Peng E. W., Jord\'an A., Blakeslee J. P., Mieske S.,  C\^ot\'e P., Ferrarese L.,  Harris W., Madrid J.,  Gerhardt R. Meurer, 2009, ApJ, 703, 42

\bibitem[\protect\citeauthoryear{Pierce et al.}{2006a}]{PBFBGFFZSHP2006} Pierce M., Beasley M., Forbes D.A., Bridges T., Gebhardt K., Faifer F. R., Forte J. C.,  Zepf S., Sharples R., Hanes D.,  2006a, MNRAS, 366, 1253

\bibitem[\protect\citeauthoryear{Pierce et al.}{2006b}]{PBFPBGFFZSH2006} Pierce M., Bridges T., Forbes D., Proctor R., Beasley M., Gebhardt K., Faifer F., Forte J. C., Zepf  S., Sharples R., Hanes D, 2006b, MNRAS, 368, 325

\bibitem[\protect\citeauthoryear{Puzia et al.}{2002}]{PZKHMG2002} Puzia T. H., Zepf S. E., Kissler-Patig M., Hilker M., Minniti D., Goudfrooij P., 2002, A\&A, 391, 453

\bibitem[\protect\citeauthoryear{Puzia et al.}{2004}]{PKTMSBRGH2004} Puzia T. H., Kissler-Patig M., Thomas D., Maraston C., Saglia1 R. P., Bender R., Richtler T., Goudfrooij P., Hempel M., A\&A, 415, 123 

\bibitem[\protect\citeauthoryear{Quinn}{1984}]{Q1984} Quinn P., 1984, ApJ, 279, 596

\bibitem[\protect\citeauthoryear{Randall et al.}{Randall, Sarazin \& Irwin}{2004}]{RSI2004} Randall S. W., Sarazin C. L., \& Irwin J. A., 2004, ApJ, 600, 729

\bibitem[\protect\citeauthoryear{Randall et al.}{Randall, Sarazin \& Irwin}{2006}]{RSI2006} Randall S. W., Sarazin C. L., Irwin J. A., 2006, ApJ,636, 200


\bibitem[\protect\citeauthoryear{Revnivtsev et al.}{2008}]{RCSFJ2008} Revnivtsev M., Churazov E., Sazonov S., Forman W., Jones C., 2008, A\&A, 490, 37

\bibitem[\protect\citeauthoryear{Rhode \& Zepf}{2004}]{RZ2004} Rhode K., Zepf S.,  2004, AJ, 127, 302

\bibitem[\protect\citeauthoryear{Robin et al.}{2003}]{RRD2003} Robin A.C., Reyl\'e C., Derri\`ere S., et al., 2003, A\&AS, 409, 523

\bibitem[\protect\citeauthoryear{Sarazin et al.}{2003}]{SKISBR2003} Sarazin C., Kundu A., Irwin J., Sivakoff G., Blanton E., Randall S., 2003, ApJ, 595, 743

\bibitem[\protect\citeauthoryear{Schweizer \& Seitzer}{1992}]{SchwS92} Schweizer F., Seitzer P., 1992, AJ, 104, 1039

\bibitem[\protect\citeauthoryear{Schlegel et al.}{Schlegel, Finkbeiner \& Davis}{1998}]{SFD98} Schlegel D.J., Finkbeiner D.P., Davis M., 1998, ApJ, 500, 525

\bibitem[\protect\citeauthoryear{Searle \& Zinn}{1978}]{SZ78} Searle L., Zinn R., 1978, ApJ, 205, 357

\bibitem[\protect\citeauthoryear{Sikkema et al.}{2006}]{SPCVB2006} Sikkema G., Peletier R. F., Carter D., Valentijn E. A., Balcells M., 2006, A\&A, 458, 53

\bibitem[\protect\citeauthoryear{Sil'chenko et al.}{Sil'chenko, Afanasiev \& Vlasyuk}{1992}]{SAV92} Sil'chenko O. K., Afanasiev V. L., Vlasyuk V. V., 1992, AZh, 69, 1121
\bibitem[\protect\citeauthoryear{Sil'chenko}{2000}]{S2000} Sil'chenko O. K., 2000, AJ, 120, 741

\bibitem[\protect\citeauthoryear{Sinnott et al.}{2010}]{SHAHW10} Sinnott B., Hou A., Anderson R., Harris W., Woodley K., 2010, AJ, 140, 2101
            
\bibitem[\protect\citeauthoryear{Smith et al.}{2002}]{S2002} Smith el al., 2002, AJ, 123, 2121

\bibitem[\protect\citeauthoryear{Spitler et al.}{2006}]{SLSBFB06} Spitler, L. R., Larsen, S. S., Strader, J., Brodie, J. P., Forbes, D. A., \& Beasley, M. A. 2006, AJ, 132, 1593

\bibitem[\protect\citeauthoryear{Spitler et al.}{2010}]{S2010} Spitler L, 2010, MNRAS, 406, 1125

\bibitem[\protect\citeauthoryear{Stetson}{1987}]{S87} Stetson P. B., 1987, PASP, 99, 191

\bibitem[\protect\citeauthoryear{Strader et al.}{2006}]{SBSB06} Strader J., Brodie, J. P., Spitler L., Beasley M., AJ, 132, 2333

\bibitem[\protect\citeauthoryear{Strader \& Smith}{2008}]{SS08} Strader J.,  Smith G.H., 2008, AJ, 136, 1828

\bibitem[\protect\citeauthoryear{Terlevich \& Forbes}{2002}]{TF2002} Terlevich A., Forbes D., 2002, MNRAS, 330, 547

\bibitem[\protect\citeauthoryear{Tonry et al.}{2001}]{TDB2001} Tonry J. L., Dressler A., Blakeslee J. P., et al. 2001, ApJ, 546, 681

\bibitem[\protect\citeauthoryear{Thomas et al.}{2004}]{TMK04} Thomas D., Maraston C., Korn A., 2004, MNRAS, 351, 19

\bibitem[\protect\citeauthoryear{Thomas et al.}{2005}]{TMBO05} Thomas D., Maraston C., Bender R., Mendes de Oliveira C., 2005, ApJ, 621, 673

\bibitem[\protect\citeauthoryear{Thomson \& Wright}{1990}]{TW1990} Thomson R., Wright A., 1990, MNRAS, 247, 122

\bibitem[\protect\citeauthoryear{Tortora et al.}{2010}]{TNCCJM10} Tortora C., Napolitano N. R., Cardone V. F., Capaccioli M., Jetzer Ph., Molinaro R., 2010, MNRAS, 407, 144

\bibitem[\protect\citeauthoryear{Villegas et al.}{2010}]{V2010} Villegas et al., 2010, ApJ, 717, 603

\bibitem[\protect\citeauthoryear{Waters et al.}{2009}]{WZLB09} Waters C., Zepf S., Lauer T., Baltz E., 2009,ApJ, 693, 463

\bibitem[\protect\citeauthoryear{Wehner \& Harris}{2007}]{WH2007} Wehner E., Harris W., 2007, ApJ, 668L, 35

\bibitem[\protect\citeauthoryear{Wehner et al.}{2008}]{WHWRW08} Wehner E., Harris W., Whitmore B., Rothberg B., Woodley K., 2008, ApJ, 681, 1233

\bibitem[\protect\citeauthoryear{White et al.}{White, Keel \& Conselice}{2000}]{WKC2000} White R. E., Keel W. C., Conselice C. J., 2000, ApJ, 542, 761

\bibitem[\protect\citeauthoryear{Whitlock et al.}{Whitlock, Forbes \& Beasley}{2003}]{WFB2003}  Whitlock S., Forbes D. A., Beasley M. A., 2003,  2003, MNRAS, 345, 949

\bibitem[\protect\citeauthoryear{Zepf et al.}{Zepf, Geisler \& Ashman}{1994}]{ZGA94} Zepf S.E., Geisler D., Ashman K. M., 1994, ApJ, 435, 117

\bibitem[\protect\citeauthoryear{Zepf et al.}{Zepf, Ashman \& Geisler}{1995}]{ZAG95} Zepf S.E., Ashman K. M., Geisler D., 1995, ApJ, 443, 570
\end{thebibliography}
\end{document}